\documentclass[11pt]{report}
\usepackage{amsmath}
\usepackage{sw20mitt}
\usepackage{graphicx}
\usepackage{amsfonts}
\usepackage{amssymb}
\newtheorem{theorem}{Theorem}
\newtheorem{acknowledgement}[theorem]{Acknowledgement}

\begin{document}

\title{Studies in the Theory of Quantum Games}
\author{Azhar Iqbal}
\prevdegrees{B.Sc. (Hons), University of Sheffield, UK, 1995}
\department{Department of Electronics }
\thisdegree{Doctor of Philosophy}
\university{Quaid-i-Azam University, Islamabad, Pakistan}
\degreemonth{September}
\degreeyear{2004}
\date{3rd September 2004}
\chairmanname{Professor Dr. S. Azhar Abbas Rizvi}
\chairmantitle{Head of Department}
\super{Dr. Abdul Hameed Toor}
\supertitle{Associate Professor}
\maketitle
\tableofcontents
\listoffigures
\begin{abstract}
Theory of quantum games is a new area of investigation that has gone through
rapid development during the last few years. Initial motivation for playing
games, in the quantum world, comes from the possibility of re-formulating
quantum communication protocols, and algorithms, in terms of games between
quantum and classical players. The possibility led to the view that quantum
games have a potential to provide helpful insight into working of quantum
algorithms, and even in finding new ones. This thesis analyzes and compares
some interesting games when played classically and quantum mechanically. A
large part of the thesis concerns investigations into a refinement notion of
the Nash equilibrium concept. The refinement, called an evolutionarily stable
strategy (ESS), was originally introduced in 1970s by mathematical biologists
to model an evolving population using techniques borrowed from game theory.
Analysis is developed around a situation when quantization changes ESSs
without affecting corresponding Nash equilibria. Effects of quantization on
solution-concepts other than Nash equilibrium are presented and discussed. For
this purpose the notions of value of coalition, backwards-induction outcome,
and subgame-perfect outcome are selected. Repeated games are known to have
different information structure than one-shot games. Investigation is
presented into a possible way where quantization changes the outcome of a
repeated game. Lastly, two new suggestions are put forward to play quantum
versions of classical matrix games. The first one uses the association of De
Broglie's waves, with travelling material objects, as a resource for playing a
quantum game. The second suggestion concerns an EPR type setting exploiting
directly the correlations in Bell's inequalities to play a bi-matrix game.
\end{abstract}

\vspace*{3.35in}

\begin{center}
{\Large Dedication}
\end{center}

\textit{This thesis is dedicated to my wife Ayesha and son Emmad. Both
suffered and were kept waiting for long hours during the years of research. I
greatly appreciate Ayesha's warm and continuous support and Emmad's patience,
without which the task was plainly impossible. The thesis is also dedicated to
the loving memory of my late parents.}

\begin{description}
\item \bigskip
\end{description}

\newpage

This thesis is based on the following publications:

\begin{itemize}
\item  A. Iqbal and A. H. Toor, \textit{Evolutionarily stable strategies in
quantum games}. Physics Letters, A \textbf{280}/5-6, pp 249-256 (2001).

\item  A. Iqbal and A. H. Toor, \textit{Entanglement and dynamic stability of
Nash equilibria in a symmetric quantum game}. Physics Letters, A
\textbf{286}/4, pp 245-250 (2001).

\item  A. Iqbal and A. H. Toor, \textit{Quantum mechanics gives stability to a
Nash equilibrium}. Physical Review, A \textbf{65}, 022306 (2002). {\small This
article is also selected to be reproduced in the February 1, 2002 issue of the
Virtual Journal of Biological Physics Research: http://www.vjbio.org.}

\item  A. Iqbal and A. H. Toor, \textit{Quantum cooperative games}. Physics
Letters, A \textbf{293}/3-4 pp 103-108 (2002).

\item  A. Iqbal and A. H. Toor, \textit{Darwinism in quantum systems?} Physics
Letters, A \textbf{294}/5-6 pp 261-270 (2002).

\item  A. Iqbal and A. H. Toor, \textit{Backwards-induction outcome in a
quantum game}. Physical Review, A \textbf{65}, 052328 (2002). {\small This
article is also selected to be reproduced in the May 2002 issue of the Virtual
Journal of Quantum Information: http://www.vjquantuminfo.org.}

\item  A. Iqbal and A. H. Toor, \textit{Quantum repeated games}. Physics
Letters, A \textbf{300}/6, pp 537-542 (2002).

\item  A. Iqbal and A. H. Toor, \textit{Stability of mixed Nash equilibria in
symmetric quantum games}. Communications in Theoretical Physics, Vol.
\textbf{42}, No. 3, pp 335-338 (2004).

\item  A. Iqbal, \textit{Quantum games with a multi-slit electron diffraction
set-up.} Nuovo Cimento B, Vol. \textbf{118}, Issue 5, pp 463-468 (2003).

\item  A. Iqbal, \textit{Quantum correlations and Nash equilibria of a
bi-matrix game}. Journal of Physics A: Mathematical and General \textbf{37},
L353-L359 (2004).
\end{itemize}

\pagebreak 

\chapter{Introduction}

Game theory \cite{Rasmusen89} is a branch of mathematics that presents formal
analysis of the interaction among a group of rational players. The players
have choices available to them, so as to select particular course of action.
They are supposed to behave strategically and are motivated to increase their
utilities that depend on the collective course of action.

Modern game theory started with the work of John von Neumann and Oskar
Morgenstern \cite{Neumann} in 1930s. During the same years von Neumann
\cite{NeumannQM} also made important contributions in quantum mechanics, a
branch of physics developed in 1920s to understand the microworld of atoms and
molecules. However, game theory and quantum mechanics were developed as
separate fields with apparently different domains of applications.

The early years of development in both of these fields could not find some
common ground, or physical situation, that could motivate an interplay between
the two fields. More than fifty years afterwards, quantum computation
\cite{Feynman1, Deutsch,Shor,Shor1,Grover,Grover1}\ was developed in 1980s as
a new field of research that combined elements from quantum mechanics and the
theory of computation \cite{Turing}. Computer science extensively uses the
theory of information and communication \cite{Shannon}. Quantum computation
motivated the development of quantum information \cite{Schumacher}; thus
providing an environment where the two distinct interests of von Neumann, i.e.
quantum mechanics and game theory, could be shown to have some sort of
interplay. Certain quantum communication protocols, and algorithms, were
reformulated in the language of game theory
\cite{Wiesner,Goldenberg,Vaidman,Ekert,Gisin,WernerRF}. It was not long before
the first systematic procedures \cite{MeyerDavid,Eisert} were proposed to
quantize well-known classical games \cite{Rasmusen89}.

Classical bits are the entities that are used to physically implement
classical information. Quantum information \cite{Nielsen}, on the other hand,
uses quantum bits (qubits) for its physical implementation. It is known that
the problems of classical game theory can be translated into physical set-ups
that use classical bits. It immediately motivates the question of how games
can be transformed when implemented with qubits. Like it is the case with
quantum information, it helps to quantize classical games when qubits, instead
of classical bits, are used in physical set-ups to play games.

This thesis follows a particular approach in the theory of quantum games. The
thesis builds up on proposed convincing procedures telling how to quantize
well-known games from the classical game theory. A large part of this thesis
concerns studying the concept of an Evolutionarily Stable Strategy (ESS)
\cite{Smith} from mathematical population biology within the context of
quantum games. The thesis argues that importing a population setting towards
quantum games is not unusual though it may give such an impression. It is
observed that even John Nash \cite{JohnNash,JohnNash1} had a population
setting in his mind when he introduced his solution concept of the Nash
equilibrium for non-cooperative games. The study of evolutionary stability in
quantum games is presented with the view that importing the concept of an ESS,
and its associated population setting, to quantum games is natural to an equal
extent as it is to study Nash equilibrium in quantum games.

Game theory \cite{Rasmusen89} also offers solution concepts that are relevant
to certain types of games. The notions of value of coalition,
backwards-induction outcome and subgame-perfect outcome present a few
examples. The types of games for which these concepts are appropriate are
known to be the cooperative, the sequential (with moves made in order) and
repeated (with moves made simultaneously in one stage) games, respectively. To
show how quantization affects solutions of these games, the relevant solution
concepts are investigated in relation to quantization of these games. This
study shows that quantum versions of these games may have outcomes that are
often extraordinary and sometimes may even be counter-intuitive, from the
point of view of classical game theory.

Motivated by our preferred approach towards quantum games, i.e. to rely on
proposed convincing procedures to quantize known games from the classical game
theory, two new suggestions are put forward about quantum versions of
two-player two-strategy games. The first suggestion presents a set-up that
uses the association of De Broglie waves with travelling material objects to
play a quantum version of a two-player two-strategy game. The second
suggestion uses an EPR type setting in which spatially separated players make
measurements along chosen directions to play a game.

The concluding chapter collects together main results obtained in the thesis.

\chapter{Elements of game theory}

\section{Introduction}

Many decision making problems in sociology, politics and economics deal with
situations in which the results depend not only on the action of one
individual but also on the actions of others. Game theory is a branch of
mathematics which is used in modelling situations in which many individuals
with conflicting interests interact, such that the results depend on the
actions of all the participants. It is considered a formal way to analyze
interaction among a group of individuals who behave rationally and
strategically. The participants in a game strive to maximize their (expected)
utilities by choosing particular courses of action. Because the actions of the
others matter, a player's final utility depends on the profile of courses of
action chosen by all the individuals. A game deals with the following concepts:

\begin{itemize}
\item \emph{Players}. These are the individuals who compete in the game. A
player can be an individual or a set of individuals.

\item  A \emph{move}\textit{\ }will be a player's action.

\item  A player's (pure) \emph{strategy}\textit{\ }will be a rule (or
function) that associates a player's move with the information available to
her at the time when she decides which move to choose.

\item  A player's \emph{mixed strategy} is a probability measure on the
player's space of pure strategies.

\item \emph{Payoffs} are real numbers representing the players' utilities.
\end{itemize}

Although first attempts to analyze such problems are apparently rather old
\cite{Cournot}, modern game theory started with the work of John von Neumann
and Oskar Morgenstern who wrote the book \textit{Theory of Games and Economic
Behaviour }\cite{Neumann}. Game theory is now widely used in research in
diverse areas ranging from economics, social science, to evolutionary biology
and population dynamics.

\section{Representations of games}

There are different ways to represent a strategic interaction between players.
In game theory \cite{Rasmusen89} two representations are well known:

\subsection{Normal form}

A \textit{normal} (strategic) \textit{form} of a game consists of:

\begin{enumerate}
\item  A finite set of $N$ agents or players

\item  Strategy sets $S_{1},S_{2},...S_{N}$ for the $N$ players

\item  Payoff functions $P_{i}$, $i=1,2,...N$, are mappings from the set
$S_{1}\times S_{2}\times...\times S_{N}$ to the set of real numbers
$\mathbf{R}$.
\end{enumerate}

The set $S_{1}\times S_{2}\times...\times S_{N}$ is called the strategy space
$S$. A member $s\in S$ is known as a strategy profile with $s=(s_{1}%
,s_{2},...s_{N})$ and $s_{i}\in S_{i}$.

\subsection{Extensive form}

The \textit{extensive form} of a game is a complete description of:

\begin{enumerate}
\item  the set of players

\item  who moves when and what their choices are

\item  the players' payoffs as a function of the choices that are made

\item  what players know when they move
\end{enumerate}

The extensive form of a game, as opposed to the normal (or strategic) form,
provides a more appropriate framework for the analysis of strategic
interactions that involve sequential moves. It gives a richer specification of
a strategic interaction by specifying who moves when, doing what and with what
information. The easiest way to represent an extensive form game is to use a
\textit{game tree}, which is multi-person generalization of a \textit{decision
tree }\cite{Rasmusen89}.

\section{Information structure in games}

The information at the disposal of a player, when she has to select a move, is
described by the \textit{information structure }in the game\textit{. }Based on
this structure games can usually be put in either one of the following two
broad classes, which also form the two main branches of game theory.

\subsection{Cooperative games}

In cooperative games the players are allowed to form \textit{binding
agreements. }These are restrictions on the possible actions decided by two or
more players. To be binding an agreements usually requires an outside
authority that can monitor the agreement at no cost and impose on violators
sanctions so severe that cheating is prevented. For players in a binding
agreement there is a strong incentive to work together to receive the largest
total payoff. The agreements may include, for example, \textit{commitments}
and \textit{threats}.

\subsection{Non-cooperative games}

In non-cooperative games the players may not form binding agreements. Neither
do the players cooperate nor do they enter into negotiation for achieving a
common course of action. However the players know how the actions, their own
and the actions of the other players, will determine the payoffs of every player.

\section{Matrix games}

One way to describe a game is to list the players participating in the game,
and to list the alternative choices or moves available to each player. In the
case of a two-player game, the moves of the first player form the rows, and
the moves of the second player the columns of a \textit{matrix}. The entries
in the matrix are two numbers representing the payoff to the first and second
player, respectively. Such a description of a game makes possible to
completely represent the players' payoffs by a matrix. In game theory these
games are recognized as \textit{matrix games. }The example below is a matrix
game between two players:%

\begin{equation}%
\begin{array}
[c]{c}%
\text{Alice}%
\end{array}%
\begin{array}
[c]{c}%
S_{1}\\
S_{2}\\
...\\
S_{N}%
\end{array}
\overset{\overset{%
\begin{array}
[c]{c}%
\text{Bob}%
\end{array}
}{%
\begin{array}
[c]{ccccccc}%
S_{1} &  & S_{2} &  & ... &  & S_{N}%
\end{array}
}}{\left(
\begin{array}
[c]{cccc}%
(a_{11},b_{11}) & (a_{12},b_{12}) & ... & (a_{1N},b_{1N})\\
(a_{21},b_{21}) & (a_{22},b_{22}) & ... & (a_{2N},b_{2N})\\
... & ... & ... & ...\\
(a_{N1},b_{N1}) & (a_{N2},b_{N2}) & ... & (a_{NN},b_{NN})
\end{array}
\right)  }%
\end{equation}

\subsection{Constant-sum games}

In a \textit{constant-sum game}, the sum of all players' payoffs is the same
for any outcome. Hence, a gain for one participant is always at the expense of
another, such as in most sporting events.

\subsection{Zero-sum game}

A \textit{zero-sum game} is a special case of a constant sum game in which all
outcomes involve a sum of all player's payoffs of $0$. Since payoffs can
always be normalized, a constant sum game may be represented as (and is
equivalent to) a zero-sum game.

\subsection{Bi-matrix games}

A class of games that have attracted much attention because of the relative
simplicity of their mathematical analysis involve two players Alice and Bob.
Each player has his own payoff matrix written as $a_{ij}$ and $b_{ij}$,
respectively. Games of this kind are called \textit{bi-matrix games.}

\section{Examples of matrix games}

The following examples describe some well-known matrix games.

\subsection{Prisoners' Dilemma}

The most popular bi-matrix game is the so-called the \textit{Prisoners'
Dilemma} (PD) describing the following situation:

\begin{itemize}
\item  Two criminals are arrested after having committed a crime together and
wait for their trial.

\item  Each suspect is placed in a separate cell and offered the opportunity
to confess to the crime.

\item  Each suspect may choose between two strategies namely confessing ($D$)
and not confessing ($C$), where $C$ and $D$ stand for cooperation and defection.

\item  If neither suspect confesses, i.e. $(C,C),$ they go free, and split the
proceeds of their crime which we represent by $3$ units of payoff for each suspect.

\item  However, if one prisoner confesses ($D$) and the other does not ($C$),
the prisoner who confesses testifies against the other in exchange for going
free and gets the entire $5$ units of payoff, while the prisoner who did not
confess goes to prison and gets nothing.

\item  If both prisoners confess, i.e. ($D,D$), then both are given a reduced
term, but both are convicted, which we represent by giving each $1$ unit of
payoff: better than having the other prisoner confess, but not so good as
going free.
\end{itemize}

The game can be represented by the following matrix of payoffs:%

\begin{equation}%
\begin{array}
[c]{c}%
\text{Alice}%
\end{array}%
\begin{array}
[c]{c}%
C\\
D
\end{array}
\overset{\overset{%
\begin{array}
[c]{c}%
\text{Bob}%
\end{array}
}{%
\begin{array}
[c]{cc}%
C & D
\end{array}
}}{\left(
\begin{array}
[c]{cc}%
(3,3) & (0,5)\\
(5,0) & (1,1)
\end{array}
\right)  } \label{PDmatrix1}%
\end{equation}
where the first and the second entry correspond to Alice's and Bob's payoff, respectively.

For either choice of the opponent it is hence advantageous to defect ($D$). On
the other hand, if both defect ($D,D$) the payoff remains less than in the
case when both cooperate ($C,C$). This is the origin of dilemma.

A generalized matrix for the PD is given as:%

\begin{equation}%
\begin{array}
[c]{c}%
\text{Alice}%
\end{array}%
\begin{array}
[c]{c}%
C\\
D
\end{array}
\overset{\overset{%
\begin{array}
[c]{c}%
\text{Bob}%
\end{array}
}{%
\begin{array}
[c]{cc}%
C & D
\end{array}
}}{\left(
\begin{array}
[c]{cc}%
(r,r) & (s,t)\\
(t,s) & (u,u)
\end{array}
\right)  } \label{PDmatrix}%
\end{equation}
where $s<u<r<t$.

\subsection{Battle of Sexes}

Battle of Sexes (BoS) is a bi-matrix game that can be described as follows:

\begin{itemize}
\item  Alice and Bob agree to meet in the evening, but cannot recall if they
will be attending the opera or a boxing match.

\item  Alice prefers the opera and Bob prefers the boxing match.

\item  Both prefer being together to being apart.
\end{itemize}

Thus, while both parties prefer to find themselves at the same place, Alice
and Bob cannot agree which event to attend. The game has the following matrix representation:%

\begin{equation}%
\begin{array}
[c]{c}%
\text{Alice}%
\end{array}%
\begin{array}
[c]{c}%
S_{1}\\
S_{2}%
\end{array}
\overset{\overset{%
\begin{array}
[c]{c}%
\text{Bob}%
\end{array}
}{%
\begin{array}
[c]{cc}%
S_{1} & S_{2}%
\end{array}
}}{\left(
\begin{array}
[c]{cc}%
(\alpha,\beta) & (\gamma,\gamma)\\
(\gamma,\gamma) & (\beta,\alpha)
\end{array}
\right)  } \label{BoSMatrix}%
\end{equation}
where $\alpha>\beta>\gamma$.

\subsection{Matching Pennies}

Matching Pennies is a zero-sum game with two players Alice and Bob. Each shows
either heads or tails from a coin. If both are heads or both are tails then
Alice wins, otherwise Bob wins. The payoff matrix is given as%

\begin{equation}%
\begin{array}
[c]{c}%
\text{Alice}%
\end{array}%
\begin{array}
[c]{c}%
H\\
T
\end{array}
\overset{\overset{%
\begin{array}
[c]{c}%
\text{Bob}%
\end{array}
}{%
\begin{array}
[c]{cc}%
H & T
\end{array}
}}{\left(
\begin{array}
[c]{cc}%
(1,-1) & (-1,1)\\
(-1,1) & (1,-1)
\end{array}
\right)  }%
\end{equation}
with a winner getting a reward of $1$ against the loser getting $-1$.

\subsection{Rock-Scissors-Paper}

Two children, Alice and Bob, simultaneously make one of three symbols with
their fists - a rock, paper, or scissors (RSP). Simple rules of ``rock breaks
scissors, scissors cut paper, and paper covers rock'' dictate which symbol
beats the other. If both symbols are the same, the game is a tie:%

\begin{equation}%
\begin{array}
[c]{c}%
\text{Alice}%
\end{array}%
\begin{array}
[c]{c}%
R\\
S\\
P
\end{array}
\overset{\overset{%
\begin{array}
[c]{c}%
\text{Bob}%
\end{array}
}{%
\begin{array}
[c]{ccc}%
R & S & P
\end{array}
}}{\left(
\begin{array}
[c]{ccc}%
0 & 1 & -1\\
-1 & 0 & 1\\
1 & -1 & 0
\end{array}
\right)  }%
\end{equation}

\section{Solution concepts}

Solving a game means finding a set of moves for the players which represent
their rational choices. Unlike in other fields, the notion of a ``solution''
is more tenuous in game theory. In game theory a solution is generally thought
of as a systematic description\textit{ }of the outcomes that may emerge during
the play of a game.

\subsection{Rational ``solution'' of Prisoners' Dilemma}

For the bi-matrix PD it is self-evident how an intelligent individual should
behave. No matter what a suspect believes his partner is going to do, it is
always best to confess ($D$):

\begin{itemize}
\item  If the partner in the other cell is not confessing ($C$), it is
possible to get $5$ instead of $3$.

\item  If the partner in the other cell is confessing ($D$), it is possible to
get $1$ instead of $0$.
\end{itemize}

Yet the pursuit of individually sensible behavior results in each player
getting only $1$ unit of payoff, much less than the $3$ units each that they
would get if neither confessed ($C,C$). This conflict between the pursuit of
individual goals and the common good is at the heart of many game theoretic
problems. For PD the rational choice for both players is to defect.

\subsection{Nash equilibrium}

A \textit{Nash equilibrium (}NE), named after John Nash, is a set of
strategies, one for each player, such that no player has an incentive to
unilaterally change her action. Players are in equilibrium if a change in
strategies by any one of them would lead that player to earn less than if she
remained with her current strategy.

The implicit assumption behind the concept of a NE is that players make their
choices simultaneously and independently. This idea also assumes that each
player participating in a game behaves rational and searches to maximize
his/her own payoff. A strategy profile $s=(s_{1}^{\ast},s_{2}^{\ast}%
,...s_{N}^{\ast})$ is a NE if none of them is left with a motivation to
deviate unilaterally. Suppose $P_{i}$ is the $i$th-player's payoff then the
following condition defines the NE:%

\begin{equation}
P_{i}(s_{1}^{\ast},s_{2}^{\ast},...s_{i-1}^{\ast},s_{i}^{\ast},s_{i+1}^{\ast
}...,s_{N}^{\ast})\geq P_{i}(s_{1}^{\ast},s_{2}^{\ast},...s_{i-1}^{\ast}%
,s_{i},s_{i+1}^{\ast}...,s_{N}^{\ast})
\end{equation}
When the $N$ players are playing the strategy profile $s=(s_{1}^{\ast}%
,s_{2}^{\ast},...s_{N}^{\ast})$ the $i$th player's decision to play $s_{i} $
instead of $s_{i}^{\ast}$ cannot increase his/her payoff. A NE thus defines a
set of strategies that represents a \textit{best choice} for each single
player if all the other players take their best decisions too.

The well-known \emph{Nash Theorem} \cite{JohnNash} in game theory guarantees
the existence of a set of mixed strategies for finite non-cooperative games of
two or more players in which no player can improve his payoff by unilaterally
changing his/her strategy.

\subsection{Nash equilibrium in the Prisoners' Dilemma}

Let Alice play $C$ with probability $p$ and play $D$ with probability $(1-p)$.
Similarly, let Bob play $C$ with probability $q$ and play $D$ with probability
$(1-q)$. The players' payoffs for the PD matrix (\ref{PDmatrix1}) are%

\begin{align}
P_{A}(p,q)  &  =pq(3)+p(1-q)(0)+(1-p)q(5)+(1-p)(1-q)(1)\nonumber\\
&  =-p+4q-pq+1\\
P_{B}(p,q)  &  =pq(3)+p(1-q)(5)+(1-p)q(0)+(1-p)(1-q)(1)\nonumber\\
&  =4p-q-pq+1
\end{align}
The inequalities defining the NE in PD can be written as%

\begin{align}
P_{A}(p^{\ast},q^{\ast})-P_{A}(p,q^{\ast})  &  =-(p^{\ast}-p)(1+q^{\ast}%
)\geq0\nonumber\\
P_{B}(p^{\ast},q^{\ast})-P_{B}(p^{\ast},q)  &  =-(q^{\ast}-q)(1+p^{\ast})\geq0
\end{align}
which produces a unique NE in the PD: $p^{\ast}=q^{\ast}=0$. The NE
corresponds to both players playing the pure strategy $D$.

\subsection{Nash equilibrium in the Battle of Sexes}

Similar to the case of PD we assume that the numbers $p,q\in\lbrack0,1]$ are
the probabilities with which Alice and Bob play the strategy $S_{1}$,
respectively. They then play $S_{2}$ with the probabilities $(1-p)$ and
$(1-q)$, respectively. Players' payoffs for the BoS matrix (\ref{BoSMatrix})
are \cite{Marinatto1}:%

\begin{align}
P_{A}(p,q)  &  =p\left[  q(\alpha-2\gamma+\beta)+\gamma-\beta\right]
+q(\gamma-\beta)+\beta\nonumber\\
P_{A}(p,q)  &  =q\left[  p(\alpha-2\gamma+\beta)+\gamma-\alpha\right]
+p(\gamma-\alpha)+\alpha
\end{align}
The NE $(p^{\ast},q^{\ast})$ is then found from the inequalities:%

\begin{align}
P_{A}(p^{\ast},q^{\ast})-P_{A}(p,q^{\ast})  &  =(p^{\ast}-p)\left[  q^{\ast
}(\alpha+\beta-2\gamma)-\beta+\gamma\right]  \geq0\nonumber\\
P_{B}(p^{\ast},q^{\ast})-P_{B}(p^{\ast},q)  &  =(q^{\ast}-q)\left[  p^{\ast
}(\alpha+\beta-2\gamma)-\alpha+\gamma\right]  \geq0
\end{align}
Three NE arise:

\subsubsection{1. $p_{1}^{\ast}=q_{1}^{\ast}=1,$}

Both players play the pure strategy $S_{1}$. The Nash inequalities are%

\begin{align}
P_{A}(1,1)-P_{A}(p,1)  &  =(1-p)(\alpha-\gamma)\geq0\nonumber\\
P_{B}(1,1)-P_{B}(1,q)  &  =(1-q)(\beta-\gamma)\geq0
\end{align}
and the payoffs they obtain are%

\begin{equation}
P_{A}(1,1)=\alpha\text{ \ \ }P_{B}(1,1)=\beta
\end{equation}

\subsubsection{2. $p_{2}^{\ast}=q_{2}^{\ast}=0,$}

Both players now play the pure strategy $S_{2}$ and the Nash inequalities are%

\begin{align}
P_{A}(0,0)-P_{A}(p,0)  &  =p(\beta-\gamma)\geq0\nonumber\\
P_{B}(0,0)-P_{B}(0,q)  &  =q(\alpha-\gamma)\geq0
\end{align}
The players get%

\begin{equation}
P_{A}(0,0)=\beta\text{ \ \ }P_{B}(0,0)=\alpha
\end{equation}

\subsubsection{3. $p_{3}^{\ast}=\frac{\alpha-\gamma}{\alpha+\beta-2\gamma},$
$\ \ q_{3}^{\ast}=\frac{\beta-\gamma}{\alpha+\beta-2\gamma}.$}

Players play a \emph{mixed} strategy because $p_{3}^{\ast},q_{3}^{\ast}%
\in(0,1)$. The players' payoffs are%

\begin{equation}
P_{A}(p_{3}^{\ast},q_{3}^{\ast})=P_{B}(p_{3}^{\ast},q_{3}^{\ast})=\frac
{\alpha\beta-\gamma^{2}}{\alpha+\beta-2\gamma}%
\end{equation}
Compared to the equilibria $(p_{1}^{\ast},q_{1}^{\ast})$ and $(p_{2}^{\ast
},q_{2}^{\ast})$ the players now get strictly smaller payoffs because%

\begin{equation}
\gamma<P_{A(B)}(p_{3}^{\ast},q_{3}^{\ast})<\beta<\alpha
\end{equation}

\section{Evolutionary game theory\label{EGT}}

Game theory suggests static `solutions' obtained by analyzing the behavior of
`rational agents'. Such models are obviously unrealistic because real life
behavior is shaped by trial and error. Real life `players' are subjected to
the pressures of adaptation and are forced to learn individually. In
situations where players do not have the capacity to learn individually,
natural selection favors better players through step-wise adaptation. John von
Neumann and Oskar Morgenstern, in their pioneering work on game theory
\cite{Neumann}, also realized the need for such a dynamic approach towards
game theory. After all, the word game itself suggests `motion' in one way or
the other.

In 1970's Maynard Smith developed game-theoretic models of evolution in a
population which is subjected to Darwinian selection. In his book
\textit{Evolution and the Theory of Games} \cite{Smith} he diverted attention
away from the prevalent view -- treating players as rational beings -- and
presented an evolutionary approach in game theory. This approach can be seen
as a large population model of adjustment to a NE i.e. an adjustment of
population segments by evolution as opposed to learning. Maynard Smith's model
consisted of strategic interaction among the members of a population
continuing over time in which higher payoff strategies gradually displace
strategies with lower payoffs. To distinguish evolutionary from revolutionary
changes some inertia is involved, guaranteeing that aggregate behavior does
not change too abruptly.

Most important feature of evolutionary game theory is that the assumption of
rational players -- borrowed from game theory -- does not remain crucial. It
is achieved when players' payoffs are equated to success in terms of their
survival. Players in an evolutionary model are programmed to play only one
strategy. Step-wise selection assures survival of better players at the
expense of others. In other words, an initial collection of strategies play a
tournament and the average scores are recorded. Successful strategies increase
their share of the population. Changing the population mix changes the
expected payoff. Again successful strategies increase in the population, and
the expected payoff is calculated. A population equilibrium occurs when the
population shares are such that the expected payoffs for all strategies are equal.

Many successful applications of evolutionary game theory appeared in
mathematical biology \cite{Broom et al} to predict the behavior of bacteria
and insects, that can hardly be said to think at all.

Economists too did not like game theory that mostly concerned itself with
hyper-rational players who are always trying to maximize their payoffs. Hence
the population setting of game theory, invented by mathematical biologists,
was welcomed by the economists too. Even John Nash himself, as it was found
later \cite{Hofbauer}, had a population setting in his mind when he introduced
his equilibrium notion. In his unpublished thesis he wrote `\textit{it is
unnecessary to assume that the participants have...... the ability to go
through any complex reasoning process. But the participants are supposed to
accumulate empirical information on the various pure strategies at their
disposal.......We assume that there is a population .......of
participants......and that there is a stable average frequency with which a
pure strategy is employed by the ``average member'' of the appropriate
population}'\cite{JohnNash1,Leonard}.

\subsection{Evolutionarily stable strategies}

Maynard Smith introduced the idea of an Evolutionarily Stable Strategy (ESS)
in a seminal paper `The logic of animal conflict' \cite{Smith Price}. In rough
terms \cite{MarkBroom3} an ESS is a strategy which, if played by almost all
the members of a population, cannot be displaced by a small invading group
that plays any alternative strategy. So that, a population playing an ESS can
withstand invasion by a small group. The concept was developed by combining
ingredients from game theory and some work on the evolution of the sex ratio
\cite{CanningsOrive}.

Maynard Smith considers a large population in which members are matched
repeatedly and randomly in pairs to play a bi-matrix game. The players are
anonymous, that is, any pair of players plays the same symmetric bi-matrix
game. Also the players are identical with respect to their set of strategies
and their payoff functions. The symmetry of a bi-matrix game means that for a
strategy set $S$ Alice's payoff when she plays $S_{1}\in S$ and Bob plays
$S_{2}\in S$ is the same as Bob's payoff when he plays $S_{1}$ and Alice plays
$S_{2}$. In game theory \cite{Rasmusen89}\ a symmetric bi-matrix game is
represented by an expression $G=(M,M^{T})$ where $M$ is the first player's
payoff matrix and $M^{T}$, its transpose, is the second players' payoff
matrix. In a symmetric pair-wise contest $P(x,y)$ gives the payoff to a
$x$-player against a $y$-player. In such contest exchange of strategies by the
two players also exchanges their respective payoffs. Hence, a player's payoff
is defined by his/her strategy and \emph{not} by his/her identity.

Mathematically speaking, \cite{Weibull} $x$ is an ESS when for each strategy
$y\neq x$ the inequality:%

\begin{equation}
P[x,(1-\epsilon)x+\epsilon y]>P[y,(1-\epsilon)x+\epsilon y] \label{ESSDefIneq}%
\end{equation}
should hold for all sufficiently small $\epsilon>0$. The left side of
(\ref{ESSDefIneq}) is the payoff to the strategy $x$ when played against the
strategy $(1-\epsilon)x+\epsilon y$ where $\epsilon\in\left[  0,\epsilon
_{0}\right)  $. For $\epsilon$ becoming greater than $\epsilon_{0}$ the
inequality (\ref{ESSDefIneq}) does not hold and $x$ does not remain an ESS.
The situation when $\epsilon>\epsilon_{0}$ is also known as the
\emph{invasion} by the mutant strategy. The quantity $\epsilon_{0}$ is called
the \emph{invasion barrier}.

To be precise \cite{Hofbauer} a strategy $x$ is an ESS:

\begin{itemize}
\item  If for each mutant strategy $y$ there exists a positive invasion barrier\textit{.}

\item  The invasion barrier exists such that if the population share of
individuals playing the mutant strategy $y$ falls below this barrier, then $x$
earns a higher expected payoff than $y$.
\end{itemize}

This condition for an ESS can be shown \cite{Smith} equivalent to the
following two requirements:%

\begin{align}
1.\text{ \ \ \ }P(x,x)  &  >P(y,x)\nonumber\\
2.\text{ If\ }P(x,x)  &  =P(y,x)\ \text{then}\ P(x,y)>P(y,y) \label{DefESS}%
\end{align}
An ESS, therefore, is a symmetric NE which also possesses a property of
stability against small mutations. Condition $1$ in the definition
(\ref{DefESS}) shows $(x,x)$ is a NE for the bi-matrix game $G=(M,M^{T})$ if
$x$ is an ESS. Nevertheless, the converse is not true. That is, if $(x,x)$ is
a NE then $x$ is an ESS only if $x$ satisfies condition $2$ in the definition.

In evolutionary game theory the concept of \emph{fitness} \cite{Prestwich} of
a strategy is considered crucial. Suppose $x$ and $y$ are pure strategies
played in a population setting. Their fitnesses are defined as:%

\begin{align}
W(x)  &  =P(x,x)F_{x}+P(x,y)F_{y}\nonumber\\
W(y)  &  =P(y,x)F_{x}+P(y,y)F_{y} \label{fitnesses}%
\end{align}
where $F_{x}$ and $F_{y}$\ are frequencies (the relative proportions) of the
pure strategies $x$ and $y$ respectively.

The concept of evolutionary stability provided much of the motivation for the
development of evolutionary game theory. Presently, the ESS concept is
considered as the central model of evolutionary dynamics of a populations of
interacting individuals. It asks, and finds answer to it, a basic question:
Which states of a population -- during the course of a selection process that
favors better performing strategies -- are stable against perturbations
induced by mutations? The theory is inspired by Darwinian natural selection
which is formulated as an algorithm called \emph{replicator dynamics}.
Iterations of selections from randomly mutating replicators is the important
feature of the dynamics. The dynamics is a mathematical statement saying that
in a population the proportion of players which play better strategies
increase with time. With replicator dynamics being the underlying selection
mechanism in a population, ESSs come out \cite{TaylorJonker}\ as stable
strategies against small perturbations. In other words ESSs are \emph{rest
points} of the replicator dynamics.

\subsection{ESS as a refinement of Nash equilibrium}

In the history of game theory elaborate definitions of rationality, on the
behalf of the players, led to many refinements \cite{MyersonRB} of the NE
concept. In situations where multiple NE appear as potential solutions to a
game, a refinement is required to prefer some over the others. Refinements of
NE are popular as well as numerous in classical game theory. Speaking
historically, the set of refinements became so large that eventually almost
any NE could be justified in terms of someone or other's refinement. The
concept of an ESS is a refinement on the set of symmetric Nash equilibria
\cite{Weibull}. Apart from being a symmetric NE it has robustness against
small mutations \cite{Cressman}. For symmetric bi-matrix games this
relationship is described as \cite{Gerard van}:%

\begin{equation}
\bigtriangleup^{ESS}\subset\bigtriangleup^{PE}\subset\bigtriangleup^{NE}%
\end{equation}
where $\bigtriangleup^{PE}\neq\Phi$ and $\bigtriangleup^{NE}$, $\bigtriangleup
^{PE}$, $\bigtriangleup^{ESS}$ are the sets of symmetric Nash equilibria,
symmetric proper equilibrium, and ESSs respectively.

\chapter{Review of quantum mechanics}

\textit{Quantum mechanics: Real Black Magic Calculus}

\textit{-- Albert Einstein}

\section{Introduction}

Quantum theory is the theoretical basis of modern physics that explains the
nature and behavior of matter and energy on the atomic and subatomic level.
The physical systems at these levels are known as \emph{quantum systems}. Thus
quantum mechanics is a mathematical model of the physical world that describes
the behavior of quantum systems. A physical model is characterized by how it
represents \emph{physical states}, \emph{observables}, \emph{measurements},
and \emph{dynamics} of the system under consideration. A quantum description
of a physical model is based on the following concepts:

\section{Fundamental concepts}

A \emph{state} is a complete description of a physical system. Quantum
mechanics associates a ray in \emph{Hilbert space} to the physical state of a
system. What is Hilbert space?

\begin{itemize}
\item  Hilbert space is a complex linear vector space. In Dirac's ket-bra
notation states are denoted by \emph{ket vectors} $\left|  \psi\right\rangle $
in Hilbert space. Any two state vectors differing only by an overall phase
factor $e^{i\theta}$ ($\theta$ real) represent the same state.

\item  Corresponding to a ket vector $\left|  \psi\right\rangle $ there is
another kind of state vector called \emph{bra vector}, which is denoted by
$\left\langle \psi\right|  $. The \emph{inner product} of a bra $\left\langle
\psi\right|  $ and ket $\left|  \phi\right\rangle $ is defined as follows:%

\begin{align}
\left\langle \psi\right|  \left\{  \left|  \phi_{1}\right\rangle +\left|
\phi_{2}\right\rangle \right\}   &  =\left\langle \psi\mid\phi_{1}%
\right\rangle +\left\langle \psi\mid\phi_{2}\right\rangle \nonumber\\
\left\langle \psi\right|  \left\{  c\left|  \phi_{1}\right\rangle \right\}
&  =c\left\langle \psi\mid\phi_{1}\right\rangle
\end{align}

for any $c\in\mathbf{C}$, the set of complex numbers. There is a one-to-one
correspondence between the bras and the kets. Furthermore%

\begin{align}
\left\langle \psi\mid\phi\right\rangle  &  =\left\langle \phi\mid
\psi\right\rangle ^{\ast}\nonumber\\
\left\langle \psi\mid\psi\right\rangle  &  >0\text{ for }\left|
\psi\right\rangle \neq0
\end{align}

\item  The state vectors in Hilbert space are normalized which means that the
inner product of a state vector with itself gives unity, i.e.,
\end{itemize}%

\begin{equation}
\left\langle \psi\mid\psi\right\rangle =1
\end{equation}

\begin{itemize}
\item  Operations can be performed on a ket $\left|  \psi\right\rangle $ and
transform it to another ket $\left|  \chi\right\rangle $. There are operations
on kets which are called \emph{linear operators}, which have the following
properties. For a linear operator $\hat{\alpha}$ we have
\end{itemize}%

\begin{align}
\hat{\alpha}\left\{  \left|  \psi\right\rangle +\left|  \chi\right\rangle
\right\}   &  =\hat{\alpha}\left|  \psi\right\rangle +\hat{\alpha}\left|
\chi\right\rangle \nonumber\\
\hat{\alpha}\left\{  c\left|  \psi\right\rangle \right\}   &  =c\hat{\alpha
}\left|  \psi\right\rangle
\end{align}
for any $c\in\mathbf{C}$.

\begin{itemize}
\item  The sum and product of two linear operators $\hat{\alpha}$ and
$\hat{\beta}$ are defined as:%

\begin{align}
\left\{  \hat{\alpha}+\hat{\beta}\right\}  \left|  \psi\right\rangle  &
=\hat{\alpha}\left|  \psi\right\rangle +\hat{\beta}\left|  \psi\right\rangle
\nonumber\\
\left\{  \hat{\alpha}\hat{\beta}\right\}  \left|  \psi\right\rangle  &
=\hat{\alpha}\left\{  \hat{\beta}\left|  \psi\right\rangle \right\}
\end{align}
Generally speaking $\hat{\alpha}\hat{\beta}$ is not necessarily equal to
$\hat{\beta}\hat{\alpha}$, i.e. $\left[  \hat{\alpha},\hat{\beta}\right]  \neq0$

\item  The \emph{adjoint} $\hat{\alpha}^{\dagger}$ of an operator $\hat
{\alpha}$ is defined by the requirement:%

\begin{equation}
\left\langle \psi\mid\hat{\alpha}\chi\right\rangle =\left\langle \hat{\alpha
}^{\dagger}\psi\mid\chi\right\rangle
\end{equation}

for all kets $\left|  \psi\right\rangle $, $\left|  \chi\right\rangle $ in the
Hilbert space.

\item  An operator $\hat{\alpha}$ is said to be \emph{self-adjoint} or
\emph{Hermitian} if:%

\begin{equation}
\hat{\alpha}^{\dagger}=\hat{\alpha}%
\end{equation}
\end{itemize}

Hermitian operators are the counterparts of real numbers in operators. In
quantum mechanics, the dynamical variables of physical systems are represented
by Hermitian operators. More specifically, every experimental arrangement in
quantum mechanics is associated with a set of operators describing the
dynamical variables that can be observed. These operators are usually called
\emph{observables}.

\section{Postulates of quantum mechanics}

For an isolated quantum system, quantum theory is based on the following postulates:

\begin{itemize}
\item  A ket vector $\left|  \psi\right\rangle $ in Hilbert space gives a
\emph{complete description} of the state of the physical system.

\item  Dynamics are specified by \emph{Hermitian operators} and time evolution
is given by \emph{Schr\"{o}dinger's equation:}
\end{itemize}%

\begin{equation}
i\hbar\frac{\partial\left|  \psi\right\rangle }{\partial t}=\hat{H}\left|
\psi\right\rangle \label{SchrodingerEq}%
\end{equation}
where $\hat{H}$ is the \emph{Hamiltonian operator}. Schr\"{o}dinger's equation
is a \textit{deterministic equation of motion} that allows one to determine
the state vector at any time once the initial conditions are provided.

Classical games can be played when players share coins. Coins are physical
systems that represent classical bits which takes one of the two possible
values $\left\{  0,1\right\}  $, or simply head and tail. A bit is also the
indivisible unit of classical information. For example in non-cooperative
games coins are distributed among the players and they do their actions on
them. At the end of the game the coins are collected by a referee who rewards
the players, after observing the collected coins. The earliest suggestions for
playing quantum games can be thought of letting players act on \emph{qubits},
which are a quantum generalization of classical two level systems like a coin.

\section{Qubits}

In two-dimensional Hilbert space an orthonormal basis can be written as
$\left\{  \left|  0\right\rangle ,\left|  1\right\rangle \right\}  $. A
general qubit state is then%

\begin{equation}
\left|  \psi\right\rangle =a\left|  0\right\rangle +b\left|  1\right\rangle
\label{QubitState}%
\end{equation}
where $a,b\in\mathbf{C}$ satisfying $\left|  a\right|  ^{2}+\left|  b\right|
^{2}=1$. In other words, $\left|  \psi\right\rangle $ is a unit vector in
two-dimensional complex vector space for which a particular basis has been
fixed. One of the simplest physical examples of a qubit is the spin $1/2$ of
an electron. The spin-up and spin-down states of an electron can be taken as
the states $\left|  0\right\rangle $, $\left|  1\right\rangle $ of a qubit.

A non-cooperative classical game can be played by coin distribution and
players' rewards are decided after observing the coins. Likewise, a
non-cooperative quantum game can be played by distributing qubits among the
players. After the players' moves the qubits are brought together for an
observation which is known as \emph{quantum measurement}.

\section{Quantum measurement}

Unlike observation of coins by a referee who organizes a classical game, the
concept of measurement of a quantum state of many qubits is subtle and lies at
the heart of quantum theory. The \emph{measurement postulate} of quantum
mechanics states \cite{Preskill}:

\begin{itemize}
\item  Mutually exclusive measurement outcomes correspond to orthogonal
\emph{projection operators} $\left\{  \hat{P}_{0},\text{ }\hat{P}%
_{1},...\right\}  $ and the probability of a particular outcome $i$ is
$\left\langle \psi\mid\hat{P}_{i}\mid\psi\right\rangle $. If the outcome $i$
is attained the (normalized) quantum state after the measurement becomes
\end{itemize}%

\begin{equation}
\frac{\hat{P}_{i}\left|  \psi\right\rangle }{\sqrt{\left\langle \psi\mid
P_{i}\mid\psi\right\rangle }}%
\end{equation}

Consider a measurement made on a qubit whose state vector resides in
two-dimensional Hilbert space. A measuring device has associated an
\emph{orthonormal basis} with respect to which the quantum measurement takes
place. Measurement transforms the state of the qubit into one of measuring
device's associated \emph{basis vectors}. Assume the measurement is performed
on the qubit that has the state (\ref{QubitState}). The measurement projects
the state (\ref{QubitState}) to the basis $\left\{  \left|  0\right\rangle
,\left|  1\right\rangle \right\}  $. Now in this case the measurement
postulate says that the outcome $\left|  0\right\rangle $ will happen with
probability $\left|  a\right|  ^{2}$ and the outcome $\left|  1\right\rangle $
with probability $\left|  b\right|  ^{2}$.

Furthermore, measurement of a quantum state changes the state according to the
result of the measurement. That is, if the measurement of $\left|
\psi\right\rangle =a\left|  0\right\rangle +b\left|  1\right\rangle $ results
in $\left|  0\right\rangle $, then the state $\left|  \psi\right\rangle $
changes to $\left|  0\right\rangle $ and a second measurement, with respect to
the same basis, will return $\left|  0\right\rangle $ with probability $1$.
Thus, unless the original state happened to be one of the basis vectors,
measurement will change that state, and it is \emph{not} possible to determine
what the original state was.

Although a qubit can be put in infinitely many superposition states, only a
single classical bit's worth of information can be extracted from it. It is
because the measurement changes the state of the qubit to one of the basis states.

Measurement made with orthogonal projection operators $\left\{  \hat{P}%
_{0},\text{ }\hat{P}_{1},...\right\}  $ is also called \emph{projective
measurement}.

\subsection{Positive Operator-Valued Measure}

Apart from projective measurement the quantum theory also uses another
important concept of measurement, whose implementation can be useful. It is
the concept of positive operator-valued measure (POVM). A POVM consists of a
set of non-negative quantum mechanical Hermitian operators that add up to the
identity. The probability that a quantum system is in a particular state is
given by the expectation value of the POVM operator corresponding to that
state. POVMs are sometimes also referred to as the ``generalized measurements''.

Nielsen and Chuang \cite{Nielsen}\ have discussed a simple example showing the
utility of POVM formalism. Suppose Alice gives Bob a qubit prepared in one of
two states, $\left|  \psi_{1}\right\rangle =\left|  0\right\rangle $ or
$\left|  \psi_{2}\right\rangle =(\left|  0\right\rangle +\left|
1\right\rangle )/\sqrt{2}$. It can be shown that there is no quantum
measurement capable of distinguishing the two states with perfect reliability.
However, using a POVM Bob can perform a measurement that distinguishes the two
states some of the time, but never makes an error of misidentification.

In this connection Neumark's theorem \cite{AsherPeres} needs to be mentioned
here that states that, at least in principle, any POVM can be implemented by
the adjunction of an ancilla \footnote{Ancilla bits are extra scratch qubits
that quantum operations often use.} in a known state, followed by a standard
measurement in the enlarged Hilbert space.

\section{Pure and mixed states}

In quantum mechanics a \emph{pure state} is defined as a quantum state that
can be described by a ket vector:%

\begin{equation}
\left|  \psi\right\rangle =\underset{k}{\sum}c_{k}\left|  \psi_{k}%
\right\rangle
\end{equation}
Such a state evolves in time according to the time-dependent Schr\"{o}dinger
equation (\ref{SchrodingerEq}). A \emph{mixed quantum state} is a statistical
mixture of pure states. In such a state the exact quantum-mechanical state of
the system is not known and only the probability of the system being in a
certain state can be given, which is accomplished by the \emph{density matrix}.

\section{Density matrix}

A quantum game involves two or more players having access to parts of a
quantum system. These parts are usually the subsystems of a bigger quantum
system. To use the system for playing a game one must know it detailed
statistical state. Quantum mechanics uses the concept of a density matrix to
describe the statistical state of a quantum system. It is the
quantum-mechanical analogue to a phase-space density (probability distribution
of position and momentum) in classical statistical mechanics. Suppose the
quantum state of a system is expressed in terms of a denumerable orthonormal
basis $\left\{  \left|  \phi_{n}\right\rangle ,\text{ }n=1,2,3...\right\}  $.
The state $\left|  \psi(t)\right\rangle $ of the system at time $t$ in the
basis is given as%

\begin{equation}
\left|  \psi(t)\right\rangle =\underset{n}{\sum}a_{n}(t)\left|  \phi
_{n}\right\rangle
\end{equation}
Let $\left|  \psi(t)\right\rangle $ be normalized%

\begin{equation}
\left\langle \psi(t)\mid\psi(t)\right\rangle =1=\underset{n}{\sum}\underset
{m}{\sum}a_{n}(t)a_{m}^{\ast}(t)\left\langle \phi_{m}\mid\phi_{n}\right\rangle
=\underset{n}{\sum}\left|  a_{n}(t)\right|  ^{2}%
\end{equation}
The matrix elements of a self-adjoint operator $\hat{O}$ in the basis are%

\begin{equation}
\hat{O}_{mn}=\left\langle \phi_{m}\mid\hat{O}\phi_{n}\right\rangle
=\left\langle \hat{O}\phi_{m}\mid\phi_{n}\right\rangle =\left\langle \phi
_{m}\right|  \hat{O}\left|  \phi_{n}\right\rangle
\end{equation}
The average (expectation) value of $\hat{O}$ at time $t$ for the system in
state $\left|  \psi(t)\right\rangle $ is%

\begin{equation}
\left\langle \hat{O}\right\rangle =\left\langle \psi(t)\mid\hat{O}%
\psi(t)\right\rangle =\underset{n}{\sum}\underset{m}{\sum}a_{m}^{\ast}%
(t)a_{n}(t)\hat{O}_{mn}%
\end{equation}
Consider the operator $\left|  \psi(t)\right\rangle \left\langle
\psi(t)\right|  $. It has matrix elements%

\begin{equation}
\left\langle \phi_{m}\mid\psi(t)\right\rangle \left\langle \psi(t)\mid\phi
_{n}\right\rangle =a_{m}(t)a_{n}^{\ast}(t)
\end{equation}
The calculation of $\left\langle \hat{O}\right\rangle $ involves these matrix
elements. Hence define%

\begin{equation}
\rho(t)=\left|  \psi(t)\right\rangle \left\langle \psi(t)\right|
\label{DensityMatrixProj}%
\end{equation}
which is known as the \emph{density matrix} of the pure state $\left|
\psi(t)\right\rangle $. It is a Hermitian operator, acting on the Hilbert
space of the system in question, with matrix elements%

\begin{equation}
\rho_{mn}=\left\langle \phi_{m}\mid\rho(t)\phi_{n}\right\rangle =a_{m}%
(t)a_{n}^{\ast}(t)
\end{equation}
Eq. (\ref{DensityMatrixProj}) shows that for a pure state the density matrix
is given by the projection operator of this state.

Since $\left|  \psi(t)\right\rangle $ is normalized, we also have%

\begin{equation}
1=\underset{n}{\sum}\left|  a_{n}(t)\right|  ^{2}=\underset{n}{\sum}\rho
_{nn}(t)=\text{Tr}\left[  \rho(t)\right]
\end{equation}
The expectation value of the observable $\hat{O}$ can now be re-expressed
using the density operator:%

\begin{align}
\left\langle \hat{O}\right\rangle  &  =\underset{m}{\sum}\underset{n}{\sum
}a_{m}(t)a_{n}^{\ast}(t)\hat{O}_{mn}=\underset{m}{\sum}\underset{n}{\sum}%
\rho_{nm}(t)\hat{O}_{mn}\nonumber\\
&  =\underset{n}{\sum}\left[  \rho(t)\hat{O}\right]  _{nm}=\text{Tr}\left[
\rho(t)\hat{O}\right]
\end{align}
For a mixed state, where a quantum system is in the state $\left|  \psi
_{j}(t)\right\rangle $ with probability $p_{j}$, the density matrix is the sum
of the projectors weighted with the appropriate probabilities:%

\begin{equation}
\rho(t)=\underset{j}{\sum}p_{j}\left|  \psi_{j}(t)\right\rangle \left\langle
\psi_{j}(t)\right|
\end{equation}
Density matrix is a powerful tool in quantum games because a game usually
involves a multi-partite quantum system. Compared to the description of a
quantum game based on state vectors, density matrix provides much compact notation.

\section{Quantum Entanglement}

Some of the most interesting investigations in quantum games concern the
relationship between game-theoretic solution concepts and entanglement present
within the quantum system that players are using to play the game. The
phenomenon of \emph{entanglement} can be traced back to Einstein, Podolsky and
Rosen (EPR)'s famous paper \cite{EPR}\ of 1935. EPR argued that quantum
mechanical description of \emph{physical reality} can not be considered
\emph{complete} because of its rather strange predictions about two particles
that once have interacted but now are separate from one another and do not
interact. Quantum mechanics predicts that the particles can be
\emph{entangled} even after separation. Entangled particles have correlated
properties and these correlations are at the heart of the EPR paradox.

Consider a system that can be divided into two subsystems. Assume $H_{A}$ and
$H_{B}$ to be the Hilbert spaces corresponding to the subsystems. Let $\left|
i\right\rangle _{A}$ $($where $i=1,2,...)$ be a complete orthonormal basis for
$H_{A}$, and $\left|  j\right\rangle _{B}$ $($where $j=1,2,...)$ be a complete
orthonormal basis for $H_{B}$. In quantum mechanics the Hilbert space
$H_{A}\otimes H_{B}$ (tensor product) is associated to the two subsystems
taken together. The tensor product Hilbert space $H_{A}\otimes H_{B}$ is
spanned by the states $\left|  i\right\rangle _{A}\otimes\left|
j\right\rangle _{B}$. By dropping the tensor product sign $\left|
i\right\rangle _{A}\otimes\left|  j\right\rangle _{B}$ is also written as
$\left|  i\right\rangle _{A}\left|  j\right\rangle _{B}$. Any state of the
system $\left|  \Psi\right\rangle _{AB}$ is a linear combination of the basis
states $\left|  i\right\rangle _{A}\left|  j\right\rangle _{B}$ i.e.%

\begin{equation}
\left|  \Psi\right\rangle _{AB}=\underset{i,j}{\sum c_{ij}}\left|
i\right\rangle _{A}\left|  j\right\rangle _{B}%
\end{equation}
where $c_{ij}$ are complex coefficients. State $\left|  \Psi\right\rangle
_{AB}$ is usually taken to be normalized%

\begin{equation}
\underset{i,j}{\sum\left|  c_{ij}\right|  ^{2}}=1
\end{equation}
A state $\left|  \Psi\right\rangle _{AB}$ is a \emph{direct product} state
when it factors into a normalized state $\left|  \psi^{(A)}\right\rangle
_{A}=\underset{i}{\sum}c_{i}^{(A)}\left|  i\right\rangle _{A}$ in $H_{A}$ and
a normalized state $\left|  \psi^{(B)}\right\rangle _{B}=\underset{j}{\sum
}c_{j}^{(B)}\left|  j\right\rangle _{B}$ in $H_{B}$ i.e.%

\begin{equation}
\left|  \Psi\right\rangle _{AB}=\left|  \psi^{(A)}\right\rangle _{A}\left|
\psi^{(B)}\right\rangle _{B}=\left(  \underset{i}{\sum}c_{i}^{(A)}\left|
i\right\rangle _{A}\right)  \left(  \underset{j}{\sum}c_{j}^{(B)}\left|
j\right\rangle _{B}\right)
\end{equation}
Now, interestingly, there exist some states in $H_{A}\otimes H_{B}$ that can
not be written as product states. The state $\left(  \left|  1\right\rangle
_{A}\left|  1\right\rangle _{B}+\left|  2\right\rangle _{A}\left|
2\right\rangle _{B}\right)  \diagup\sqrt{2}$ is one example. When $\left|
\Psi\right\rangle _{AB}$ is not a product state it is called \emph{entangled
}\cite{Hoi-Kwong,Nielsen}.

Quantum games have extensively used entangled states to see the resulting
affects on solutions of a game. However, it is considered a usual requirement
in quantum games that players' access to product states leads to the classical game.

\chapter{Quantum games}

\section{Introduction}

It is difficult to trace back the earliest work on quantum games. Many
situations in quantum theory can be reformulated in terms of game theory.
Several works in the literature of quantum physics can be identified having
game-like underlying structure. For example:

\begin{itemize}
\item  Wiesner's work on quantum money \cite{Wiesner}.

\item  Mermin's account \cite{Mermin}\ of Greeberger, Horne, and Zeilinger
(GHZ)'s \cite{GHZ} version of the Bell's theorem \cite{Bell,AsherPeres}
without inequalities.

\item  Elitzur-Vaidman bomb detector \cite{Elitzur-Vaidman}, suggesting an
interferometer which splits a photon in two and then puts it back together
again (interaction-free measurement).

\item  Vaidman's illustration \cite{Vaidman}\ of GHZ's version of the Bell's theorem.

\item  Meyer's demonstration \cite{MeyerDavid} of a quantum version of a
penny-flip game.

\item  Eisert, Wilkens, and Lewenstein's \cite{Eisert} quantization of the
famous game of Prisoners' Dilemma (PD).
\end{itemize}

In general, a quantum game can be thought of as strategic manoeuvreing of a
quantum system by parties who have necessary means for such actions. Some of
its essential parts can be recognized as follows:

\begin{itemize}
\item  A definition of the physical system which can be analyzed using the
tools of quantum mechanics.

\item  Existence of one or more parties, usually referred to as players, who
are able to manipulate the quantum system.

\item  Players' knowledge about the quantum system on which they will make
their moves or actions.

\item  A definition of what constitutes a strategy for a player.

\item  A definition of strategy space for the players, which is the set of all
possible actions that players can take on the quantum system.

\item  A definition of the pay-off functions or utilities associated with the
players' strategies.
\end{itemize}

A two-player quantum game, for example, is a set \cite{Eisert1}:%

\begin{equation}
\Gamma=(\mathcal{H},\rho,S_{A},S_{B},P_{A},P_{B})
\end{equation}
consisting of an underlying Hilbert space $\mathcal{H}$ of the physical
system, the initial state $\rho$, the sets $S_{A}$ and $S_{B}$ of allowed
quantum operations for two players, and the pay-off functions or utilities
$P_{A}$ and $P_{B}$. In most of the existing set-ups to play quantum games the
initial state $\rho$ is the state of one or more qubits. More complex quantum
systems like qutrits (three-dimensional quantum systems) or even qudits
(d-dimensional quantum system) can also be used to play quantum games.

\section{Why games in the quantum world?}

The question why game theory can be interesting in the quantum world has been
addressed in the earliest suggestions for quantum games. Some of the stated
reasons \cite{MeyerDavid,Eisert} are:

\begin{itemize}
\item  Classical game theory is based on probability to a large extent.
Generalizing it to quantum probability is fundamentally interesting.

\item  Quantum algorithms may be thought of as games between classical and
quantum agents. Only a few quantum algorithms are known to date. It appears
reasonable that an analysis of quantum games may help finding new quantum algorithms.

\item  There is an intimate connection between the theory of games and theory
of quantum communication. Eavesdropping \cite{Ekert,Gisin} and optimal cloning
\cite{WernerRF} can readily be conceived as games between players.

\item  Quantum mechanics may assure fairness in remote gambling
\cite{Goldenberg}.

\item  If the `Selfish Gene' \cite{Dawkins} is a reality then the games of
survival are already being played at molecular level, where quantum mechanics
dictates the rules.
\end{itemize}

\section{Examples of quantum games}

As the subject of quantum games has developed during recent years, many
examples have been put forward illustrating how such game can be different
from their classical analogues. Here are some of the well known quantum games:

\subsection{Vaidman's game}

Vaidman \cite{Vaidman} presented an example of a game for a team of three
players that can \emph{only} be won if played in the quantum world. Three
players are sent to remote locations $A,$ $B$ and $C$. At a certain time $t$
each player is asked one of the two possible questions:

\begin{enumerate}
\item  What $X$?

\item  What $Y$?
\end{enumerate}

Both of these questions have $+1$ or $-1$ as possible answers. Rules of the
game are such that:

\begin{itemize}
\item  Either all players are asked the $X$ question or

\item  Only one player is asked the $X$ question and the other two are asked
the $Y$ question.
\end{itemize}

The team wins if

\begin{itemize}
\item  The product of their three answers is $-1$ in the case of three $X$
questions or

\item  The product of three answers is $1$ in the case of one $X$ and two $Y$ questions.
\end{itemize}

What should the team do? Let $X_{A}$ be the answer of player $A$ to the $X$
question. Similarly, one can define $X_{B},X_{C}$ etc. The winning condition
can now be written as%

\begin{align}
X_{A}X_{B}X_{C}  &  =-1\nonumber\\
X_{A}Y_{B}Y_{C}  &  =1\nonumber\\
Y_{A}X_{B}Y_{C}  &  =1\nonumber\\
Y_{A}Y_{B}X_{C}  &  =1 \label{Vaidmanproduct}%
\end{align}
The product of all left hand sides of Eqs. (\ref{Vaidmanproduct}) is
$X_{A}^{2}X_{B}^{2}X_{C}^{2}Y_{A}^{2}Y_{B}^{2}Y_{C}^{2}=1$, because each of
the $X$ or $Y$ take the values $\pm1$ only. The product of right sides is
$-1$, which leads to a contradiction. Therefore, the game cannot be won, with
a success probability of $1$, by a team of classical players. Eqs.
(\ref{Vaidmanproduct}) show that the probability of winning the game by
classical players can not exceed $3/4$. However, Vaidman showed that a quantum
solution exists for the team. Three particles are prepared in a correlated
state (GHZ):%

\begin{equation}
\left|  GHZ\right\rangle =\frac{1}{\sqrt{2}}\left\{  \left|  \uparrow
_{Z}\right\rangle _{A}\left|  \uparrow_{Z}\right\rangle _{B}\left|
\uparrow_{Z}\right\rangle _{C}-\left|  \downarrow_{Z}\right\rangle _{A}\left|
\downarrow_{Z}\right\rangle _{B}\left|  \downarrow_{Z}\right\rangle
_{C}\right\}
\end{equation}
If a member \ of the team is asked the $X$ question, she measures $\hat
{\sigma}_{x}$. If she is asked the $Y$ question, she measures $\hat{\sigma
}_{y}$ instead. Quantum mechanics implies that for the GHZ state one gets \cite{Vaidman}%

\begin{align}
\hat{\sigma}_{A_{x}}\hat{\sigma}_{B_{x}}\hat{\sigma}_{C_{x}}  &
=-1\nonumber\\
\hat{\sigma}_{A_{x}}\hat{\sigma}_{B_{y}}\hat{\sigma}_{C_{y}}  &  =1\nonumber\\
\hat{\sigma}_{A_{y}}\hat{\sigma}_{B_{x}}\hat{\sigma}_{C_{y}}  &  =1\nonumber\\
\hat{\sigma}_{A_{y}}\hat{\sigma}_{B_{y}}\hat{\sigma}_{C_{x}}  &  =1
\end{align}
The Vaidman's game can, therefore, be won by a group of quantum players with
$100\%$ success probability.

There remains a subtle point in Vaidman's argument. The contradiction obtained
by comparing the four equations in (\ref{Vaidmanproduct}) assume that all four
equations hold simultaneously. In fact, the four equations represent four
incompatible situations.

\subsection{Meyer's PQ Penny-Flip}

Two players can play a simple game if they share a coin having two possible
states, head or tail. The first strong argument for quantum games was
presented by Meyer \cite{MeyerDavid} as a coin flip game played by two
characters, Captain Picard and Q, from the popular American science fiction
series Star Trek. In a quantum version of the game the flipping action is
performed on a ``quantum coin'', which can be thought of as an electron that,
on measurement, is found to exist either in spin-up ($H$) or in spin-down
($T$) state.

In Meyer's interesting description of a quantum game, the story starts when
starship Enterprise faces some imminent catastrophe. Q appears on the bridge
and offers Picard to rescue the ship if he can beat him in a penny-flip game.
Q asks Picard to place the penny in a box, head up. Then Q, Picard, and
finally Q play their moves. Q wins if the penny is head up when the box is
opened. For classical version of this game the payoff matrix can be
constructed as%

\begin{equation}%
\begin{array}
[c]{c}%
\text{Picard}%
\end{array}%
\begin{array}
[c]{c}%
N\\
F
\end{array}
\overset{\overset{%
\begin{array}
[c]{c}%
\text{Q}%
\end{array}
}{%
\begin{array}
[c]{cccc}%
NN & NF & FN & FF
\end{array}
}}{\left(
\begin{array}
[c]{cccc}%
-1 & 1 & 1 & -1\\
1 & -1 & -1 & 1
\end{array}
\right)  } \label{PQPennyFlipMatrix}%
\end{equation}
where rows and columns are Picard's and Q's pure strategies respectively. Let
$(H,T)$ be the basis of a $2$-dimensional vector space. The players' moves can
be represented by a sequence of $2\times2$ matrices. In the matrix
(\ref{PQPennyFlipMatrix}) the moves `to flip' and `not to flip' are
represented by $F$ and $N$, respectively:%

\begin{equation}
F:%
\begin{array}
[c]{c}%
H\\
T
\end{array}
\overset{%
\begin{array}
[c]{cc}%
H & T
\end{array}
}{\left(
\begin{array}
[c]{cc}%
0 & 1\\
1 & 0
\end{array}
\right)  }\text{ \ \ \ \ \ \ \ \ \ \ }N:%
\begin{array}
[c]{c}%
H\\
T
\end{array}
\overset{%
\begin{array}
[c]{cc}%
H & T
\end{array}
}{\left(
\begin{array}
[c]{cc}%
1 & 0\\
0 & 1
\end{array}
\right)  }%
\end{equation}
defined to act, on left multiplication, on a vector representing the state of
the coin. A general mixed strategy is described by the matrix:%

\begin{equation}%
\begin{array}
[c]{c}%
H\\
T
\end{array}
\overset{%
\begin{array}
[c]{cc}%
H & T
\end{array}
}{\left(
\begin{array}
[c]{cc}%
1-p & p\\
p & 1-p
\end{array}
\right)  } \label{MixedStrategy}%
\end{equation}
where $p\in\left[  0,1\right]  $ is the probability with which the player
flips the coin. A sequence of mixed actions puts the state of the coin into a
convex linear combination $aH+(1-a)T$ where $0\leq a\leq1$. The coin is then
in $H$ state with probability $a$. Q plays his move first, after Picard puts
the coin in the $H$ state.

Now Meyer presents a look at a quantum version of this game. Q has studied
quantum theory and implements his strategy as a sequence of unitary rather
than stochastic matrices. Such action requires a description of the state of
the coin in two-dimensional Hilbert space. Let its basis be the kets $\left|
H\right\rangle $ and $\left|  T\right\rangle $, in Dirac notation. A pure
state of the coin is $a\left|  H\right\rangle +b\left|  T\right\rangle $ where
$a,b\in\mathbf{C}$ and $aa^{\ast}+bb^{\ast}=1$.

Given the coin is initially in the state $\left|  H\right\rangle $, the
following unitary action $U(a,b)$ by $Q$ puts the coin into the state
$a\left|  H\right\rangle +b\left|  T\right\rangle $:%

\begin{equation}%
\begin{array}
[c]{c}%
U(a,b)=
\end{array}%
\begin{array}
[c]{c}%
H\\
T
\end{array}
\overset{%
\begin{array}
[c]{cc}%
H & T
\end{array}
}{\left(
\begin{array}
[c]{cc}%
a & b\\
b^{\ast} & -a^{\ast}%
\end{array}
\right)  } \label{Q's action}%
\end{equation}
Using the density matrix notation, the initial state of the coin can be
written as%

\begin{equation}
\rho_{0}=\left|  H\right\rangle \left\langle H\right|
\end{equation}
Q's unitary action $U(a,b)$ changes the state $\rho_{0}$ to%

\begin{equation}
\rho_{1}=U\rho_{0}U^{\dagger}=\left(
\begin{array}
[c]{cc}%
aa^{\ast} & ab^{\ast}\\
ba^{\ast} & bb^{\ast}%
\end{array}
\right)
\end{equation}
because unitary transformations act on density matrices by conjugation. Picard
is restricted to use only a classical mixed strategy (\ref{MixedStrategy}) by
flipping the coin with probability $p$. After his action the coin is in the
pure state $b\left|  H\right\rangle +a\left|  T\right\rangle $ with
probability $p$ and in the pure state $a\left|  H\right\rangle +b\left|
T\right\rangle $ with probability $(1-p)$. Picard's action acts on the density
matrix $\rho_{1}$, not as a stochastic matrix on a probabilistic state, but as
a convex linear combination of unitary (deterministic) transformations:%

\begin{align}
\rho_{2}  &  =pF\rho_{1}F^{\dagger}+(1-p)N\rho_{1}N^{\dagger}\nonumber\\
&  =\left(
\begin{array}
[c]{cc}%
pbb^{\ast}+(1-p)aa^{\dagger} & pba^{\ast}+(1-p)ab^{\dagger}\\
pab^{\ast}+(1-p)ba^{\ast} & paa^{\ast}+(1-p)bb^{\dagger}%
\end{array}
\right)
\end{align}
Interestingly, Q has at his disposal a move:%

\begin{equation}
U_{1}=U(\frac{1}{\sqrt{2}},\frac{1}{\sqrt{2}})=\frac{1}{\sqrt{2}}\left(
\begin{array}
[c]{cc}%
1 & 1\\
1 & -1
\end{array}
\right)
\end{equation}
that can put the coin into a simultaneous eigenstate with eigenvalue $1$ of
both $F$ and $N,$ which then becomes an invariant under any mixed strategy
$pF+(1-p)N$ of Picard. In his second action Q acts again with $U(\frac
{1}{\sqrt{2}},\frac{1}{\sqrt{2}})$ and gets back the state $\rho_{0}=\left|
H\right\rangle \left\langle H\right|  $ and wins. The game can also be
understood with the following chart.%

\begin{equation}
\left|  H\right\rangle \text{ \ \ }\overset{%
\begin{array}
[c]{c}%
\text{Q}%
\end{array}
}{\underset{%
\begin{array}
[c]{c}%
\hat{H}%
\end{array}
}{\longrightarrow}}\text{ \ \ }\frac{1}{\sqrt{2}}\left(  \left|
H\right\rangle +\left|  T\right\rangle \right)  \text{ \ \ }\overset{%
\begin{array}
[c]{c}%
\text{Picard}%
\end{array}
}{\underset{%
\begin{array}
[c]{c}%
\sigma_{x}\text{ or }\hat{I}%
\end{array}
}{\longrightarrow}}\text{ \ \ }\frac{1}{\sqrt{2}}\left(  \left|
H\right\rangle +\left|  T\right\rangle \right)  \text{ \ \ }\overset{%
\begin{array}
[c]{c}%
\text{Q}%
\end{array}
}{\underset{%
\begin{array}
[c]{c}%
\hat{H}%
\end{array}
}{\longrightarrow}}\text{ \ \ }\left|  H\right\rangle
\end{equation}
where $\hat{H}=\frac{1}{\sqrt{2}}\left(
\begin{array}
[c]{cc}%
1 & 1\\
1 & -1
\end{array}
\right)  $ is a Hadamard transformation and $\sigma_{x}=\left(
\begin{array}
[c]{cc}%
0 & 1\\
1 & 0
\end{array}
\right)  $ is the flipping operator. $\left|  H\right\rangle $ is the head
state of the coin and $\hat{I}$ is the identity operator. Q plays a quantum
strategy by putting the coin into a symmetric superposition of head and tail.
Now, whether Picard flips the coin or not, it remains in the symmetric
superposition which Q can rotate back to head applying $\hat{H}$ again since
$\hat{H}=$ $\hat{H}^{-1}$.

\subsection{Eisert, Wilkens and Lewenstein's quantum Prisoners' Dilemma}

Eisert, Wilkens, and Lewenstein \cite{Eisert} gave a physical model of the PD
and suggested that the players can escape the dilemma if they both resort to
quantum strategies. Their physical model consists of

\begin{itemize}
\item  A source making available two bits, one for each player.

\item  Physical instruments enabling the players to manipulate, in a strategic
manner, their own bits.

\item  A measurement device that determines the players' payoffs from the
final state of the two bits.
\end{itemize}

In a quantum formulation the classical strategies $C$ and $D$ are assigned two
basis vectors $\left|  C\right\rangle $ and $\left|  D\right\rangle $ in
Hilbert space of a qubit. A vector in the tensor product space, which is
spanned by the classical game basis $\left|  CC\right\rangle ,\left|
CD\right\rangle ,\left|  DC\right\rangle $ and $\left|  DD\right\rangle $
describes the state of the game.

The game's initial state is $\left|  \psi_{in}\right\rangle =\hat{J}\left|
CC\right\rangle $ where $\hat{J}$ is a unitary operator known to both players.
Alice and Bob's strategic moves are associated with unitary operators $\hat
{U}_{A}$ and $\hat{U}_{B}$ respectively, chosen from a strategic space $S$.
The players' actions are local i.e. each operates on his/her qubit. After
players' moves the state of the game changes to $(\hat{U}_{A}\otimes\hat
{U}_{B})\hat{J}\left|  CC\right\rangle $. Measurements are now performed to
determine the players' payoffs. Measurement consists of applying a reverse
unitary operator $\hat{J}^{\dagger}$ followed by a pair of Stern-Gerlach type
detectors. Before detection the final state of the game is given by%

\begin{equation}
\left|  \psi_{f}\right\rangle =\hat{J}^{\dagger}(\hat{U}_{A}\otimes\hat{U}%
_{B})\hat{J}\left|  CC\right\rangle
\end{equation}
Eisert et al. \cite{Eisert} define Alice's expected payoff as%

\begin{equation}
P_{A}=r\left|  \left\langle CC\mid\psi_{f}\right\rangle \right|  ^{2}+s\left|
\left\langle CD\mid\psi_{f}\right\rangle \right|  ^{2}+t\left|  \left\langle
DC\mid\psi_{f}\right\rangle \right|  ^{2}+u\left|  \left\langle DD\mid\psi
_{f}\right\rangle \right|  ^{2} \label{PDAlice'sPayoff}%
\end{equation}
where the quantities $r,s,t$ and $u$ are from the PD matrix (\ref{PDmatrix}).
Bob's payoff $P_{B}$ is obtained by interchanging $s\leftrightarrow t$ in Eq.
(\ref{PDAlice'sPayoff}). Eisert and Wilkens \cite{Eisert1} use following
matrix representations of unitary operators of their one- and two-parameter
strategies, respectively:%

\begin{align}
U(\theta)  &  =\left(
\begin{array}
[c]{cc}%
\cos(\theta/2) & \sin(\theta/2)\\
\text{-}\sin(\theta/2) & \cos(\theta/2)
\end{array}
\right) \label{OneParameterSet}\\
U(\theta,\phi)  &  =\left(
\begin{tabular}
[c]{ll}%
e$^{i\phi}\cos(\theta/2)$ & $\sin(\theta/2)$\\
$\text{-}\sin(\theta/2)$ & e$^{-i\phi}\cos(\theta/2)$%
\end{tabular}
\right)  \label{TwoParameterSet}%
\end{align}
where $0\leq\theta\leq\pi$ and $0\leq\phi\leq\pi/2$. To ensure that the
ordinary PD is faithfully represented in its quantum version, Eisert et al.
imposed additional conditions on $\hat{J}$:%

\begin{equation}
\left[  \hat{J},\hat{D}\otimes\hat{D}\right]  =0,\left[  \hat{J},\hat
{D}\otimes\hat{C}\right]  =0,\left[  \hat{J},\hat{C}\otimes\hat{D}\right]  =0
\label{condition1}%
\end{equation}
where $\hat{C}$ and $\hat{D}$ are the operators corresponding to the
strategies of cooperation and defection respectively. A unitary operator
satisfying the condition (\ref{condition1}) is%

\begin{equation}
\hat{J}=\exp\left\{  i\gamma\hat{D}\otimes\hat{D}/2\right\}
\end{equation}
where $\gamma\in\left[  0,\pi/2\right]  $. $\hat{J}$ can be called a measure
of the game's entanglement. At $\gamma=0$ the game reduces to its classical
form. For a maximally entangled game $\gamma=\pi/2$ the classical NE $\hat
{D}\otimes\hat{D}$ is replaced by a different unique equilibrium $\hat
{Q}\otimes\hat{Q}$ with $\hat{Q}\sim\hat{U}(0,\pi/2).$ The new equilibrium is
also found to be \emph{Pareto optimal}, that is, a player cannot increase
his/her payoff by deviating from this pair of strategies without reducing the
other player's payoff. Classically ($C,C$) is Pareto optimal, but is not an
equilibrium. Eisert et al. claimed that in its quantum version the dilemma in
PD disappears from the game and quantum strategies give a superior performance
if entanglement is present.%

\begin{figure}
[ptb]
\begin{center}
\includegraphics[
height=3.3555in,
width=5.1517in
]%
{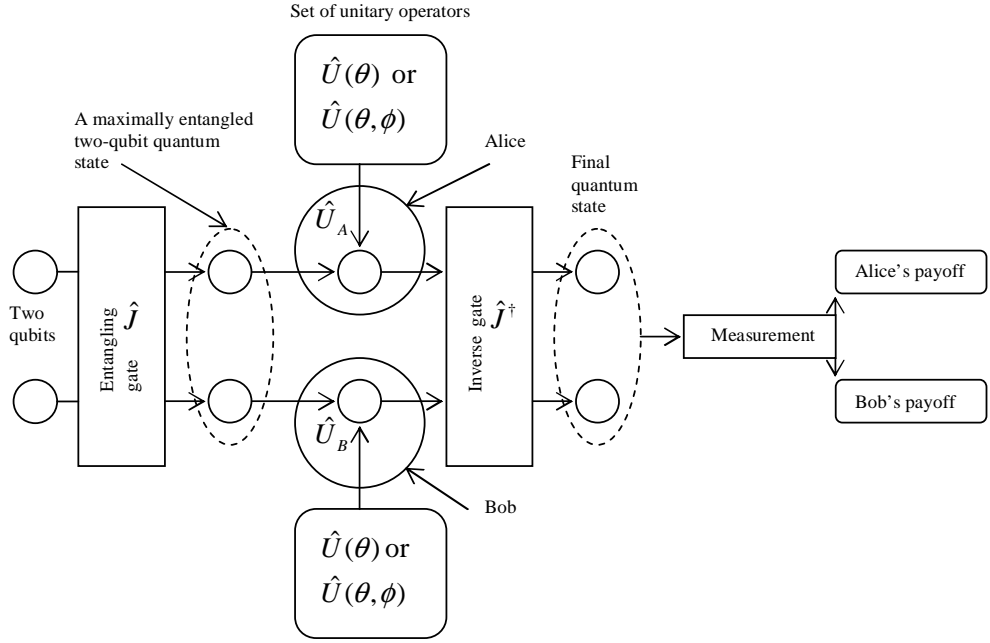}%
\caption{Eisert et al.'s scheme to play a quantum game.}%
\label{Fig1}%
\end{center}
\end{figure}

In density matrix notation, the players' actions change the initial state
$\rho$ to%

\begin{equation}
\hat{\sigma}=(\hat{U}_{A}\otimes\hat{U}_{B})\rho(\hat{U}_{A}\otimes\hat{U}%
_{B})^{\dagger}%
\end{equation}
The arbiter applies the following operators on $\sigma$:%

\begin{align}
\hat{\pi}_{CC}  &  =\left|  \psi_{CC}\right\rangle \left\langle \psi
_{CC}\right|  \text{, \ \ \ \ \ \ }\hat{\pi}_{CD}=\left|  \psi_{CD}%
\right\rangle \left\langle \psi_{CD}\right| \nonumber\\
\hat{\pi}_{DC}  &  =\left|  \psi_{DC}\right\rangle \left\langle \psi
_{DC}\right|  \text{, \ \ \ \ \ }\hat{\pi}_{DD}=\left|  \psi_{DD}\right\rangle
\left\langle \psi_{DD}\right|
\end{align}
The expected payoffs are%

\begin{align}
P_{A,B}  &  =[P_{CC}]_{A,B}\text{Tr}[\hat{\pi}_{CC}\hat{\sigma}]+[P_{CD}%
]_{A,B}\text{Tr}[\hat{\pi}_{CD}\hat{\sigma}]\nonumber\\
&  +[P_{DC}]_{A,B}\text{Tr}[\hat{\pi}_{DC}\hat{\sigma}]+[P_{DD}]_{A,B}%
\text{Tr}[\hat{\pi}_{DD}\hat{\sigma}]
\end{align}
Where, for example, $[P_{CD}]_{A}$ is the Alice's classical payoff when she
plays $C$ and Bob plays $D$. For one-parameter strategies the classical pure
strategies $C$ and $D$ are realized as $\hat{U}(0)$ and $\hat{U}(\pi)$,
respectively; while for two-parameter strategies the classical pure strategies
$C$ and $D$ are realized as $\hat{U}(0,0)$ and $\hat{U}(\pi,0)$, respectively.
Fig. (\ref{Fig1}) shows Eisert et al.'s scheme to play a quantum game.

Many recent investigations \cite{FlitneyAbbott4,Flitney5,Flitney6,Flitney7}%
\ in quantum games have been motivated by the Eisert et. al.'s scheme.

\subsection{Quantum Battle of Sexes}

Motivated by the Eisert et al.'s proposal, Marinatto and Weber
\cite{Marinatto1} introduced a new scheme for quantizing bi-matrix games by
presenting a quantized version of the BoS. In this scheme a state in a
$2\otimes2$ dimensional Hilbert space is referred to as a \emph{strategy}. At
the start of the game the players are supplied with this strategy. The players
manipulate the strategy, in the next phase, by playing their \emph{tactics}.
The state is finally measured and the payoffs are rewarded depending on the
results of the measurement. A player can do actions within a two-dimensional
subspace. Tactics are therefore \emph{local actions} on a player's qubit. The
final measurement, made independently on each qubit, takes into consideration
the local nature of players' manipulations. It is achieved by selecting a
measurement basis that respects the division of Hilbert space into two equal parts.

Essentially, the scheme differs from the earlier proposed scheme of Eisert et
al. \cite{Eisert} by the absence of reverse gate $J^{\dagger}$. The gate makes
sure that the classical game remains a subset of its quantum version. In
Marinatto and Weber's scheme the state is measured without passing it through
the reverse gate. They showed that the classical game still remains a subset
of the quantum game if the players' tactics are limited to a probabilistic
choice between applying the identity $\hat{I}$ and the Pauli spin-flip
operator $\hat{\sigma}_{x}$. Also the classical game results when the players
are forwarded an initial strategy $\left|  \psi_{in}\right\rangle =\left|
00\right\rangle $.

Suppose $\rho_{in}$ is the initial strategy, which the players Alice and Bob
receive at the start of the game. Let Alice acts with identity $\hat{I}$ on
$\rho_{in}$ with probability $p$ and with $\hat{\sigma}_{x}$ with probability
$(1-p)$. Similarly, let Bob acts with identity $\hat{I}$ with probability $q$
and with $\hat{\sigma}_{x}$ with probability $(1-q)$. After the players'
actions the state changes to%

\begin{align}
\rho_{fin}  &  =pq\hat{I}_{A}\otimes\hat{I}_{B}\rho_{in}\hat{I}_{A}^{\dagger
}\otimes\hat{I}_{B}^{\dagger}+p(1-q)\hat{I}_{A}\otimes\hat{\sigma}_{xB}%
\rho_{in}\hat{I}_{A}^{\dagger}\otimes\hat{\sigma}_{xB}^{\dagger}+\nonumber\\
&  q(1-p)\hat{\sigma}_{xA}\otimes\hat{I}_{B}\rho_{in}\hat{\sigma}%
_{xA}^{\dagger}\otimes\hat{I}_{B}^{\dagger}+\nonumber\\
&  (1-p)(1-q)\hat{\sigma}_{xA}\otimes\hat{\sigma}_{xB}\rho_{in}\hat{\sigma
}_{xA}^{\dagger}\otimes\hat{\sigma}_{xB}^{\dagger}%
\end{align}
For the bi-matrix:%

\begin{equation}%
\begin{array}
[c]{c}%
\text{Alice}%
\end{array}%
\begin{array}
[c]{c}%
S_{1}\\
S_{2}%
\end{array}
\overset{\overset{%
\begin{array}
[c]{c}%
\text{Bob}%
\end{array}
}{%
\begin{array}
[c]{cc}%
S_{1} & S_{2}%
\end{array}
}}{\left(
\begin{array}
[c]{cc}%
(\alpha_{A},\alpha_{B}) & (\beta_{A},\beta_{B})\\
(\gamma_{A},\gamma_{B}) & (\delta_{A},\delta_{B})
\end{array}
\right)  }%
\end{equation}
Marinatto and Weber defined the following payoff operators%

\begin{align}
(P_{A})_{oper}  &  =\alpha_{A}\left|  00\right\rangle \left\langle 00\right|
+\beta_{A}\left|  01\right\rangle \left\langle 01\right|  +\gamma_{A}\left|
10\right\rangle \left\langle 10\right|  +\delta_{A}\left|  11\right\rangle
\left\langle 11\right| \nonumber\\
(P_{B})_{oper}  &  =\alpha_{B}\left|  00\right\rangle \left\langle 00\right|
+\beta_{B}\left|  01\right\rangle \left\langle 01\right|  +\gamma_{B}\left|
10\right\rangle \left\langle 10\right|  +\delta_{B}\left|  11\right\rangle
\left\langle 11\right|
\end{align}
where the states $\left|  0\right\rangle $ and $\left|  1\right\rangle $ are
used for the measurement basis, corresponding to the pure strategies $S_{1}$
and $S_{2}$, respectively. The payoff functions are then obtained as mean
values of these operators:%

\begin{equation}
P_{A,B}=\text{Tr}\left\{  (P_{A,B})_{oper}\rho_{fin}\right\}
\end{equation}%

\begin{figure}
[ptb]
\begin{center}
\includegraphics[
height=3.2984in,
width=5.5988in
]%
{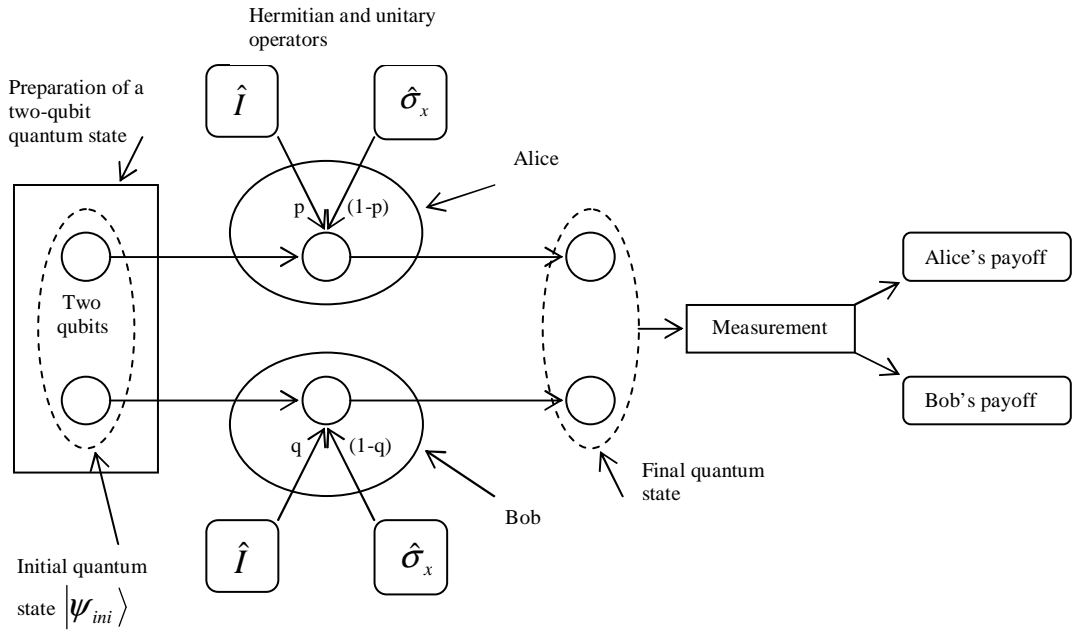}%
\caption{Marinatto and Weber's scheme to play a quantum game.}%
\label{Fig2}%
\end{center}
\end{figure}

Fig. (\ref{Fig2}) sketches the idea of playing a quantum game in Marinatto and
Weber's scheme. The scheme was developed for the BoS given by the matrix
(\ref{BoSMatrix}). On receiving an initial strategy:%

\begin{equation}
\psi_{ini}=\frac{1}{\sqrt{2}}(\left|  00\right\rangle +\left|  11\right\rangle
)
\end{equation}
the players' tactics cannot change it, and the final strategy remains
identical to the initial one. The players' expected payoffs are maximized for
the tactics $p^{\ast}=q^{\ast}=0$ or $p^{\ast}=q^{\ast}=1$, that is, both
players either apply $\hat{\sigma}_{x}$ with certainty or $\hat{I}$ with
certainty. In either case the expected payoff is $(\alpha+\beta)/2$ to each
player. Marinatto and Weber suggested that a unique solution thus exists in
the game.

\subsection{Quantum version of the Monty Hall problem}

Monty Hall is a game in which Alice secretly selects one door out of three to
place a prize there. It is now Bob's turn who picks a door. Alice then opens a
different door showing that the prize is not behind it. Bob now has the option
of changing to the untouched door or sticking with his current selection. In
classical version of the game Bob's optimum strategy is to alter his choice of
door and it doubles his chances of winning.

Li et al. \cite{Li et al}, Flitney and Abbot \cite{FlitneyAbbott} and D'Ariano
et al. \cite{DAriano}\ have proposed various quantum versions of the Monty
Hall problem.

\subsection{Quantum market games}

During recent years Piotrowski and Sladkowski
\cite{Piotrowski1,Piotrowski2,Piotrowski3} have proposed quantum-like
description of markets and economics. This development can be shown having
roots in quantum game theory and is considered part of new field of
econophysics. In econophysics \cite{Econophysics} mathematical techniques
developed by physicists are used to analyze the complex financial and economic
systems. Developments in econophysics have motivated some authors
\cite{Blankmeyer} to ask about the possibility of a meaning of Heisenberg
uncertainty principle in economics. Others have even claimed that quantum
mechanics and mathematical economics are isomorphic \cite{Lambertini}.

\subsection{Quantum Parrondo's Games}

A Parrondo's game is an interesting problem in game theory. Two games that are
losing when played individually can be combined to produce a winning game. The
game can be put into the form of a gambling utilizing a set of biased coins.

Flitney and Abbott \cite{FlitneyAbbott1,FlitneyAbbott2} studied a quantum
version of the Parrondo's game where the rotation operators representing the
toss of a classical biased coin are replaced by general SU(2) operators to
transform the game into the quantum domain. They found that superposition of
qubits can couple the two games and produce interference leading to different
payoffs than in the classical case.

\chapter{Comments on proposed set-ups to play quantum games}

Meyer \cite{MeyerDavid} demonstrated with the example of a penny-flip game how
quantum mechanics can affect game theory. He introduced a game where a
suitable quantum strategy can beat any classical strategy. Comments and
criticism followed soon after this demonstration of the power of quantum
strategies, which are reviewed in the following.

\section{Enk's comment on Meyer's quantum Penny-Flip}

Though agreeing that Meyer reached a correct conclusion, Enk \cite{Enk}
commented that Meyer's particular example is flawed for the following reasons:

\begin{itemize}
\item  Q's quantum strategy can also be implemented classically.

\item  Meyer's game only shows the superiority of an extended set of
strategies over a restricted one, which is not surprising.

\item  A single qubit is not truly a quantum system because its dynamics and
its response to measurements can also be described by a classical
hidden-variable model. Bell's inequalities, or the Kochen-Specker theorem, do
not exist for a two-dimensional system, thus making it possible to explicitly
construct classical models for such systems.
\end{itemize}

\subsection{Meyer's reply}

Meyer replied \cite{Meyer's Reply} and disagreed with Enk's claim that the
existence of classical models for Q's strategy necessarily prevents it from
being called quantum mechanical. He argued as the following:

\begin{itemize}
\item  Enk's claim implies that P's strategy is also not classical because
quantum models exist for flipping a two-state system.

\item  Though classical models do indeed exist for qubit systems but they
scale exponentially as the number of qubits increase.
\end{itemize}

Entangled qubits do not possess classical models but entanglement itself has
been shown unnecessary to outperform a quantum algorithm from a classical one.
For example, Grover's algorithm \cite{Grover}, although discovered in the
context of quantum computation, can be implemented using a system allowing
superposition of states, like classical coupled simple harmonic oscillators
\cite{ApporvaPatel}. It does not seem fair to claim that such an
implementation prohibits calling it a quantum algorithm.

Related to the third point in Enk's comment, it seems related to mention that
recently Khrennikov \cite{Khrennikov} proved an analogue of Bell's inequality
for conditional probabilities. Interestingly the inequality can be applied not
only to pairs of correlated particles, but also to a single particle. The
inequality is violated for spin projections of the single particle. Khrennikov
concludes that a realistic pre-quantum model does not exist even for the
two-dimensional Hilbert space.

\section{Benjamin and Hayden's comment on quantization of Prisoners' Dilemma}

Eisert et al. obtained $\hat{Q}$ as the new quantum equilibrium in PD, when
both players have access to a two-parameter set (\ref{TwoParameterSet}) of
unitary $2\times2$ matrices. Benjamin and Hayden \cite{Benjamin1} observed
that when their two-parameter set is extended to all local unitary operations
(i.e. all of $SU(2)$) the strategy $\hat{Q}$ does not remain an equilibrium.
They showed that in the full space of deterministic quantum strategies there
exists no equilibrium for Eisert et al.'s quantum PD. They also observed that
Eisert's set of two-parameter quantum strategies is not closed under
composition, which is reasonable requirement for a set of quantum strategies.

\section{Benjamin's comment on Marinatto and Weber's quantum Battle of Sexes}

In his comment Benjamin \cite{Benjamin2} made two observations about Marinatto
and Weber's quantum battle of sexes:

\begin{itemize}
\item  The overall quantization scheme is fundamentally very similar to the
Eisert et al.'s previously proposed scheme \cite{Eisert}.

\item  The quantum BoS does not have a unique solution. Though the dilemma may
be easier to resolve in its quantum version, the players still face it as they
do in the traditional game.
\end{itemize}

In the quantum BoS the players's expected payoffs are maximized when their
tactics consist of either both applying $\hat{I}$ with certainty ($p^{\ast
}=1,q^{\ast}=1$) or both applying $\hat{\sigma}_{x}$ with certainty ($p^{\ast
}=0,q^{\ast}=0$). Marinatto and Weber concluded that an entangled initial
strategy $(\left|  00\right\rangle +\left|  11\right\rangle )/\sqrt{2}$,
therefore, gives a unique solution in the game. Given that the players'
tactics are independent, the players are faced with a dilemma once again in
opting for ($p^{\ast}=1,q^{\ast}=1$) or ($p^{\ast}=0,q^{\ast}=0$). Mismatched
tactics, i.e. ($p^{\ast}=1,q^{\ast}=0$) or ($p^{\ast}=0,q^{\ast}=1$), both
lead to a worst-case situation.

Benjamin \cite{Benjamin2} also pointed out a \emph{difference} in terminology.
In Marinatto and Weber's set-up an initial strategy, in the form of a quantum
state, is forwarded to the players who then apply their tactics to modify the
state. In the Eisert et al.'s scheme, on the other hand, players' `moves' are
their manipulations, and their overall act of choosing what move to play is
their strategy.

Recently Nawaz and Toor \cite{NawazToor} showed that by using a more general
initial quantum state the dilemma in the classical BoS can be resolved, and a
unique solution can be found.

\subsection{Marinatto and Weber's reply}

In a reply Marinatto and Weber \cite{Marinatto's reply} defended their choice
of calling strategies the quantum states instead of operators used to
manipulate them. They claimed that their choice is very natural and consistent
with the spirit of classical game theory, where at the start of a game each
player has at her disposal an ensemble of strategies.

Regarding Benjamin's claim that the dilemma persists in quantum BoS since
players cannot decide between the two options, i.e. ($p^{\ast}=0,q^{\ast}=0 $)
and ($p^{\ast}=1,q^{\ast}=1$), Marinatto and Weber replied that the second
option of doing nothing ($p^{\ast}=1,q^{\ast}=1$) amounts to the most rational
behavior of the two players. According to them no incentive exists for a
player for doing something (i.e. $p^{\ast}=0$ or $q^{\ast}=0$) because:

\begin{itemize}
\item  It cannot lead to a better payoff and each player knows that.

\item  It only generates the extra risk of incurring a loss.

\item  It is more expensive than doing nothing, in terms of resources needed
to operate on the strategies.
\end{itemize}

\subsection{`Quantum form' of a matrix game and initial quantum states}

In Eisert et al.'s set-up when the parameter $\gamma$ of the initial quantum
state is different from zero, the players' payoffs are generally
non-classical, except for special moves available to them that can result in
the classical payoffs. Eisert et al. allow a range of values to the parameter
$\gamma$ and found how it affects the equilibria of the game.

Marinatto and Weber \cite{Marinatto1} forward an initial strategy to the two
players who then apply their `tactics' on it. In their scheme the classical
game corresponds to the initial state $\left|  00\right\rangle $.

Suppose the players receive pure two-qubit states, different from $\left|
00\right\rangle $, but the measurement uses the \emph{same} payoff operators.
The payoff operators used in measurement in Marinatto and Weber's scheme
contains \emph{all} the information about what matrix game is being played.
Given the measurement apparatus remains the same, a `quantum form' of the
matrix game can be obtained by \emph{only }choosing among different initial
states. Hence, this approach translates the problem of finding a quantum form
of a matrix game to the problem of finding a pure initial state.

The approach should be seen from the view that the only restriction on a
`quantum form' of a game is that the corresponding classical game must be
reproducible as a special case. Because a product initial state results in a
classical game, therefore, the above approach is within the mentioned restriction.

The Eisert et al.'s set-up suggests studying the behavior of equilibria in
relation to the parameter $\gamma$. The above approach, on the other hand,
suggests studying the behavior of equilibria in relation to different pure
initial states.

\section{\label{EnkPikeComment}Enk and Pike's comment on quantum Prisoners' Dilemma}

More recently Enk and Pike \cite{EnkPike} have argued that the quantum
solutions of PD, found by Eisert et al. \cite{Eisert}, are neither quantum
mechanical nor do they solve the classical game. Their argument is based on
the observation that it is possible to capture the essence of quantized PD by
simply extending the payoff matrix of the classical game, by \emph{only}
including an additional purely classical move corresponding to $\hat{Q}$,
which Eisert et al. obtained as a new quantum-mechanical `solution-move' that
could remove the dilemma inherent in the game. Enk and Pike maintained that
when Eisert's quantum solution to PD can be reconstructed in a classical way,
the only defense that remains for its quantum solution is its efficiency,
which does not play a role in PD.

Enk and Pike also suggested that a quantum game that exploits non-classical
correlations in entangled states, similar to those that violate the Bell's
inequality, should be worthy of investigation. Such correlations are without a
role in Eisert et al.'s set-up, and other quantization procedures derived from
it, even though entangled states may be present. It is because various qubits,
after their local unitary manipulations, are brought together during the final
stage of the game to make the payoffs-generating measurement.

\chapter{Evolutionary stability in quantum games}

\section{Introduction}

As discussed in Section (\ref{EGT}), the concept of an ESS was introduced in
classical game theory for two reasons:

\begin{enumerate}
\item \label{Reason1}Two player games can have multiple Nash equilibria and
ESS offers its refinement notion.

\item \label{Reason2}Population biology problems can be modelled with the help
of the ESS concept.
\end{enumerate}

The reasons for claim that (\ref{Reason1}) holds for quantum as well as
classical games are not far from obvious. In our opinion the reason
(\ref{Reason2}) also has a meaning in a quantum context. Like NE, the ESS is a
game-theoretic concept. The concept assumes a population setting which is
relevant to problems in evolutionary biology. As a game-theoretic concept, the
ESS is equally worthy of investigation as the concept of NE is in relation to
quantization of games. The view that a population setting of evolutionary
biology can not be relevant in quantum games is based on the assumption that
the participants in a quantum game \emph{must} always be rational agents. We
believe that when the rewards for players, forming a population, not only
depend on their individual moves but also on whether the game they play is
classical or quantum in nature, then the concepts fundamentally developed for
a population setting also become relevant in a quantum context. As mentioned
in the Section (\ref{EGT}), John Nash himself had a population setting in his
mind when he introduced his equilibrium notion. His equilibrium notion is
well-known from the early known studies in quantum games. The fact that a
population setting was behind the notion of a NE provides an almost natural
relevance of this setting for the quantum games as well. The idea of a
population of `quantum players' itself is not very much beyond imagination.
Such a population may, for example, consist of a large number of interacting
molecules where `decisions' are taken in individual quantum interactions.
These interactions can easily be imagined pair-wise and also random, which are
the fundamental assumptions behind the concept of an ESS.

In quantum setting the players' payoffs become sensitive to quantum affects.
Which direction evolution drives the population of quantum players now? The
direction should, of course, be decided by the nature of quantum affects.

\section{Quantization as a refinement notion of Nash equilibrium?}

Research in quantum games \cite{Eisert,Marinatto1}\ has shown appearance of
entirely new equilibria on quantization of a game. The next logical question
is to ask whether quantization can provide another refinement to the NE
concept? Such a question is relevant in a situation where an equilibrium is
retained, whether the game is played classically or quantum mechanically, but
some property of the equilibrium changes during such a switch-over. ESS, being
a refinement notion of the NE concept, is a symmetric NE with an extra
property of stability against small perturbations. We believe the question
whether quantization can affect stability of a symmetric NE is equally
interesting as the question how quantization leads to different equilibria.

\section{Quantization changing evolutionary stability?}

Our motivation is how game-theoretic models, of evolutionary dynamics in a
population, shape themselves in the new setting recently provided to game
theory by quantum mechanics? This motivation is, in a sense, a portion of a
bigger question: Can quantum mechanics have a role in directing, or even
dictating, the dynamics of evolution? To study evolution in a quantum setting
we have chosen the ESS concept firstly for its beauty and simplicity.
Secondly, because ESS is a game-theoretical concept, the new developments in
quantum games themselves provide a motivation to look at the resulting effects
on such concepts. Following questions arise immediately:

\begin{itemize}
\item  How ESSs are affected when a classical game, played by a population,
changes itself to one of its quantum forms?

\item  How pure and mixed ESSs are distinguished from one another when such a
change in the form of a game takes place?
\end{itemize}

And most importantly

\begin{itemize}
\item  How and if evolutionary dynamics can be related to quantum entanglement?
\end{itemize}

Imagine a population of players in which a classical strategy has established
itself as an ESS. We ask:

\begin{itemize}
\item  What happens when `mutants' of ESS theory come up with quantum
strategies and try to invade the classical ESS?

\item  What happens if such an invasion is successful and a new ESS is
established -- an ESS that is quantum in nature?

\item  Suppose afterwards another small group of mutants appears which is
equipped with some other quantum strategy. Would it be successful now to
invade the quantum ESS?
\end{itemize}

In the following we present an analysis based on these questions considering a
population in which symmetric pair-wise contests are taking place.

In trying to extend an idea, originally proposed for problems in population
biology, to quantum domain we give an analysis using Eisert et al.'s
quantization of the symmetric bi-matrix game of PD.

\section{ESSs in Eisert, Wilkens and Lewenstein's scheme}

For PD Cooperation ($C$) and Defection ($D$) are the pure classical
strategies. Which strategies are likely to be stable and persistent when the
game is played by a population engaged in pair-wise contests? In each such
contest PD is played. Straightforward analysis \cite{Prestwich} shows that $D$
will be the pure classical strategy prevalent in the population and hence the
classical ESS.

Eisert et al. used the matrix (\ref{PDmatrix1}) in their quantum version of
PD. Assume a population setting where in each pair-wise encounter the players
play PD with the \emph{same} matrix. Consider the following three situations:

\begin{enumerate}
\item \label{caseA}A small group of mutants appear equipped with one-parameter
quantum strategy $\hat{U}(\theta)$ when $D$ exists as a classical ESS.

\item \label{caseB}The mutants are equipped with two-parameter quantum
strategy $\hat{U}(\theta,\phi)$ against the classical ESS.

\item \label{caseC}The mutants have successfully invaded and a two-parameter
quantum strategy $\hat{Q}\sim\hat{U}(0,\pi/2)$ has established itself as a new
quantum ESS. Again another small group of mutants appear, using some other
two-parameter quantum strategy, and try to invade the quantum ESS, that is
$\hat{Q}$.
\end{enumerate}

\subsection{Case (\ref{caseA})}

In quantum PD with the matrix (\ref{PDmatrix1}):%

\begin{equation}
\left(
\begin{array}
[c]{cc}%
(3,3) & (0,5)\\
(5,0) & (1,1)
\end{array}
\right)  \label{PDMatrix2}%
\end{equation}
the players are anonymous and one can denote, for example, $P(\hat{U}%
(\theta),D)$ to represent the payoff to $\hat{U}(\theta)$-player against the
$D$-player. Here $\hat{U}(\theta)$ is the Eisert et al.'s one-parameter
quantum strategy set (\ref{OneParameterSet}). Players' payoffs can be found as%

\begin{align}
P(\hat{U}(\theta),D)  &  =\sin^{2}(\theta/2)\nonumber\\
P(\hat{U}(\theta),\hat{U}(\theta))  &  =2\cos^{2}(\theta/2)+5\cos^{2}%
(\theta/2)\sin^{2}(\theta/2)+1\nonumber\\
P(D,\hat{U}(\theta))  &  =5\cos^{2}(\theta/2)+\sin^{2}(\theta/2)\nonumber\\
P(D,D)  &  =1
\end{align}
Now $P(D,D)>P(\hat{U}(\theta),D)$ for all $\theta\in\lbrack0,\pi)$. Hence the
first condition for an ESS holds and $D\sim\hat{U}(\pi)$ is an ESS. The case
$\theta=\pi$ corresponds to one-parameter mutant strategy coinciding with the
ESS, which is ruled out. If $D\sim\hat{U}(\pi)$ is played by almost all the
members of the population -- which corresponds to high frequency $F_{D}$ for
$D$ -- we then have $W(D)>W(\theta)$ for all $\theta\in\lbrack0,\pi)$. The
fitness of a one-parameter quantum strategy\footnote{In Eisert et al.'s set-up
one-parameter quantum strategies correspond to mixed (randomized) classical
strategies.}, therefore, cannot be greater than that of a classical ESS. And a
one-parameter quantum strategy cannot invade a classical ESS.

\subsection{Case (\ref{caseB})}

Let $\hat{U}(\theta,\phi)$ be a two-parameter strategy from the set
(\ref{TwoParameterSet}). The expected payoffs are%

\begin{align}
P(D,D)  &  =1\nonumber\\
P(D,\hat{U}(\theta,\phi))  &  =5\cos^{2}(\phi)\cos^{2}(\theta/2)+\sin
^{2}(\theta/2)\nonumber\\
P(\hat{U}(\theta,\phi),D)  &  =5\sin^{2}(\phi)\cos^{2}(\theta/2)+\sin
^{2}(\theta/2)\nonumber\\
P(\hat{U}(\theta,\phi),\hat{U}(\theta,\phi))  &  =3\left|  \cos(2\phi)\cos
^{2}(\theta/2)\right|  ^{2}+5\cos^{2}(\theta/2)\sin^{2}(\theta/2)\left|
\sin(\phi)-\cos(\phi)\right|  ^{2}+\nonumber\\
&  \left|  \sin(2\phi)\cos^{2}(\theta/2)+\sin^{2}(\theta/2)\right|  ^{2}%
\end{align}
Here $P(D,D)>P(\hat{U}(\theta,\phi),D)$ if $\phi<\arcsin(1/\sqrt{5})$ and if
$P(D,D)=P(\hat{U}(\theta,\phi),D)$ then $P(D,\hat{U}(\theta,\phi))>P(\hat
{U}(\theta,\phi),\hat{U}(\theta,\phi))$. Therefore $D$ is an ESS if
$\phi<\arcsin(1/\sqrt{5})$ otherwise the strategy $\hat{U}(\theta,\phi)$ will
be in position to invade $D$. Alternatively if most of the members of the
population play $D\sim\hat{U}(\pi,0)$ -- meaning high frequency $F_{D}$ for
$D$ -- then the fitness $W(D)$ will remain greater than the fitness $W[\hat
{U}(\theta,\phi)]$ if $\phi<\arcsin(1/\sqrt{5})$. For $\phi>\arcsin(1/\sqrt
{5})$ the strategy $\hat{U}(\theta,\phi)$ can invade the strategy $D$, which
is the classical ESS.

In this analysis the possession of a richer strategy by the mutants leads to
invasion of $D$ when $\phi>\arcsin(1/\sqrt{5})$. Such an invasion may seem not
so unusual given the mutants exploiting richer strategies. But it leads to the
third case when `quantum mutants' have successfully invaded and a
two-parameter strategy $\hat{U}$ has established itself. Can now some new
mutants coming up with $\hat{Q}\sim\hat{U}(0,\pi/2)$ and invade the `quantum ESS'?

\subsection{Case (\ref{caseC})}

Eisert et al. \cite{Eisert,Eisert1} showed that in their quantum PD the
quantum strategy $\hat{Q}$, played by both the players, is the unique NE. How
mutants playing $\hat{Q}$ come up against $\hat{U}(\theta,\phi)$ which already
exists as an ESS? To find it following payoffs are obtained.%

\begin{align}
P(\hat{Q},\hat{Q})  &  =3\nonumber\\
P(\hat{U}(\theta,\phi),\hat{Q})  &  =[3-2\cos^{2}(\phi)]\cos^{2}%
(\theta/2)\nonumber\\
P(\hat{Q},\hat{U}(\theta,\phi))  &  =[3-2\cos^{2}(\phi)]\cos^{2}%
(\theta/2)+5\sin^{2}(\theta/2)
\end{align}
Now the inequality $P(\hat{Q},\hat{Q})>P(\hat{U}(\theta,\phi),\hat{Q})$ holds
for all $\theta\in\lbrack0,\pi]$ and $\phi\in\lbrack0,\pi/2]$ except when
$\theta=0$ and $\phi=\pi/2$, which is the case when the mutant strategy
$\hat{U}(\theta,\phi)$ is the same as $\hat{Q}$. This case is obviously ruled
out. The first condition for $\hat{Q}$ to be an ESS, therefore, holds. The
condition $P(\hat{Q},\hat{Q})=P(\hat{U}(\theta,\phi),\hat{Q})$ implies
$\theta=0$ and $\phi=\pi/2$. Again we have the situation of mutant strategy
same as $\hat{Q}$ and the case is neglected. If $\hat{Q}$ is played by most of
the players, meaning high frequency $F_{\hat{Q}}$ for $\hat{Q}$, then
$W(\hat{Q})>W[\hat{U}(\theta,\phi)]$ for all $\theta\in(0,\pi]$ and $\phi
\in\lbrack0,\pi/2)$. A two parameter quantum strategy $\hat{U}(\theta,\phi)$,
therefore, cannot invade the quantum ESS (i.e. the strategy $\hat{Q}\sim
\hat{U}(0,\pi/2)$). Mutants' access to richer strategies, as it happens in the
case (B), does not continue to be an advantage and most of the population also
have access to it. Hence $\hat{Q}$ comes out as the unique NE and ESS of the game.

\section{ESSs in Marinatto and Weber's scheme}

What happens to PD, from the point of view of evolutionary stability, when it
is played via Marinatto and Weber's scheme \cite{Marinatto1}? In our view this
scheme is more suitable for consideration of evolutionary stability in quantum
regime for the following reasons:

\begin{itemize}
\item  In a symmetric bi-matrix game, played in a population setting, players
have access to two pure strategies. Players can also play a mixed strategy by
combining the pure strategies with certain probabilities. In a similar way
players in Marinatto and Weber's scheme can be said to play a mixed strategy
when they apply the two unitary operators, on the initial state, with a
probabilistic combination.

\item  Definition (\ref{fitnesses}) of fitness of a pure strategy in
evolutionary games \cite{Prestwich}\ can be given a straightforward extension
in Marinatto and Weber's scheme. It corresponds to a situation when, in the
quantum game, a player uses only one unitary operator out of the two.

\item  Theory of ESSs, in the classical domain, deals with anonymous players
possessing a discrete number of pure strategies. Eisert's scheme involves
players possessing a continuum of pure quantum strategies. The concept of an
ESS as a stable equilibrium is confronted with problems \cite{Oechssler} when
players possess a continuum of pure strategies.
\end{itemize}

\subsection{Example of quantum Prisoners' Dilemma}

Assume the PD, defined by the matrix (\ref{PDMatrix2}), is played with
Marinatto and Weber's scheme. The initial state made available to the players is%

\begin{equation}
\left|  \psi_{in}\right\rangle =a\left|  CC\right\rangle +b\left|
DD\right\rangle \text{, \ \ \ \ with }\left|  a\right|  ^{2}+\left|  b\right|
^{2}=1 \label{IniStatQuantumPD}%
\end{equation}
where $\left|  C\right\rangle \sim\left|  0\right\rangle $ and $\left|
D\right\rangle \sim\left|  1\right\rangle $. Payoffs to Alice and Bob can be
found as%

\begin{align}
P_{A}(p,q)  &  =3\{pq\left|  a\right|  ^{2}+(1-p)(1-q)\left|  b\right|
^{2}\}+5\{p(1-q)\left|  b\right|  ^{2}+q(1-p)\left|  a\right|  ^{2}%
\}+\nonumber\\
&  \{pq\left|  b\right|  ^{2}+(1-p)(1-q)\left|  a\right|  ^{2}\}\nonumber\\
P_{B}(p,q)  &  =3\{pq\left|  a\right|  ^{2}+(1-p)(1-q)\left|  b\right|
^{2}\}+5\{p(1-q)\left|  a\right|  ^{2}+q(1-p)\left|  b\right|  ^{2}%
\}+\nonumber\\
&  \{pq\left|  b\right|  ^{2}+(1-p)(1-q)\left|  a\right|  ^{2}\}
\end{align}
where $p$ and $q$ are the probabilities for Alice and Bob, respectively, to
act with the operator $\hat{I}$. We look for symmetric Nash equilibria from
the Nash inequalities while using only the parameter $b\in\mathbf{C}$ of the
initial state $\left|  \psi_{in}\right\rangle $. For the state
(\ref{IniStatQuantumPD}) the game is reduced to the classical game when
$\left|  b\right|  ^{2}=0$, i.e. when it is a product state. Nash inequalities
are then%

\begin{align}
P_{A}(\overset{\star}{p},\overset{\star}{q})-P_{A}(p,\overset{\star}{q})  &
=(\overset{\star}{p}-p)\{3\left|  b\right|  ^{2}-(\overset{\star}{q}%
+1)\}\geq0\nonumber\\
P_{B}(\overset{\star}{p},\overset{\star}{q})-P_{B}(\overset{\star}{p},q)  &
=(\overset{\star}{q}-q)\{3\left|  b\right|  ^{2}-(\overset{\star}{p}+1)\}\geq0
\end{align}
Here the parameter $b$ decides what should be the Nash equilibria of the game.
Three symmetric equilibria arise:%

\begin{align}
1.\qquad\overset{\star}{p}  &  =\overset{\star}{q}=0\text{ \ when \ }3\left|
b\right|  ^{2}\leq1\nonumber\\
2.\qquad\overset{\star}{p}  &  =\overset{\star}{q}=1\text{ \ when \ }3\left|
b\right|  ^{2}\geq2\nonumber\\
3.\qquad\overset{\star}{p}  &  =\overset{\star}{q}=3\left|  b\right|
^{2}-1\text{ \ when \ }1<3\left|  b\right|  ^{2}<2
\end{align}
The first two equilibria are independent of the parameter $b$ while the third
depends on it. We ask which of these equilibria are evolutionary stable
assuming that an equilibrium exists for initial states of the form
(\ref{IniStatQuantumPD}). Because the players play a symmetric game, the
payoff to a player using $\hat{I}$ with probability $p$, when opponent uses it
with the probability $q$, can be written as%

\begin{align}
P(p,q)  &  =3\{pq\left|  a\right|  ^{2}+(1-p)(1-q)\left|  b\right|
^{2}\}+5\{p(1-q)\left|  b\right|  ^{2}+q(1-p)\left|  a\right|  ^{2}%
\}+\nonumber\\
&  \{pq\left|  b\right|  ^{2}+(1-p)(1-q)\left|  a\right|  ^{2}\}
\end{align}
which can also be identified as the payoff to the $p$-player against the
$q$-player. For the strategy pair $\overset{\star}{p}=\overset{\star}{q}=0$
one gets $P(0,0)>P(p,0)$ when $3\left|  b\right|  ^{2}<1$ and $P(0,0)=P(p,0)$
imply $3\left|  b\right|  ^{2}=1$. Also $P(q,q)=-q^{2}+\frac{5}{3}(q+1)$ and
$P(0,q)=\frac{5}{3}(q+1)$. Now $P(0,q)>P(q,q)$ when $q\neq0.$ Therefore the
pair $\overset{\star}{p}=\overset{\star}{q}=0$ is an ESS when $3\left|
b\right|  ^{2}\leq1$.

For the pair $\overset{\star}{p}=\overset{\star}{q}=1$ we have $P(1,1)>P(p,1)$
which means $3\left|  b\right|  ^{2}>2$ if $p\neq1$. And $P(1,1)=P(p,1)$ means
for$\ p\neq1$ we have $3\left|  b\right|  ^{2}=2$. In such case $P(q,q)=-q^{2}%
+\frac{1}{3}(q+7)$ and $P(1,q)=\frac{5}{3}(2-q)$. Now $P(1,q)>P(q,q)$ because
$(1-q)^{2}>0$ for $q\neq1$. Therefore $\overset{\star}{p}=\overset{\star}%
{q}=1$ is an ESS when $3\left|  b\right|  ^{2}\geq2 $.

For the third pair, $\overset{\star}{p}=\overset{\star}{q}=3\left|  b\right|
^{2}-1$, we get $P(3\left|  b\right|  ^{2}-1,3\left|  b\right|  ^{2}%
-1)=-36\left|  b\right|  ^{6}+36\left|  b\right|  ^{4}-5\left|  b\right|
^{2}+6$. Also we find $P(p,3\left|  b\right|  ^{2}-1)=-21\left|  b\right|
^{4}+21\left|  b\right|  ^{2}-3$. Hence, the condition $P(3\left|  b\right|
^{2}-1,3\left|  b\right|  ^{2}-1)>P(p,3\left|  b\right|  ^{2}-1)$ holds and
the pair $\overset{\star}{p}=\overset{\star}{q}=3\left|  b\right|  ^{2}-1$ is
an ESS for $1<3\left|  b\right|  ^{2}<2$.

All three symmetric equilibria, definable for different ranges of $\left|
b\right|  ^{2}$, are also ESSs. Each of the three sets of initial states
$\left|  \psi_{in}\right\rangle $\ give a unique equilibrium which is an ESS
too. Switching from one to the other set of initial states also changes the
equilibrium and ESS accordingly.

A question arises here: Is it possible that a particular equilibrium
switches-over between `ESS' and `not ESS' when the initial state changes
between some of its possible choices? The question is relevant given the fact
that transition between classical and quantum game is also achieved by a
change in the initial state: classical payoffs are obtained when the initial
state is a product state. It implies that it may be possible for a symmetric
NE to switch-over between being `ESS' and being `not ESS' when a game changes
between its `classical' and `quantum' forms. This possibility makes the ESS
concept interesting also from the point of view of quantum games. Because the
quantum PD, in the form considered above, does not allow such a possibility,
therefore, asymmetric bi-matrix games are investigated now.

\subsection{ESSs in two-player two-strategy asymmetric games}

ESS for an asymmetric bi-matrix game, i.e. $G=(M,N)$ when $N\neq M^{T}$, is
defined as a strict NE \cite{Weibull}. A strategy pair $(\overset{\star}%
{x},\overset{\star}{y})\in S$ is an ESS of the game $G$ if it is a strict NE:%

\begin{align}
1.\qquad P_{A}(\overset{\star}{x},\overset{\star}{y})  &  >P_{A}%
(x,\overset{\star}{y})\text{ for all }x\neq\overset{\star}{x}\nonumber\\
2.\qquad P_{B}(\overset{\star}{x},\overset{\star}{y})  &  >P_{B}%
(\overset{\star}{x},y)\text{ for all }y\neq\overset{\star}{y}%
\end{align}
For example, consider BoS with the matrix:%

\begin{equation}
\left(
\begin{array}
[c]{cc}%
(\alpha,\beta) & (\gamma,\gamma)\\
(\gamma,\gamma) & (\beta,\alpha)
\end{array}
\right)  \label{AsymmetricGame}%
\end{equation}
where $\alpha>\beta>\gamma$. It is a asymmetric game with three classical NE
\cite{Marinatto1}:%

\begin{align}
1.\qquad\overset{\star}{p_{1}}  &  =\overset{\star}{q_{1}}=0\nonumber\\
2.\qquad\overset{\star}{p_{2}}  &  =\overset{\star}{q_{2}}=1\nonumber\\
3.\qquad\overset{\star}{p_{3}}  &  =\frac{\alpha-\gamma}{\alpha+\beta-2\gamma
}\qquad\overset{\star}{q_{3}}=\frac{\beta-\gamma}{\alpha+\beta-2\gamma}%
\end{align}
The equilibria ($1$) and ($2$) are also ESS's but ($3$) is not because it is
not a strict NE. The asymmetric game (\ref{AsymmetricGame}) played with an
initial state $\left|  \psi_{in}\right\rangle =a\left|  S_{1}S_{1}%
\right\rangle +b\left|  S_{2}S_{2}\right\rangle $, where $S_{1}$ and $S_{2}$
are players' pure classical strategies, has the following three Nash
equilibria \cite{Marinatto1}:%

\begin{align}
1.\qquad\overset{\star}{p_{1}}  &  =\overset{\star}{q_{1}}=1\nonumber\\
2.\qquad\overset{\star}{p_{2}}  &  =\overset{\star}{q_{2}}=0\nonumber\\
3.\qquad\overset{\star}{p_{3}}  &  =\frac{(\alpha-\gamma)\left|  a\right|
^{2}+(\beta-\gamma)\left|  b\right|  ^{2}}{\alpha+\beta-2\gamma}\qquad
\overset{\star}{q_{3}}=\frac{(\alpha-\gamma)\left|  b\right|  ^{2}%
+(\beta-\gamma)\left|  a\right|  ^{2}}{\alpha+\beta-2\gamma}\nonumber\\
&
\end{align}
Similar to the classical case, the equilibria ($1$) and ($2$) are ESSs while
($3$) is not. First two ESSs do not depend on the parameters $a$ and $b$ of
the initial state while the third NE does. Interestingly, playing BoS game
with a different initial state:%

\begin{equation}
\left|  \psi_{in}\right\rangle =a\left|  S_{1}S_{2}\right\rangle +b\left|
S_{2}S_{1}\right\rangle \label{antisymmetricState}%
\end{equation}
changes the scene. The payoffs to Alice and Bob are:%

\begin{align}
P_{A}(p,q)  &  =p\left\{  -q(\alpha+\beta-2\gamma)+\alpha\left|  a\right|
^{2}+\beta\left|  b\right|  ^{2}-\gamma\right\}  +\nonumber\\
&  q\left\{  \alpha\left|  b\right|  ^{2}+\beta\left|  a\right|  ^{2}%
-\gamma\right\}  +\gamma\nonumber\\
P_{B}(p,q)  &  =q\left\{  -p(\alpha+\beta-2\gamma)+\beta\left|  a\right|
^{2}+\alpha\left|  b\right|  ^{2}-\gamma\right\}  +\nonumber\\
&  p\left\{  \beta\left|  b\right|  ^{2}+\alpha\left|  a\right|  ^{2}%
-\gamma\right\}  +\gamma
\end{align}
and there is only one NE i.e.%

\begin{equation}
\overset{\star}{p}=\frac{\beta\left|  a\right|  ^{2}+\alpha\left|  b\right|
^{2}-\gamma}{\alpha+\beta-\gamma}\qquad\overset{\star}{q_{3}}=\frac
{\alpha\left|  a\right|  ^{2}+\beta\left|  b\right|  ^{2}-\gamma}{\alpha
+\beta-\gamma}%
\end{equation}
which is not an ESS. So that no ESS exists when BoS is played with the state
(\ref{antisymmetricState}).

An essential requirement on a quantum version of a game is that the
corresponding classical game must be its subset. Suppose for a quantum game,
corresponding to an asymmetric bi-matrix classical game, a particular strategy
pair $(\overset{\star}{x},\overset{\star}{y})$ is an ESS for all initial
states $\left|  \psi_{in}\right\rangle $ given in some particular form. That
is, it remains an ESS for all $a$ and $b$ when $\left|  \psi_{in}\right\rangle
$ is given in terms of $a$ and $b$. Because classical game is a subset of the
quantum game, the strategy pair $(\overset{\star}{x},\overset{\star}{y})$
\emph{must} then also be an ESS in the classical game. On the other hand a
strategy pair $(\overset{\star}{x},\overset{\star}{y})$ which is an ESS in a
classical game \emph{may not} remain so in its quantum version.

Quantization of an asymmetric classical game can thus make disappear the
classical ESSs but it cannot make appear new ESSs, provided an ESS in quantum
version remains so for every possible choice of the parameters $a$ and $b$.
However when an ESS is defined as a strict NE existing only for a set of
initial states for which that NE exists, the statement that quantization can
only make disappear classically available ESSs may not remain valid. In such a
case quantization may make appear new ESSs definable for certain ranges of the
parameters $a$ and $b$.

To find games with the property that a particular NE switches over between
`ESS' and `not ESS' when the initial state changes between its possible
choices, we now look at the following asymmetric quantum game:%

\begin{equation}
\left(
\begin{array}
[c]{cc}%
(\alpha_{1},\alpha_{2}) & (\beta_{1},\beta_{2})\\
(\gamma_{1},\gamma_{2}) & (\sigma_{1},\sigma_{2})
\end{array}
\right)
\end{equation}
where%

\begin{equation}
\left(
\begin{array}
[c]{cc}%
\alpha_{1} & \beta_{1}\\
\gamma_{1} & \sigma_{1}%
\end{array}
\right)  \neq\left(
\begin{array}
[c]{cc}%
\alpha_{2} & \beta_{2}\\
\gamma_{2} & \sigma_{2}%
\end{array}
\right)  ^{T}%
\end{equation}
For the initial state $\left|  \psi_{in}\right\rangle =a\left|  S_{1}%
S_{1}\right\rangle +b\left|  S_{2}S_{2}\right\rangle $ the players' payoffs are:%

\begin{align}
P_{A,B}(p,q)  &  =\alpha_{1,2}\left\{  pq\left|  a\right|  ^{2}%
+(1-p)(1-q)\left|  b\right|  ^{2}\right\}  +\nonumber\\
&  \beta_{1,2}\left\{  p(1-q)\left|  a\right|  ^{2}+q(1-p)\left|  b\right|
^{2}\right\}  +\nonumber\\
&  \gamma_{1,2}\left\{  p(1-q)\left|  b\right|  ^{2}+q(1-p)\left|  a\right|
^{2}\right\}  +\nonumber\\
&  \sigma_{1,2}\left\{  pq\left|  b\right|  ^{2}+(1-p)(1-q)\left|  a\right|
^{2}\right\}
\end{align}
The NE conditions are%

\begin{gather}
P_{A}(\overset{\star}{p},\overset{\star}{q})-P_{A}(p,\overset{\star}%
{q})=\nonumber\\
(\overset{\star}{p}-p)\left[  \left|  a\right|  ^{2}(\beta_{1}-\sigma
_{1})+\left|  b\right|  ^{2}(\gamma_{1}-\alpha_{1})-\overset{\star}{q}\left\{
(\beta_{1}-\sigma_{1})+(\gamma_{1}-\alpha_{1})\right\}  \right]  \geq0\\
P_{B}(\overset{\star}{p},\overset{\star}{q})-P_{B}(\overset{\star}%
{p},q)=\nonumber\\
(\overset{\star}{q}-q)\left[  \left|  a\right|  ^{2}(\gamma_{2}-\sigma
_{2})+\left|  b\right|  ^{2}(\beta_{2}-\alpha_{2})-\overset{\star}{p}\left\{
(\gamma_{2}-\sigma_{2})+(\beta_{2}-\alpha_{2})\right\}  \right]  \geq0
\end{gather}
Let now $\overset{\star}{p}=\overset{\star}{q}=0$ be a NE:%

\begin{align}
P_{A}(0,0)-P_{A}(p,0)  &  =-p\left[  (\beta_{1}-\sigma_{1})+\left|  b\right|
^{2}\left\{  (\gamma_{1}-\alpha_{1})-(\beta_{1}-\sigma_{1})\right\}  \right]
\geq0\nonumber\\
P_{B}(0,0)-P_{B}(0,q)  &  =-q\left[  (\gamma_{2}-\sigma_{2})+\left|  b\right|
^{2}\left\{  (\beta_{2}-\alpha_{2})-(\gamma_{2}-\sigma_{2})\right\}  \right]
\geq0\nonumber\\
&
\end{align}
When the strategy pair $(0,0)$ is an ESS in the classical game $($i.e.
$\left|  b\right|  ^{2}=0)$ we should have%

\begin{align}
P_{A}(0,0)-P_{A}(p,0)  &  =-p(\beta_{1}-\sigma_{1})>0\text{ for all }%
p\neq0\nonumber\\
P_{B}(0,0)-P_{B}(0,q)  &  =-q(\gamma_{2}-\sigma_{2})>0\text{ for all }q\neq0
\end{align}
It implies $(\beta_{1}-\sigma_{1})<0$ and $(\gamma_{2}-\sigma_{2})<0$.

For the pair $(0,0)$ to be not an ESS for some $\left|  b\right|  ^{2}\neq0$,
let take $\gamma_{1}=\alpha_{1\text{ }}$and $\beta_{2}=\alpha_{2}$. We have%

\begin{align}
P_{A}(0,0)-P_{A}(p,0)  &  =-p(\beta_{1}-\sigma_{1})\left\{  1-\left|
b\right|  ^{2}\right\} \nonumber\\
P_{B}(0,0)-P_{B}(0,q)  &  =-q(\gamma_{2}-\sigma_{2})\left\{  1-\left|
b\right|  ^{2}\right\}
\end{align}
and the pair $(0,0)$ doesn't remain an ESS at $\left|  b\right|  ^{2}=1$. A
game with these properties is given by the matrix:%

\begin{equation}
\left(
\begin{array}
[c]{cc}%
(1,1) & (1,2)\\
(2,1) & (3,2)
\end{array}
\right)
\end{equation}
For this game the pair $(0,0)$ is an ESS when $\left|  b\right|  ^{2}=0$
(classical game) but it is not when for example $\left|  b\right|  ^{2}%
=\frac{1}{2}$, though it remains a NE in both the cases. The example shows a
NE can be switched between ESS and `not ESS' by adjusting the parameters $a$
and $b$ of the initial state. Opposite to the previous case, the initial
states - different from the one corresponding to the classical game - can also
make a strategy pair an ESS. An example of a game for which it happens is%

\begin{equation}%
\begin{array}
[c]{c}%
S_{1}\\
S_{2}%
\end{array}
\overset{%
\begin{array}
[c]{cc}%
S_{1} & S_{2}%
\end{array}
}{\left(
\begin{array}
[c]{cc}%
(2,1) & (1,0)\\
(1,0) & (1,0)
\end{array}
\right)  } \label{ExampleGame1}%
\end{equation}
Playing this game again via $\left|  \psi_{in}\right\rangle =a\left|
S_{1}S_{1}\right\rangle +b\left|  S_{2}S_{2}\right\rangle $ gives the
following payoff differences for the strategy pair $(0,0)$:%

\begin{equation}
P_{A}(0,0)-P_{A}(p,0)=p\left|  b\right|  ^{2}\ \ \text{and}\ \ P_{B}%
(0,0)-P_{B}(0,q)=q\left|  b\right|  ^{2}%
\end{equation}
for Alice and Bob respectively. Therefore (\ref{ExampleGame1}) is an example
of a game for which the pair $(0,0)$ is not an ESS when the initial state
corresponds to the classical game. But the pair is an ESS for other initial
states for which $0<\left|  b\right|  ^{2}<1$.

\subsection{ESSs in two-player two-strategy symmetric games}

To explore a possible relation between evolutionary stability and quantization
consider the following symmetric bi-matrix game:%

\begin{equation}%
\begin{array}
[c]{c}%
\text{Alice}%
\end{array}%
\begin{array}
[c]{c}%
S_{1}\\
S_{2}%
\end{array}
\overset{\overset{%
\begin{array}
[c]{c}%
\text{Bob}%
\end{array}
}{%
\begin{array}
[c]{cc}%
S_{1} & S_{2}%
\end{array}
}}{\left(
\begin{array}
[c]{cc}%
(\alpha,\alpha) & (\beta,\gamma)\\
(\gamma,\beta) & (\delta,\delta)
\end{array}
\right)  } \label{PayoffMatrixGen2Player}%
\end{equation}
which is played by an initial state:%

\begin{equation}
\left|  \psi_{in}\right\rangle =a\left|  S_{1}S_{1}\right\rangle +b\left|
S_{2}S_{2}\right\rangle \text{, \ \ with }\left|  a\right|  ^{2}+\left|
b\right|  ^{2}=1 \label{IniStatGen2Player}%
\end{equation}
Let Alice's strategy consists of applying the identity operator $\hat{I}$ with
probability $p$ and the operator $\hat{\sigma}_{x}$ with probability $(1-p)$,
on the initial state written $\rho_{in}$ in density matrix notation. Similarly
Bob applies the operators $\hat{I}$ and $\hat{\sigma}_{x} $ with the
probabilities $q$ and $(1-q)$ respectively. The final state is%

\begin{equation}
\rho_{fin}=\underset{\hat{U}=\hat{I},\hat{\sigma}_{x}}{\sum}\Pr(\hat{U}%
_{A})\Pr(\hat{U}_{B})[\hat{U}_{A}\otimes\hat{U}_{B}\rho_{in}\hat{U}%
_{A}^{\dagger}\otimes\hat{U}_{B}^{\dagger}]
\end{equation}
where unitary and Hermitian operator $\hat{U}$ is either $\hat{I}$ or
$\hat{\sigma}_{x}$. $\Pr(\hat{U}_{A})$, $\Pr(\hat{U}_{B})$ are the
probabilities, for Alice and Bob respectively, to apply the operator on the
initial state. The matrix $\rho_{fin}$ is obtained from $\rho_{in}$ by a
convex combination of players' possible quantum operations. Payoff operators
for Alice and Bob are \cite{Marinatto1}%

\begin{equation}
(P_{A,B})_{oper}=\alpha,\alpha\left|  S_{1}S_{1}\right\rangle \left\langle
S_{1}S_{1}\right|  +\beta,\gamma\left|  S_{1}S_{2}\right\rangle \left\langle
S_{1}S_{2}\right|  +\gamma,\beta\left|  S_{2}S_{1}\right\rangle \left\langle
S_{2}S_{1}\right|  +\delta,\delta\left|  S_{2}S_{2}\right\rangle \left\langle
S_{2}S_{2}\right|
\end{equation}
The payoffs are then obtained as mean values of these operators i.e.
$P_{A,B}=$Tr$\left[  (P_{A,B})_{oper}\rho_{fin}\right]  $. Because the quantum
game is symmetric with the initial state (\ref{IniStatGen2Player}) and the
payoff matrix (\ref{PayoffMatrixGen2Player}), there is no need for subscripts.
We can , then, write the payoff to a $p$-player against a $q$-player as
$P(p,q)$, where the first number is the focal player's move. When
$\overset{\star}{p}$ is a NE we find the following payoff difference:%

\begin{gather}
P(\overset{\star}{p},\overset{\star}{p})-P(p,\overset{\star}{p})=(\overset
{\star}{p}-p){\LARGE [}\left|  a\right|  ^{2}(\beta-\delta)+\nonumber\\
\left|  b\right|  ^{2}(\gamma-\alpha)-\overset{\star}{p}\left\{  (\beta
-\delta)+(\gamma-\alpha)\right\}  {\LARGE ]} \label{General2PlayerDifference1}%
\end{gather}
Now the ESS conditions for the pure strategy $p=0$ are given as%

\begin{gather}
1.\text{ \ \ \ }\left|  b\right|  ^{2}\left\{  (\beta-\delta)-(\gamma
-\alpha)\right\}  >(\beta-\delta)\nonumber\\
2.\text{ If }\left|  b\right|  ^{2}\left\{  (\beta-\delta)-(\gamma
-\alpha)\right\}  =(\beta-\delta)\nonumber\\
\text{then }q^{2}\left\{  (\beta-\delta)+(\gamma-\alpha)\right\}  >0
\end{gather}
where $1$ is the NE condition. Similarly the ESS conditions for the pure
strategy $p=1$ are%

\begin{gather}
1.\text{ \ \ \ }\left|  b\right|  ^{2}\left\{  (\gamma-\alpha)-(\beta
-\delta)\right\}  >(\gamma-\alpha)\nonumber\\
2.\text{ If }\left|  b\right|  ^{2}\left\{  (\gamma-\alpha)-(\beta
-\delta)\right\}  =(\gamma-\alpha)\nonumber\\
\text{then }(1-q)^{2}\left\{  (\beta-\delta)+(\gamma-\alpha)\right\}  >0
\end{gather}
Because these conditions, for both the pure strategies $p=1$ and $p=0$, depend
on $\left|  b\right|  ^{2}$, therefore, there can be examples of two-player
symmetric games for which the evolutionary stability of pure strategies can be
changed while playing the game using initial state in the form $\left|
\psi_{in}\right\rangle =a\left|  S_{1}S_{1}\right\rangle +b\left|  S_{2}%
S_{2}\right\rangle $. However, for the mixed NE, given as $\overset{\star}%
{p}=\frac{\left|  a\right|  ^{2}(\beta-\delta)+\left|  b\right|  ^{2}%
(\gamma-\alpha)}{(\beta-\delta)+(\gamma-\alpha)}$, the corresponding payoff
difference (\ref{General2PlayerDifference1}) becomes identically zero. From
the second condition of an ESS we find for the mixed NE $\overset{\star}{p}$
the difference%

\begin{align}
&  P(\overset{\star}{p},q)-P(q,q)=\frac{1}{(\beta-\delta)+(\gamma-\alpha
)}\times\nonumber\\
&  {\LARGE [}(\beta-\delta)-q\left\{  (\beta-\delta)+(\gamma-\alpha)\right\}
-\left|  b\right|  ^{2}\left\{  (\beta-\delta)-(\gamma-\alpha)\right\}
{\LARGE ]}^{2}%
\end{align}
Therefore, the mixed strategy $\overset{\star}{p}$ is an ESS when $\left\{
(\beta-\delta)+(\gamma-\alpha)\right\}  >0$. This condition, making the mixed
NE $\overset{\star}{p}$ an ESS, is independent \footnote{An alternative
possibility is to adjust $\left|  b\right|  ^{2}$=$\frac{(\beta-\delta
)-q\left\{  (\beta-\delta)+(\gamma-\alpha)\right\}  }{\left\{  (\beta
-\delta)-(\gamma-\alpha)\right\}  }$ which makes the difference $\left\{
P(\overset{\star}{p},q)-P(q,q)\right\}  $ identically zero. The mixed strategy
$\overset{\star}{p}$ then does not remain an ESS. However such `mutant
dependent' adjustment of $\left|  b\right|  ^{2}$ is not reasonable because
the mutant strategy $q$ can be anything in the range $[0,1]$.} of $\left|
b\right|  ^{2}$. So that, in this symmetric two-player quantum game,
evolutionary stability of the mixed NE $\overset{\star}{p}$ can not be changed
when the game is played using initial quantum states of the form
(\ref{IniStatGen2Player}).

However, evolutionary stability of pure strategies can be affected, with this
form of the initial states, for two-player symmetric games. Examples of the
games with this property are easy to find. The class of games for which
$\gamma=\alpha$ and $(\beta-\delta)<0$ the strategies $p=0$ and $p=1$ remain
NE for all $\left|  b\right|  ^{2}\in\lbrack0,1]$; but the strategy $p=1$ is
not an ESS when $\left|  b\right|  ^{2}=0$ and the strategy $p=0$ is not an
ESS when $\left|  b\right|  ^{2}=1$.

\subsubsection{An example}

Consider the symmetric bi-matrix game (\ref{PayoffMatrixGen2Player}) with the
constants $\alpha,\beta,\gamma,\delta$ satisfying the conditions:%

\begin{align}
\alpha,\beta,\gamma,\delta &  \geq0\nonumber\\
(\delta-\beta)  &  >0\nonumber\\
(\gamma-\alpha)  &  \geq0\nonumber\\
(\gamma-\alpha)  &  <(\delta-\beta) \label{GameDefinition}%
\end{align}
The condition making $(p^{\star},p^{\star})$ a NE is given by
(\ref{General2PlayerDifference1}). For this game three Nash equilibria arise
i.e. two pure strategies $p^{\ast}=0$, $p^{\ast}=1$, and one mixed strategy
$p^{\ast}=\frac{(\delta-\beta)\left|  a\right|  ^{2}-(\gamma-\alpha)\left|
b\right|  ^{2}}{(\delta-\beta)-(\gamma-\alpha)}$. These three cases are
considered below.

\paragraph{Case $p^{\star}=0$}

For the strategy $p^{\star}=0$ to be a NE one requires%

\begin{equation}
P(0,0)-P(p,0)=\frac{p}{(\gamma-\alpha)+(\delta-\beta)}\left[  \left|
a\right|  ^{2}-\frac{(\gamma-\alpha)}{(\gamma-\alpha)+(\delta-\beta)}\right]
\geq0 \label{Difference1Symmetric}%
\end{equation}
and the difference $\left\{  P(0,0)-P(p,0)\right\}  >0$ when $1\geq\left|
a\right|  ^{2}>\frac{(\gamma-\alpha)}{(\gamma-\alpha)+(\delta-\beta)}$. In
this range of $\left|  a\right|  ^{2}$ the equilibrium $p^{\star}=0$ is a pure
ESS. However, when $\left|  a\right|  ^{2}=\frac{(\gamma-\alpha)}%
{(\gamma-\alpha)+(\delta-\beta)}$ we have the difference $\left\{
P(0,0)-P(p,0)\right\}  $ identically zero. The strategy $p^{\star}=0$ can be
an ESS if%

\begin{align}
&  P(0,p)-P(p,p)\nonumber\\
&  =p\left\{  (\gamma-\alpha)+(\delta-\beta)\right\}  \left\{  \left|
a\right|  ^{2}-\frac{(1-p)(\gamma-\alpha)+p(\delta-\beta)}{(\gamma
-\alpha)+(\delta-\beta)}\right\}  >0
\end{align}
that can be written as%

\begin{equation}
P(0,p)-P(p,p)=p\left\{  (\gamma-\alpha)+(\delta-\beta)\right\}  \left\{
\left|  a\right|  ^{2}-\digamma\right\}  >0
\end{equation}
where $\frac{(\gamma-\alpha)}{(\gamma-\alpha)+(\delta-\beta)}\leq\digamma
\leq\frac{(\delta-\beta)}{(\gamma-\alpha)+(\delta-\beta)}$ when $0\leq
p\leq1.$ The strategy $p^{\star}=0$ can be an ESS only when $\left|  a\right|
^{2}>\frac{(\delta-\beta)}{(\gamma-\alpha)+(\delta-\beta)}$ which is not
possible because $\left|  a\right|  ^{2}$ is fixed at $\frac{(\gamma-\alpha
)}{(\gamma-\alpha)+(\delta-\beta)}.$ Therefore the strategy $p^{\star}=0$ is
an ESS for $1\geq\left|  a\right|  ^{2}>\frac{(\gamma-\alpha)}{(\gamma
-\alpha)+(\delta-\beta)}$ and for $\left|  a\right|  ^{2}=\frac{(\gamma
-\alpha)}{(\gamma-\alpha)+(\delta-\beta)}$ this NE becomes unstable. The
classical game is obtained by taking $\left|  a\right|  ^{2}=1$ for which
$p^{\star}=0$ is an ESS or a stable NE. However this NE does not remain stable
for $\left|  a\right|  ^{2}=\frac{(\gamma-\alpha)}{(\gamma-\alpha
)+(\delta-\beta)}$ which corresponds to an entangled initial state; though the
NE remains intact in both forms of the game.

\paragraph{Case $p^{\star}=1$}

Similar to the last case the NE condition for the strategy $p^{\star}=1$ can
be written as%

\begin{equation}
P(1,1)-P(p,1)=\frac{(1-p)}{(\gamma-\alpha)+(\delta-\beta)}\left[  -\left|
a\right|  ^{2}+\frac{(\delta-\beta)}{(\gamma-\alpha)+(\delta-\beta)}\right]
\geq0 \label{Difference2Symmetric}%
\end{equation}
Now $p^{\star}=1$ is a pure ESS for $0\leq\left|  a\right|  ^{2}<\frac
{(\delta-\beta)}{(\gamma-\alpha)+(\delta-\beta)}$. For $\left|  a\right|
^{2}=\frac{(\delta-\beta)}{(\gamma-\alpha)+(\delta-\beta)}$ the difference
$\left\{  P(1,1)-P(p,1)\right\}  $ becomes identically zero. The strategy
$p^{\star}=1$ is an ESS when%

\begin{align}
&  P(1,p)-P(p,p)\nonumber\\
&  =(1-p)\left\{  (\gamma-\alpha)+(\delta-\beta)\right\}  \left\{  -\left|
a\right|  ^{2}+\frac{(1-p)(\gamma-\alpha)+p(\delta-\beta)}{(\gamma
-\alpha)+(\delta-\beta)}\right\}  >0\nonumber\\
&
\end{align}
It is possible only if $\left|  a\right|  ^{2}<\frac{(\gamma-\alpha)}%
{(\gamma-\alpha)+(\delta-\beta)}.$ Therefore the strategy $p^{\star}=1$ is a
stable NE (ESS) for $0\leq\left|  a\right|  ^{2}<\frac{(\delta-\beta)}%
{(\gamma-\alpha)+(\delta-\beta)}.$ It is not stable classically (i.e. for
$\left|  a\right|  ^{2}=1$) but becomes stable for an entangled initial state.

\paragraph{Case $p^{\star}=\frac{(\delta-\beta)\left|  a\right|  ^{2}%
-(\gamma-\alpha)\left|  b\right|  ^{2}}{(\delta-\beta)-(\gamma-\alpha)}$}

In case of the mixed strategy:%

\begin{equation}
p^{\star}=\frac{(\delta-\beta)\left|  a\right|  ^{2}-(\gamma-\alpha)\left|
b\right|  ^{2}}{(\delta-\beta)-(\gamma-\alpha)} \label{MixedStrategySmmetric}%
\end{equation}
the NE condition (\ref{General2PlayerDifference1}) turns into%

\begin{equation}
P(p^{\star},p^{\star})-P(p,p^{\star})=0 \label{balance}%
\end{equation}
The mixed strategy (\ref{MixedStrategySmmetric}) can be an ESS if%

\begin{align}
&  P(p^{\star},p)-P(p,p)\nonumber\\
&  =(p^{\star}-p)\left[  -\left|  a\right|  ^{2}(\delta-\beta)+\left|
b\right|  ^{2}(\gamma-\alpha)+p\left\{  (\delta-\beta)-(\gamma-\alpha
)\right\}  \right]  >0\nonumber\\
&  \label{Difference3Symmetric}%
\end{align}
for all $p\neq p^{\star}$. Write now the strategy $p$ as $p=p^{\star
}+\bigtriangleup$. For the mixed strategy (\ref{MixedStrategySmmetric}) the
payoff difference of the Eq. (\ref{Difference3Symmetric}) is reduced to%

\begin{equation}
P(p^{\star},p)-P(p,p)=-\bigtriangleup^{2}\left\{  (\delta-\beta)-(\gamma
-\alpha)\right\}
\end{equation}
Hence, for the game defined in the conditions (\ref{GameDefinition}), the
mixed strategy $p^{\star}=\frac{(\delta-\beta)\left|  a\right|  ^{2}%
-(\gamma-\alpha)\left|  b\right|  ^{2}}{(\delta-\beta)-(\gamma-\alpha)}$
cannot be an ESS, though it can be a NE of the symmetric game.

It is to be pointed out that above considerations apply when the game is
played with the initial state (\ref{IniStatGen2Player}).

To find examples of symmetric quantum games, where evolutionary stability of
the mixed strategies may also be affected by controlling the initial states,
the number of players are now increased from two to three.

\subsection{ESSs in three-player two-strategy symmetric games}

In extending the two-player scheme to a three-player case, we assume that
three players $A,B,$ and $C$ play their strategies by applying the identity
operator $\hat{I}$ with the probabilities $p,q$ and $r$ respectively on the
initial state $\left|  \psi_{in}\right\rangle $. Therefore, they apply the
operator $\hat{\sigma}_{x}$ with the probabilities $(1-p),(1-q)$ and $(1-r)$
respectively. The final state becomes%

\begin{equation}
\rho_{fin}=\underset{\hat{U}=\hat{I},\hat{\sigma}_{x}}{\sum}\Pr(\hat{U}%
_{A})\Pr(\hat{U}_{B})\Pr(\hat{U}_{C})\left[  \hat{U}_{A}\otimes\hat{U}%
_{B}\otimes\hat{U}_{C}\rho_{in}\hat{U}_{A}^{\dagger}\otimes\hat{U}%
_{B}^{\dagger}\otimes\hat{U}_{C}^{\dagger}\right]
\end{equation}
where the $8$ basis vectors are $\left|  S_{i}S_{j}S_{k}\right\rangle $, for
$i,j,k=1,2$. Again we use initial quantum state in the form $\left|  \psi
_{in}\right\rangle =a\left|  S_{1}S_{1}S_{1}\right\rangle +b\left|  S_{2}%
S_{2}S_{2}\right\rangle $, where $\left|  a\right|  ^{2}+\left|  b\right|
^{2}=1$. It is a quantum state in $2\otimes2\otimes2$ dimensional Hilbert
space that can be prepared from a system of three two-state quantum systems or
qubits. Similar to the two-player case, the payoff operators for the players
$A,$ $B,$ and $C$ can be defined as%

\begin{align}
&  (P_{A,B,C})_{oper}=\nonumber\\
&  \alpha_{1},\beta_{1},\eta_{1}\left|  S_{1}S_{1}S_{1}\right\rangle
\left\langle S_{1}S_{1}S_{1}\right|  +\alpha_{2},\beta_{2},\eta_{2}\left|
S_{2}S_{1}S_{1}\right\rangle \left\langle S_{2}S_{1}S_{1}\right|  +\nonumber\\
&  \alpha_{3},\beta_{3},\eta_{3}\left|  S_{1}S_{2}S_{1}\right\rangle
\left\langle S_{1}S_{2}S_{1}\right|  +\alpha_{4},\beta_{4},\eta_{4}\left|
S_{1}S_{1}S_{2}\right\rangle \left\langle S_{1}S_{1}S_{2}\right|  +\nonumber\\
&  \alpha_{5},\beta_{5},\eta_{5}\left|  S_{1}S_{2}S_{2}\right\rangle
\left\langle S_{1}S_{2}S_{2}\right|  +\alpha_{6},\beta_{6},\eta_{6}\left|
S_{2}S_{1}S_{2}\right\rangle \left\langle S_{2}S_{1}S_{2}\right|  +\nonumber\\
&  \alpha_{7},\beta_{7},\eta_{7}\left|  S_{2}S_{2}S_{1}\right\rangle
\left\langle S_{2}S_{2}S_{1}\right|  +\alpha_{8},\beta_{8},\eta_{8}\left|
S_{2}S_{2}S_{2}\right\rangle \left\langle S_{2}S_{2}S_{2}\right|
\end{align}
where $\alpha_{l},\beta_{l},\eta_{l}$ for $1\leq l\leq8$ are $24$ constants of
the matrix of this three-player game. Payoffs to the players $A,B,$ and $C$
are then obtained as mean values of these operators%

\begin{equation}
P_{A,B,C}(p,q,r)=\text{Tr}\left[  (P_{A,B,C})_{oper}\rho_{fin}\right]
\end{equation}
Here, similar to the two-player case, the classical payoffs can be obtained by
making the initial quantum state unentangled and fixing $\left|  b\right|
^{2}=0$. To get a symmetric game we define $P_{A}(x,y,z)$ as the payoff to
player $A$ when players $A$, $B$, and $C$ play the strategies $x$, $y$, and
$z$ respectively. With following relations the players' payoffs become identity-independent.%

\begin{gather}
P_{A}(x,y,z)=P_{A}(x,z,y)=P_{B}(y,x,z)\nonumber\\
=P_{B}(z,x,y)=P_{C}(y,z,x)=P_{C}(z,y,x) \label{3PlayerSymmetricConditions}%
\end{gather}
The players in the game then become anonymous and their payoffs depend only on
their strategies. The relations (\ref{3PlayerSymmetricConditions}) hold with
the following replacements for $\beta_{i}$ and $\eta_{i}$:%

\begin{align}
\beta_{1}  &  \rightarrow\alpha_{1}\qquad\beta_{2}\rightarrow\alpha_{3}%
\qquad\beta_{3}\rightarrow\alpha_{2}\qquad\beta_{4}\rightarrow\alpha
_{3}\nonumber\\
\beta_{5}  &  \rightarrow\alpha_{6}\qquad\beta_{6}\rightarrow\alpha_{5}%
\qquad\beta_{7}\rightarrow\alpha_{6}\qquad\beta_{8}\rightarrow\alpha
_{8}\nonumber\\
\eta_{1}  &  \rightarrow\alpha_{1}\qquad\eta_{2}\rightarrow\alpha_{3}%
\qquad\eta_{3}\rightarrow\alpha_{3}\qquad\eta_{4}\rightarrow\alpha
_{2}\nonumber\\
\eta_{5}  &  \rightarrow\alpha_{6}\qquad\eta_{6}\rightarrow\alpha_{6}%
\qquad\eta_{7}\rightarrow\alpha_{5}\qquad\eta_{8}\rightarrow\alpha_{8}%
\end{align}
Also, it is now necessary that we should have%

\begin{equation}
\alpha_{6}=\alpha_{7}\text{, }\alpha_{3}=\alpha_{4}%
\end{equation}
A symmetric game between three players, therefore, can be defined by only six
constants of the payoff matrix. These constants can be taken as $\alpha
_{1},\alpha_{2},\alpha_{3},\alpha_{5},\alpha_{6},$ and $\alpha_{8}$. Payoff to
a $p$-player, when other two players play $q$ and $r$, can now be written as
$P(p,q,r)$. A symmetric NE $\overset{\star}{p}$ is now found from the Nash
condition $P(\overset{\star}{p},\overset{\star}{p},\overset{\star}%
{p})-P(p,\overset{\star}{p},\overset{\star}{p})\geq0$ i.e.%

\begin{gather}
P(\overset{\star}{p},\overset{\star}{p},\overset{\star}{p})-P(p,\overset
{\star}{p},\overset{\star}{p})=(\overset{\star}{p}-p)\text{{\LARGE [}}%
\overset{\star}{p}^{2}(1-2\left|  b\right|  ^{2})(\sigma+\omega-2\eta
)+\nonumber\\
2\overset{\star}{p}\left\{  \left|  b\right|  ^{2}(\sigma+\omega-2\eta
)-\omega+\eta\right\}  +\left\{  \omega-\left|  b\right|  ^{2}(\sigma
+\omega)\right\}  \text{{\LARGE ]}}\geq0
\end{gather}
where%

\begin{align}
(\alpha_{1}-\alpha_{2})  &  =\sigma\nonumber\\
(\alpha_{3}-\alpha_{6})  &  =\eta\nonumber\\
(\alpha_{5}-\alpha_{8})  &  =\omega
\end{align}
Three possible NE are found as%

\begin{equation}
\left.
\begin{array}
[c]{c}%
\overset{\star}{p}_{1}=\frac{\left\{  (\omega-\eta)-\left|  b\right|
^{2}(\sigma+\omega-2\eta)\right\}  \pm\sqrt{\left\{  (\sigma+\omega
)^{2}-(2\eta)^{2}\right\}  \left|  b\right|  ^{2}(1-\left|  b\right|
^{2})+(\eta^{2}-\sigma\omega)}}{(1-2\left|  b\right|  ^{2})(\sigma
+\omega-2\eta)}\\%
\begin{array}
[c]{c}%
\overset{\star}{p}_{2}=0\\
\overset{\star}{p}_{3}=1
\end{array}
\end{array}
\right\}
\end{equation}
It is observed that the mixed NE $\overset{\star}{p_{1}}$ makes the difference
$\left\{  P(\overset{\star}{p},\overset{\star}{p},\overset{\star}%
{p})-P(p,\overset{\star}{p},\overset{\star}{p})\right\}  $ identically zero
and two values for $\overset{\star}{p}_{1}$ can be found for a given $\left|
b\right|  ^{2}$. Apart from $\overset{\star}{p}_{1}$ the other two NE (i.e.
$\overset{\star}{p}_{2}$ and $\overset{\star}{p}_{3}$) are pure strategies.
Also now $\overset{\star}{p}_{1}$ comes out a NE without imposing further
restrictions on the matrix of the symmetric three-player game. However, the
pure strategies $\overset{\star}{p}_{2}$ and $\overset{\star}{p}_{3}$ can be
NE when further restriction are imposed on the matrix of the game. For
example, $\overset{\star}{p}_{3}$ can be a NE provided $\sigma\geq
(\omega+\sigma)\left|  b\right|  ^{2}$ for all $\left|  b\right|  ^{2}%
\in\lbrack0,1]$. Similarly $\overset{\star}{p}_{2}$ can be NE when $\omega
\leq(\omega+\sigma)\left|  b\right|  ^{2}$.

Now we address the question: How evolutionary stability of these three NE can
be affected while playing the game via initial quantum states given in the
following form?%

\begin{equation}
\left|  \psi_{in}\right\rangle =a\left|  S_{1}S_{1}S_{1}\right\rangle
+b\left|  S_{2}S_{2}S_{2}\right\rangle \label{IniStatSymm3Player}%
\end{equation}
For the two-player asymmetric game of BoS we showed that out of three NE only
two can be evolutionary stable. In classical evolutionary game theory the
concept of an ESS is well-known \cite{MarkBroom1,BroomMutiplayer} to be
extendable to multi-player case. When mutants are allowed to play only one
strategy the definition of an ESS for the three-player case is written as \cite{MarkBroom1}%

\begin{align}
1.\text{ \ \ \ }P(p,p,p)  &  >P(q,p,p)\nonumber\\
2.\text{ If }P(p,p,p)  &  =P(q,p,p)\text{ then }P(p,q,p)>P(q,q,p)
\end{align}
Here $p$ is a NE if it satisfies the condition $1$ against all $q\neq p$. For
our case the ESS conditions for the pure strategies $\overset{\star}{p}_{2}$
and $\overset{\star}{p}_{3}$ can be written as follows. For example
$\overset{\star}{p}_{2}=0$ is an ESS when%

\begin{align}
1.\text{ \ \ \ }\sigma\left|  b\right|  ^{2}  &  >\omega\left|  a\right|
^{2}\nonumber\\
2.\text{ If }\sigma\left|  b\right|  ^{2}  &  =\omega\left|  a\right|
^{2}\text{ then }-\eta q^{2}(\left|  a\right|  ^{2}-\left|  b\right|  ^{2})>0
\label{Cond3PlayerSymme1}%
\end{align}
where $1$ is NE condition for the strategy $\overset{\star}{p}_{2}=0$.
Similarly $\overset{\star}{p}_{3}=1$ is an ESS when%

\begin{align}
1.\text{ \ \ \ }\sigma\left|  a\right|  ^{2}  &  >\omega\left|  b\right|
^{2}\nonumber\\
2.\text{ If }\sigma\left|  a\right|  ^{2}  &  =\omega\left|  b\right|
^{2}\text{ then }\eta(1-q)^{2}(\left|  a\right|  ^{2}-\left|  b\right|
^{2})>0 \label{Conditions3PlayerSymm}%
\end{align}
and both the pure strategies $\overset{\star}{p}_{2}$ and $\overset{\star}%
{p}_{3}$ are ESSs when $\left|  a\right|  ^{2}=\left|  b\right|  ^{2}$. The
conditions (\ref{Conditions3PlayerSymm}) can also be written as
\begin{align}
1.\text{ \ \ \ }\sigma &  >(\omega+\sigma)\left|  b\right|  ^{2}\nonumber\\
2.\text{ If }\sigma &  =\left|  b\right|  ^{2}(\omega+\sigma)\text{ then
}\frac{\gamma(\omega-\sigma)}{(\omega+\sigma)}>0
\end{align}
For the strategy $\overset{\star}{p}_{2}=0$ the ESS conditions
(\ref{Cond3PlayerSymme1}) reduce to%

\begin{align}
1.\text{ \ \ \ \ }\omega &  <(\omega+\sigma)\left|  b\right|  ^{2}\nonumber\\
2.\text{ If \ }\omega &  =\left|  b\right|  ^{2}(\omega+\sigma)\text{ then
}\frac{\gamma(\omega-\sigma)}{(\omega+\sigma)}>0
\end{align}
Examples of three-player symmetric games are easy to find for which a pure
strategy is a NE for the whole range $\left|  b\right|  ^{2}\in\lbrack0,1]$,
but it is not an ESS for some particular value of $\left|  b\right|  ^{2}$. An
example of a class of such games is for which $\sigma=0$, $\omega<0$, and
$\eta\leq0$. In this class the strategy $\overset{\star}{p}_{2}=0$ is a NE for
all $\left|  b\right|  ^{2}\in\lbrack0,1]$ but not an ESS when $\left|
b\right|  ^{2}=1$.

Apart from the pure strategies, the mixed strategy equilibrium $\overset
{\star}{p}_{1}$ forms the most interesting case. It makes the payoff
difference $\left\{  P(\overset{\star}{p_{1}},\overset{\star}{p_{1}}%
,\overset{\star}{p_{1}})-P(p,\overset{\star}{p_{1}},\overset{\star}{p_{1}%
})\right\}  $ identically zero for every strategy $p$. The strategy
$\overset{\star}{p_{1}}$ is an ESS when $\left\{  P(\overset{\star}{p_{1}%
},q,\overset{\star}{p_{1}})-P(q,q,\overset{\star}{p_{1}})\right\}  >0$ but%

\begin{align}
&  P(\overset{\star}{p_{1}},q,\overset{\star}{p_{1}})-P(q,q,\overset{\star
}{p_{1}})\nonumber\\
&  =\pm(\overset{\star}{p_{1}}-q)^{2}\sqrt{\left\{  (\sigma+\omega)^{2}%
-(2\eta)^{2}\right\}  \left|  b\right|  ^{2}(1-\left|  b\right|  ^{2}%
)+(\eta^{2}-\sigma\omega)} \label{3PlayerSqr}%
\end{align}
Therefore, out of the two possible roots $(\overset{\star}{p_{1}})_{1}$ and
$(\overset{\star}{p_{1}})_{2}$ of the quadratic equation\footnote{These roots
make the difference $\left\{  P(\overset{\star}{p_{1}},q,\overset{\star}%
{p_{1}})-P(q,q,\overset{\star}{p_{1}})\right\}  $ greater than and less than
zero, respectively.}:%

\begin{gather}
\overset{\star}{p_{1}}^{2}(1-2\left|  b\right|  ^{2})(\sigma+\omega
-2\eta)+\nonumber\\
2\overset{\star}{p_{1}}\left\{  \left|  b\right|  ^{2}(\sigma+\omega
-2\eta)-\omega+\eta\right\}  +\left\{  \omega-\left|  b\right|  ^{2}%
(\sigma+\omega)\right\}  =0 \label{DefQuadEqForp1Star}%
\end{gather}
only $(\overset{\star}{p_{1}})_{1}$ can exist as an ESS. When the square root
term in the equation (\ref{3PlayerSqr}) becomes zero we have only one mixed
NE, which is not an ESS. Hence, out of four possible NE in this three-player
game only three can be ESSs.

An interesting class of three-player games is the one for which $\eta
^{2}=\sigma\omega$. For these games the mixed NE are%

\begin{equation}
\overset{\star}{p_{1}}=\frac{\left\{  (w-\eta)-\left|  b\right|  ^{2}%
(\sigma+\omega-2\eta)\right\}  \pm\left|  a\right|  \left|  b\right|  \left|
\sigma-\omega\right|  }{(1-2\left|  b\right|  ^{2})(\sigma+\omega-2\eta)}%
\end{equation}
and, when played classically, we can get only one mixed NE that is not an ESS.
However for all $\left|  b\right|  ^{2}$, different from zero, we generally
obtain two NE out of which one can be an ESS.

Similar to the two-player case, the equilibria in a three-player symmetric
game where quantization affects evolutionary stability, are the ones that
survive for two initial states, one of which is unentangled and corresponds to
the classical game. Suppose $\overset{\star}{p_{1}}$ remains a NE for $\left|
b\right|  ^{2}=0$ and some other non-zero $\left|  b\right|  ^{2}$. It is
possible when%

\begin{equation}
(\sigma-\omega)(2\overset{\star}{p_{1}}-1)=0
\end{equation}
Alternatively the strategy $\overset{\star}{p}=\frac{1}{2}$ remains a NE for
all $\left|  b\right|  ^{2}\in\lbrack0,1]$. It reduces the defining quadratic
equation (\ref{DefQuadEqForp1Star}) for $\overset{\star}{p_{1}}$ to%

\begin{equation}
\sigma+\omega+2\eta=0
\end{equation}
and makes the difference $\left\{  P(\overset{\star}{p_{1}},q,\overset{\star
}{p_{1}})-P(q,q,\overset{\star}{p_{1}})\right\}  $ independent of $\left|
b\right|  ^{2}$. Therefore the strategy $\overset{\star}{p}=\frac{1}{2}$, even
when it is retained as an equilibrium for all $\left|  b\right|  ^{2}%
\in\lbrack0,1]$, cannot be an ESS in only one version of the symmetric
three-player game. For the second possibility $\sigma=\omega$ the defining
equation for $\overset{\star}{p_{1}}$ is reduced to%

\begin{equation}
(1-2\left|  b\right|  ^{2})\left\{  \overset{\star}{p_{1}}-\frac{(\eta
-\sigma)-\sqrt{\eta^{2}-\sigma^{2}}}{2(\eta-\sigma)}\right\}  \left\{
\overset{\star}{p_{1}}-\frac{(\eta-\sigma)+\sqrt{\eta^{2}-\sigma^{2}}}%
{2(\eta-\sigma)}\right\}  =0
\end{equation}
for which%

\begin{equation}
P(\overset{\star}{p_{1}},q,\overset{\star}{p_{1}})-P(q,q,\overset{\star}%
{p_{1}})=\pm2(\overset{\star}{p_{1}}-q)^{2}\left|  \left|  b\right|
^{2}-\frac{1}{2}\right|  \sqrt{\eta^{2}-\sigma^{2}}%
\end{equation}
Here the difference $\left\{  P(\overset{\star}{p_{1}},q,\overset{\star}%
{p_{1}})-P(q,q,\overset{\star}{p_{1}})\right\}  $ still depends on $\left|
b\right|  ^{2}$ and becomes zero for $\left|  b\right|  ^{2}=1/2$.

Hence, for the class of games for which $\sigma=\omega$ and $\eta>\sigma$, one
of the mixed strategies $(\overset{\star}{p_{1}})_{1},(\overset{\star}{p_{1}%
})_{2}$ remains a NE for all $\left|  b\right|  ^{2}\in\lbrack0,1]$ but not an
ESS when $\left|  b\right|  ^{2}=1/2$. In this class of three-player quantum
games the evolutionary stability of a mixed strategy can, therefore, be
changed while the game is played using initial quantum states in the form
(\ref{IniStatSymm3Player}).

\section{Quantum Rock-Scissors-Paper game}

Long played as a children's pastime, or as an odd-man-out selection process,
the Rock-Scissors-Paper (RSP) is a game for two players typically played using
the players' hands. The players opposite each others, tap their fist in their
open palm three times (saying Rock, Scissors, Paper) and then show one of
three possible gestures. The Rock wins against the scissors (crushes it) but
looses against the paper (is wrapped into it). The Scissors wins against the
paper (cuts it) but looses against the rock (is crushed by it). The Paper wins
against the rock (wraps it) but looses against the scissors (is cut by it).
The game is also played in nature like many other games. Lizards in the Coast
Range of California play this game \cite{LizardGame} using three alternative
male strategies locked in an ecological never ending process from which there
seems little escape.

\subsection{Rock-Scissors-Paper in a slightly modified version}

In a slightly modified version of the RSP game both players get a small
premium $\epsilon$\ for a draw. This game can be represented by the payoff matrix:%

\begin{equation}%
\begin{array}
[c]{c}%
R\\
S\\
P
\end{array}
\overset{%
\begin{array}
[c]{ccc}%
R & S & P
\end{array}
}{\left(
\begin{array}
[c]{ccc}%
-\epsilon & 1 & -1\\
-1 & -\epsilon & 1\\
1 & -1 & -\epsilon
\end{array}
\right)  } \label{RSPMatrix}%
\end{equation}
where $-1<\epsilon<0$. The matrix of the usual game is obtained when
$\epsilon$ is zero.

\subsection{Nash equilibrium and ESS in Rock-Scissors-Paper game}

One cannot win if one's opponent knew which strategy was going to be picked.
For example, picking Rock consistently all the opponent needs to do is pick
Paper and s/he would win. Players find soon that in case predicting opponent's
strategy is not possible the best strategy is to pick Rock, Scissors, or Paper
at random. In other words, the player selects Rock, Scissors, or Paper with a
probability of $1/3$. In case opponent's strategy is predictable picking a
strategy at random with a probability of $1/3$ is \emph{not} the best thing to
do unless the opponent is doing the same \cite{Prestwich}.

Analysis \cite{Weibull} of the modified RSP game of matrix (\ref{RSPMatrix})
shows that its NE consists of playing each of the three different pure
strategies with a fixed equilibrial probability $1/3$. However it is not an
ESS because $\epsilon$ is negative.

Here we want to study the effects of quantization on evolutionary stability
for the modified RSP game. The game is different, from others considered
earlier, because classically each player now possesses three pure strategies
instead of two. A classical mixed NE exists which is not an ESS. Our
motivation is to explore the possibility that the classical mixed NE becomes
an ESS for some quantum form of the game.

\subsection{Quantization of Rock-Scissors-Paper game}

Using simpler notation: $R\sim1,$ $S\sim2,$ $P\sim3$ we quantize this game via
Marinatto and Weber's scheme \cite{Marinatto1}. We assume the two players are
in possession of three unitary and Hermitian operators $\hat{I},\hat{C}$ and
$\hat{D}$ defined as follows.%

\begin{align}
\hat{I}\left|  1\right\rangle  &  =\left|  1\right\rangle \text{,}\qquad
\hat{C}\left|  1\right\rangle =\left|  3\right\rangle \text{,}\qquad\hat
{D}\left|  1\right\rangle =\left|  2\right\rangle \nonumber\\
\hat{I}\left|  2\right\rangle  &  =\left|  2\right\rangle \text{,}\qquad
\hat{C}\left|  2\right\rangle =\left|  2\right\rangle \text{,}\qquad\hat
{D}\left|  2\right\rangle =\left|  1\right\rangle \nonumber\\
\hat{I}\left|  3\right\rangle  &  =\left|  3\right\rangle \text{,}\qquad
\hat{C}\left|  3\right\rangle =\left|  1\right\rangle \text{,}\qquad\hat
{D}\left|  3\right\rangle =\left|  3\right\rangle
\end{align}
where $\hat{C}^{\dagger}=\hat{C}=\hat{C}^{-1}$ and $\hat{D}^{\dagger}=\hat
{D}=\hat{D}^{-1}$ and $\hat{I}$ is the identity operator.

Consider a general two-player payoff matrix when each player has three strategies:%

\begin{equation}%
\begin{array}
[c]{c}%
1\\
2\\
3
\end{array}
\overset{%
\begin{array}
[c]{ccccccc}%
1 &  &  & 2 &  &  & 3
\end{array}
}{\left(
\begin{array}
[c]{ccc}%
(\alpha_{11},\beta_{11}) & (\alpha_{12},\beta_{12}) & (\alpha_{13},\beta
_{13})\\
(\alpha_{21},\beta_{21}) & (\alpha_{22},\beta_{22}) & (\alpha_{23},\beta
_{23})\\
(\alpha_{31},\beta_{31}) & (\alpha_{32},\beta_{32}) & (\alpha_{33},\beta_{33})
\end{array}
\right)  } \label{Gen3StrategyMatrix}%
\end{equation}
where $\alpha_{ij}$ and $\beta_{ij}$ are the payoffs to Alice and Bob,
respectively, when Alice plays $i$ and Bob plays $j$ and $1\leq i,j\leq3$.
Suppose Alice and Bob apply the operators $\hat{C}$, $\hat{D}$, and $\hat{I}$
with the probabilities $p$, $p_{1}$, $(1-p-p_{1})$ and $q$, $q_{1}$,
$(1-q-q_{1})$ respectively. The initial state of the game is $\rho_{in}$.
Alice's move changes the state changes to%

\begin{equation}
\overset{A}{\rho_{in}}=(1-p-p_{1})\hat{I}_{A}\rho_{in}\hat{I}_{A}^{\dagger
}+p\hat{C}_{A}\rho_{in}\hat{C}_{A}^{\dagger}+p_{1}\hat{D}_{A}\rho_{in}\hat
{D}_{A}^{\dagger}%
\end{equation}
The final state, after Bob too has played his move, is%

\begin{equation}
\overset{A,B}{\rho_{f}}=(1-q-q_{1})\hat{I}_{B}\overset{A}{\rho_{in}}\hat
{I}_{B}^{\dagger}+q\hat{C}_{B}\overset{A}{\rho_{in}}\hat{C}_{B}^{\dagger
}+q_{1}\hat{D}_{B}\overset{A}{\rho_{in}}\hat{D}_{B}^{\dagger}%
\end{equation}
This state can be written as%

\begin{align}
\overset{A,B}{\rho_{f}} &  =(1-p-p_{1})(1-q-q_{1})\left\{  \hat{I}_{A}%
\otimes\hat{I}_{B}\rho_{in}\hat{I}_{A}^{\dagger}\otimes\hat{I}_{B}^{\dagger
}\right\}  +p(1-q-q_{1})\times\nonumber\\
&  \left\{  \hat{C}_{A}\otimes\hat{I}_{B}\rho_{in}\hat{C}_{A}^{\dagger}%
\otimes\hat{I}_{B}^{\dagger}\right\}  +p_{1}(1-q-q_{1})\left\{  \hat{D}%
_{A}\otimes\hat{I}_{B}\rho_{in}\hat{D}_{A}^{\dagger}\otimes\hat{I}_{B}%
^{\dagger}\right\}  +\nonumber\\
&  (1-p-p_{1})q\left\{  \hat{I}_{A}\otimes\hat{C}_{B}\rho_{in}\hat{I}_{A}%
^{\dagger}\otimes\hat{C}_{B}^{\dagger}\right\}  +pq\left\{  \hat{C}_{A}%
\otimes\hat{C}_{B}\rho_{in}\hat{C}_{A}^{\dagger}\otimes\hat{C}_{B}^{\dagger
}\right\}  +\nonumber\\
&  p_{1}q\left\{  \hat{D}_{A}\otimes\hat{C}_{B}\rho_{in}\hat{D}_{A}^{\dagger
}\otimes\hat{C}_{B}^{\dagger}\right\}  +(1-p-p_{1})q_{1}\left\{
\hat{I}_{A}\otimes\hat{D}_{B}\rho_{in}\hat{I}_{A}^{\dagger}\otimes\hat{D}%
_{B}^{\dagger}\right\}  \nonumber\\
&  +pq_{1}\left\{  \hat{C}_{A}\otimes\hat{D}_{B}\rho_{in}\hat{C}_{A}^{\dagger
}\otimes\hat{D}_{B}^{\dagger}\right\}  +p_{1}q_{1}\left\{  \hat{D}_{A}%
\otimes\hat{D}_{B}\rho_{in}\hat{D}_{A}^{\dagger}\otimes\hat{D}_{B}^{\dagger
}\right\}
\end{align}
The nine basis vectors of initial quantum state with three pure classical
strategies are $\left|  ij\right\rangle $ for $i,j=1,2,3$. We consider the
initial state to be a pure quantum state of two qutrits i.e.%

\begin{equation}
\left|  \psi_{in}\right\rangle =\underset{i,j=1,2,3}{\sum}c_{ij}\left|
ij\right\rangle \text{, \ \ \ \ where }\underset{i,j=1,2,3}{\sum}\left|
c_{ij}\right|  ^{2}=1 \label{RSPIniStat}%
\end{equation}
The payoff operators for Alice and Bob are \cite{Marinatto1}%

\begin{align}
(P_{A,B})_{oper}  &  =(\alpha,\beta)_{11}\left|  11\right\rangle \left\langle
11\right|  +(\alpha,\beta)_{12}\left|  12\right\rangle \left\langle 12\right|
+(\alpha,\beta)_{13}\left|  13\right\rangle \left\langle 13\right|
+\nonumber\\
&  (\alpha,\beta)_{21}\left|  21\right\rangle \left\langle 21\right|
+(\alpha,\beta)_{22}\left|  22\right\rangle \left\langle 22\right|
+(\alpha,\beta)_{23}\left|  23\right\rangle \left\langle 23\right|
+\nonumber\\
&  (\alpha,\beta)_{31}\left|  31\right\rangle \left\langle 31\right|
+(\alpha,\beta)_{32}\left|  32\right\rangle \left\langle 32\right|
+(\alpha,\beta)_{33}\left|  33\right\rangle \left\langle 33\right| \nonumber\\
&
\end{align}
The players' payoffs are then%

\begin{equation}
P_{A,B}=\text{Tr}[\left\{  (P_{A,B})_{oper}\right\}  \overset{A,B}{\rho_{f}}]
\label{3StrategyPayoff}%
\end{equation}
Payoff to Alice, for example, can be written as%

\begin{equation}
P_{A}=\Phi\Omega\Upsilon^{T} \label{RSPPayoffAlice}%
\end{equation}
where $T$ is for transpose, and the matrices $\Phi,$ $\Omega,$ and $\Upsilon$ are%

\begin{align}
\Phi &  =[
\begin{array}
[c]{ccc}%
(1-p-p_{1})(1-q-q_{1}) & p(1-q-q_{1}) & p_{1}(1-q-q_{1})
\end{array}
\nonumber\\
&
\begin{array}
[c]{cccccc}%
(1-p-p_{1})q & pq & p_{1}q & (1-p-p_{1})q_{1} & pq_{1} & p_{1}q_{1}%
\end{array}
]\nonumber\\
\Upsilon &  =[
\begin{array}
[c]{ccccccccc}%
\alpha_{11} & \alpha_{12} & \alpha_{13} & \alpha_{21} & \alpha_{22} &
\alpha_{23} & \alpha_{31} & \alpha_{32} & \alpha_{33}%
\end{array}
]\nonumber\\
\Omega &  =\left[
\begin{array}
[c]{ccccccccc}%
\left|  c_{11}\right|  ^{2} & \left|  c_{12}\right|  ^{2} & \left|
c_{13}\right|  ^{2} & \left|  c_{21}\right|  ^{2} & \left|  c_{22}\right|
^{2} & \left|  c_{23}\right|  ^{2} & \left|  c_{31}\right|  ^{2} & \left|
c_{32}\right|  ^{2} & \left|  c_{33}\right|  ^{2}\\
\left|  c_{31}\right|  ^{2} & \left|  c_{32}\right|  ^{2} & \left|
c_{33}\right|  ^{2} & \left|  c_{21}\right|  ^{2} & \left|  c_{22}\right|
^{2} & \left|  c_{23}\right|  ^{2} & \left|  c_{11}\right|  ^{2} & \left|
c_{12}\right|  ^{2} & \left|  c_{13}\right|  ^{2}\\
\left|  c_{21}\right|  ^{2} & \left|  c_{22}\right|  ^{2} & \left|
c_{23}\right|  ^{2} & \left|  c_{11}\right|  ^{2} & \left|  c_{12}\right|
^{2} & \left|  c_{13}\right|  ^{2} & \left|  c_{31}\right|  ^{2} & \left|
c_{32}\right|  ^{2} & \left|  c_{33}\right|  ^{2}\\
\left|  c_{13}\right|  ^{2} & \left|  c_{12}\right|  ^{2} & \left|
c_{11}\right|  ^{2} & \left|  c_{23}\right|  ^{2} & \left|  c_{22}\right|
^{2} & \left|  c_{21}\right|  ^{2} & \left|  c_{33}\right|  ^{2} & \left|
c_{32}\right|  ^{2} & \left|  c_{31}\right|  ^{2}\\
\left|  c_{33}\right|  ^{2} & \left|  c_{32}\right|  ^{2} & \left|
c_{31}\right|  ^{2} & \left|  c_{23}\right|  ^{2} & \left|  c_{22}\right|
^{2} & \left|  c_{21}\right|  ^{2} & \left|  c_{13}\right|  ^{2} & \left|
c_{12}\right|  ^{2} & \left|  c_{11}\right|  ^{2}\\
\left|  c_{23}\right|  ^{2} & \left|  c_{22}\right|  ^{2} & \left|
c_{21}\right|  ^{2} & \left|  c_{13}\right|  ^{2} & \left|  c_{12}\right|
^{2} & \left|  c_{11}\right|  ^{2} & \left|  c_{33}\right|  ^{2} & \left|
c_{32}\right|  ^{2} & \left|  c_{31}\right|  ^{2}\\
\left|  c_{12}\right|  ^{2} & \left|  c_{11}\right|  ^{2} & \left|
c_{13}\right|  ^{2} & \left|  c_{22}\right|  ^{2} & \left|  c_{21}\right|
^{2} & \left|  c_{23}\right|  ^{2} & \left|  c_{32}\right|  ^{2} & \left|
c_{31}\right|  ^{2} & \left|  c_{33}\right|  ^{2}\\
\left|  c_{32}\right|  ^{2} & \left|  c_{31}\right|  ^{2} & \left|
c_{33}\right|  ^{2} & \left|  c_{22}\right|  ^{2} & \left|  c_{21}\right|
^{2} & \left|  c_{23}\right|  ^{2} & \left|  c_{12}\right|  ^{2} & \left|
c_{11}\right|  ^{2} & \left|  c_{13}\right|  ^{2}\\
\left|  c_{22}\right|  ^{2} & \left|  c_{21}\right|  ^{2} & \left|
c_{23}\right|  ^{2} & \left|  c_{12}\right|  ^{2} & \left|  c_{11}\right|
^{2} & \left|  c_{13}\right|  ^{2} & \left|  c_{32}\right|  ^{2} & \left|
c_{31}\right|  ^{2} & \left|  c_{33}\right|  ^{2}%
\end{array}
\right] \nonumber\\
&  \label{RSPAliceSigma}%
\end{align}
The payoff (\ref{RSPPayoffAlice}) corresponds to the matrix
(\ref{Gen3StrategyMatrix}). Payoffs in classical mixed strategy game can be
obtained from Eq. (\ref{3StrategyPayoff}) for the initial state $\left|
\psi_{in}\right\rangle =\left|  11\right\rangle $. The game is symmetric when
$\alpha_{ij}=\beta_{ji}$ in the matrix (\ref{Gen3StrategyMatrix}). The quantum
game played using the quantum state (\ref{RSPIniStat}) is symmetric when
$\left|  c_{ij}\right|  ^{2}=\left|  c_{ji}\right|  ^{2}$ for all constants
$c_{ij}$ in the state (\ref{RSPIniStat}). These two conditions together
guarantee a symmetric quantum game. The players' payoffs $P_{A}$, $P_{B}$ then
do not need a subscript and we can simply use $P(p,q)$ to denote the payoff to
the $p$-player against the $q$-player.

The question of evolutionary stability in quantized RSP game is addressed below.

\subsection{Consideration of evolutionary stability}

Assume a strategy is defined by a pair of numbers $(p,p_{1})$ for players
playing the quantized RSP game. These numbers are the probabilities with which
the player applies the operators $\hat{C}$ and $\hat{D}$. The identity
operator $\hat{I}$ is, then, applied with probability $(1-p-p_{1})$. Similar
to the conditions $1$ and $2$ in Eq. (\ref{DefESS}), the conditions making a
strategy $(p^{\star},p_{1}^{\star})$ an ESS can be written as \cite{Smith
Price,Weibull}
\begin{align}
1\text{. \ \ \ }P\{(p^{\star},p_{1}^{\star}),(p^{\star},p_{1}^{\star})\}  &
>P\{(p,p_{1}),(p^{\star},p_{1}^{\star})\}\nonumber\\
2\text{. if }P\{(p^{\star},p_{1}^{\star}),(p^{\star},p_{1}^{\star})\}  &
=P\{(p,p_{1}),(p^{\star},p_{1}^{\star})\}\text{ then}\nonumber\\
P\{(p^{\star},p_{1}^{\star}),(p,p_{1})\}  &  >P\{(p,p_{1}),(p,p_{1})\}
\label{ESScondsRSP}%
\end{align}
Suppose $(p^{\star},p_{1}^{\star})$ is a mixed NE then%

\begin{equation}
\left\{  \frac{\partial P}{\partial p}\mid_{\substack{p=q=p^{\star}
\\p_{1}=q_{1}=p_{1}^{\star} }}(p^{\star}-p)+\frac{\partial P}{\partial p_{1}%
}\mid_{\substack{p=q=p^{\star} \\p_{1}=q_{1}=p_{1}^{\star} }}(p_{1}^{\star
}-p_{1})\right\}  \geq0
\end{equation}
Using substitutions%

\begin{equation}%
\begin{array}
[c]{cc}%
\left|  c_{11}\right|  ^{2}-\left|  c_{31}\right|  ^{2}=\bigtriangleup_{1}, &
\left|  c_{21}\right|  ^{2}-\left|  c_{11}\right|  ^{2}=\bigtriangleup_{1}^{%
\acute{}%
}\\
\left|  c_{13}\right|  ^{2}-\left|  c_{33}\right|  ^{2}=\bigtriangleup_{2}, &
\left|  c_{22}\right|  ^{2}-\left|  c_{12}\right|  ^{2}=\bigtriangleup_{2}^{%
\acute{}%
}\\
\left|  c_{12}\right|  ^{2}-\left|  c_{32}\right|  ^{2}=\bigtriangleup_{3}, &
\left|  c_{23}\right|  ^{2}-\left|  c_{13}\right|  ^{2}=\bigtriangleup_{3}^{%
\acute{}%
}%
\end{array}
\end{equation}
we get%

\begin{align}
\frac{\partial P}{\partial p}  &  \mid_{\substack{p=q=p^{\star} \\p_{1}%
=q_{1}=p_{1}^{\star}}}=p^{\star}(\bigtriangleup_{1}-\bigtriangleup
_{2})\left\{  (\alpha_{11}+\alpha_{33})-(\alpha_{13}+\alpha_{31})\right\}
+\nonumber\\
&  p_{1}^{\star}(\bigtriangleup_{1}-\bigtriangleup_{3})\left\{  (\alpha
_{11}+\alpha_{32})-(\alpha_{12}+\alpha_{31})\right\}  -\nonumber\\
&  \bigtriangleup_{1}(\alpha_{11}-\alpha_{31})-\bigtriangleup_{2}(\alpha
_{13}-\alpha_{33})-\bigtriangleup_{3}(\alpha_{12}-\alpha_{32})
\label{RSPRestPoint1}%
\end{align}%

\begin{align}
\frac{\partial P}{\partial p_{1}}  &  \mid_{\substack{p=q=p^{\star}%
\\p_{1}=q_{1}=p_{1}^{\star}}}=p^{\star}(\bigtriangleup_{3}^{%
\acute{}%
}-\bigtriangleup_{1}^{%
\acute{}%
})\left\{  (\alpha_{11}+\alpha_{23})-(\alpha_{13}+\alpha_{21})\right\}
+\nonumber\\
&  p_{1}^{\star}(\bigtriangleup_{2}^{%
\acute{}%
}-\bigtriangleup_{1}^{%
\acute{}%
})\left\{  (\alpha_{11}+\alpha_{22})-(\alpha_{12}+\alpha_{21})\right\}
+\nonumber\\
&  \bigtriangleup_{1}^{%
\acute{}%
}(\alpha_{11}-\alpha_{21})+\bigtriangleup_{2}^{%
\acute{}%
}(\alpha_{12}-\alpha_{22})+\bigtriangleup_{3}^{%
\acute{}%
}(\alpha_{13}-\alpha_{23}) \label{RSPRestPoint2}%
\end{align}
For the matrix (\ref{RSPMatrix}) the equations (\ref{RSPRestPoint1},
\ref{RSPRestPoint2}) can be written as%

\begin{align}
\frac{\partial P}{\partial p}  &  \mid_{\substack{p=q=p^{\star} \\p_{1}%
=q_{1}=p_{1}^{\star}}}=\bigtriangleup_{1}\left\{  -2\epsilon p^{\star
}-(3+\epsilon)p_{1}^{\star}+(1+\epsilon)\right\}  +\nonumber\\
&  \bigtriangleup_{2}\left\{  2\epsilon p^{\star}+(1-\epsilon)\right\}
+\bigtriangleup_{3}\left\{  (3+\epsilon)p_{1}^{\star}-2\right\} \\
\frac{\partial P}{\partial p_{1}}  &  \mid_{\substack{p=q=p^{\star}
\\p_{1}=q_{1}=p_{1}^{\star}}}=\bigtriangleup_{1}^{%
\acute{}%
}\left\{  -p^{\star}(3-\epsilon)+2\epsilon p_{1}^{\star}+(1-\epsilon)\right\}
-\nonumber\\
&  \bigtriangleup_{2}^{%
\acute{}%
}\left\{  2\epsilon p_{1}^{\star}-(1+\epsilon)\right\}  +\bigtriangleup_{3}^{%
\acute{}%
}\left\{  (3-\epsilon)p^{\star}-2\right\}
\end{align}
The payoff difference in the second condition of an ESS given in the Eq.
(\ref{ESScondsRSP}) reduces to%

\begin{align}
&  P\{(p^{\star},p_{1}^{\star}),(p,p_{1})\}-P\{(p,p_{1}),(p,p_{1}%
)\}\nonumber\\
&  =(p^{\star}-p)[-\bigtriangleup_{1}\{2\epsilon p+(3+\epsilon)p_{1}%
-(1+\epsilon)\}+\nonumber\\
&  \bigtriangleup_{2}\{2\epsilon p+(1-\epsilon)\}+\bigtriangleup
_{3}\{(3+\epsilon)p_{1}-2\}]+\nonumber\\
&  (p_{1}^{\star}-p_{1})[-\bigtriangleup_{1}^{%
\acute{}%
}\{(3-\epsilon)p-2\epsilon p_{1}-(1-\epsilon)\}-\nonumber\\
&  \bigtriangleup_{2}^{%
\acute{}%
}\{2\epsilon p_{1}-(1+\epsilon)\}+\bigtriangleup_{3}^{%
\acute{}%
}\{(3-\epsilon)p-2\}]
\end{align}
With the substitutions $(p^{\star}-p)=x$ and $(p_{1}^{\star}-p_{1})=y$ the
above payoff difference is%

\begin{align}
&  P\{(p^{\star},p_{1}^{\star}),(p,p_{1})\}-P\{(p,p_{1}),(p,p_{1}%
)\}\nonumber\\
&  =\bigtriangleup_{1}x\left\{  2\epsilon x+(3+\epsilon)y\right\}
-\bigtriangleup_{2}(2\epsilon x^{2})-\bigtriangleup_{3}xy(3+\epsilon
)-\nonumber\\
&  \bigtriangleup_{1}^{%
\acute{}%
}y\left\{  2\epsilon y-(3-\epsilon)x\right\}  +\bigtriangleup_{2}^{%
\acute{}%
}(2\epsilon y^{2})-\bigtriangleup_{3}^{%
\acute{}%
}xy(3-\epsilon) \label{2ndESSRSP}%
\end{align}
provided%

\begin{equation}
\frac{\partial P}{\partial p}\mid_{\substack{p=q=p^{\star} \\p_{1}=q_{1}%
=p_{1}^{\star}}}=0\text{ \ \ \ \ \ \ \ \ \ \ \ \ \ }\frac{\partial P}{\partial
p_{1}}\mid_{\substack{p=q=p^{\star} \\p_{1}=q_{1}=p_{1}^{\star}}}=0
\label{CondsRSP}%
\end{equation}
The conditions in Eq. (\ref{CondsRSP}) together define the mixed NE
$(p^{\star},p_{1}^{\star})$. Consider now the modified RSP game in classical
form obtained by setting $\left|  c_{11}\right|  ^{2}=1$. The Eqs.
(\ref{CondsRSP}) become%

\begin{align}
-2\epsilon p^{\star}-(\epsilon+3)p_{1}^{\star}+(\epsilon+1)  &  =0\nonumber\\
(-\epsilon+3)p^{\star}-2\epsilon p_{1}^{\star}+(\epsilon-1)  &  =0
\end{align}
and $p^{\star}=p_{1}^{\star}=\frac{1}{3}$ is obtained as a mixed NE for all
the range $-1<\epsilon<0$. From Eq. (\ref{2ndESSRSP}) we get%

\begin{align}
&  P\{(p^{\star},p_{1}^{\star}),(p,p_{1})\}-P\{(p,p_{1}),(p,p_{1}%
)\}\nonumber\\
&  =2\epsilon(x^{2}+y^{2}+xy)=\epsilon\left\{  (x+y)^{2}+(x^{2}+y^{2}%
)\right\}  \leq0 \label{differ}%
\end{align}
In the classical RSP game, therefore, the mixed NE $p^{\star}=p_{1}^{\star
}=\frac{1}{3}$ is a NE but not an ESS, because the second condition of an ESS
given in the Eq. (\ref{ESScondsRSP}) does not hold.

Now define a new initial state as%

\begin{equation}
\left|  \psi_{in}\right\rangle =\frac{1}{2}\left\{  \left|  12\right\rangle
+\left|  21\right\rangle +\left|  13\right\rangle +\left|  31\right\rangle
\right\}  \label{InistatRSP}%
\end{equation}
and use it to play the game, instead of the classical game obtained from
$\left|  \psi_{in}\right\rangle =\left|  11\right\rangle $. The strategy
$p^{\star}=p_{1}^{\star}=\frac{1}{3}$ still forms a mixed NE because the
conditions (\ref{CondsRSP}) hold true for it. However the payoff difference of
Eq. (\ref{2ndESSRSP}) is now given below, when $-1<\epsilon<0$ and $x,y\neq0$:%

\begin{align}
&  P\{(p^{\star},p_{1}^{\star}),(p,p_{1})\}-P\{(p,p_{1}),(p,p_{1}%
)\}\nonumber\\
&  =-\epsilon\left\{  (x+y)^{2}+(x^{2}+y^{2})\right\}  >0
\end{align}
Therefore the mixed NE $p^{\star}=p_{1}^{\star}=\frac{1}{3}$, not existing as
an ESS in the classical form of the RSP game, becomes an ESS when the game is
quantized and played using an initial (entangled) quantum state given by the
Eq. (\ref{InistatRSP}).

Note that from Eq. (\ref{3StrategyPayoff}) the sum of the payoffs to Alice and
Bob $(P_{A}+P_{B})$ can be obtained for both the classical mixed strategy game
(i.e. when $\left|  \psi_{in}\right\rangle =\left|  11\right\rangle $) and the
quantum game played using the quantum state of Eq. (\ref{InistatRSP}). For the
matrix (\ref{RSPMatrix}) we write these sums as $(P_{A}+P_{B})_{cl}$ and
$(P_{A}+P_{B})_{qu}$ for classical mixed strategy and quantum games
respectively. We obtain%

\begin{equation}
(P_{A}+P_{B})_{cl}=-2\epsilon\left\{  (1-p-p_{1})(1-q-q_{1})+p_{1}%
q_{1}+pq\right\}
\end{equation}
and%

\begin{equation}
(P_{A}+P_{B})_{qu}=-\left\{  \frac{1}{2}(P_{A}+P_{B})_{cl}+\epsilon\right\}
\end{equation}
In case $\epsilon=0$ both the classical and quantum games are clearly zero
sum. For the slightly modified version of the RSP game we have $-1<\epsilon<0$
and both versions of the game become non zero-sum.

\section{Stability of a mixed Nash equilibrium}

In a classical symmetric bi-matrix game, played in an evolutionary set up
involving a population, all the members of the population are
indistinguishable and each individual is equally likely to face the other. In
such a set-up individuals interact only in pair-wise encounters.

Assume a finite set of pure strategies $\left\{  1,2,...,n\right\}  $ is
available to each player. In one pair-wise encounter let a player $A$ receives
a reward $a_{ij}$ by playing strategy $i$ against another player $B$ playing
strategy $j$. In symmetric situation the player $B$, then, gets $a_{ji}$ as a
reward. The value $a_{ij}$ is an element in the $n\times n$ payoff matrix
$\mathbf{M}$. We assume that the players also have an option to play a mixed
strategy. It means he/she plays the strategy $i$ with probability $p_{i}$ for
all $i=1,2,...,n.$ A strategy vector $\mathbf{p,}$ with components $p_{i},$
represents the mixed strategy played by the player. In standard notation an
average, or expected, payoff for player $A$ playing strategy $\mathbf{p}$
against player $B$ playing $\mathbf{q}$ is written as $P(\mathbf{p,q)}$
\cite{MarkBroom2}:%

\begin{equation}
P(\mathbf{p,q)=}\sum a_{ij}p_{i}q_{j}=\mathbf{p}^{T}\mathbf{Mq}
\label{PayoffsMixed}%
\end{equation}
where $T$ is for transpose.

In evolutionary game theory mixed strategies play a significant role. The
well-known \textit{Bishop-Cannings theorem} (BCT) \cite{BishopCanning}
describes an interesting property of mixed ESSs in symmetric bi-matrix games.
Introducing the concept of \textit{support }of an ESS is helpful to understand
the BCT \cite{MarkBroom2,Vickers}. Suppose a strategy vector $\mathbf{p=(}%
p_{i}\mathbf{)}$ is an ESS. Its \emph{support} $S(\mathbf{p})$ is the set
$S(\mathbf{p})=\left\{  i:p_{i}>0\right\}  $. Hence the support of
$\mathbf{p}$ is the set of pure strategies that can be played by a $\mathbf{p}
$-player. BCT states that if $\mathbf{p}$ is an ESS with support
I\textbf{\ }and\textbf{\ }$\mathbf{r}$\textbf{\ }$\neq\mathbf{p}$ is an ESS
with support $J$, then $I$ $\nsupseteq$ $J$. For bi-matrix games the
BCT\ shows that \textit{no pure strategy can be evolutionary stable when a
mixed ESS exists} \cite{MarkBroom2}. Naturally one, then, asks about the
classical pure ESSs when a switch-over to a quantum form of a classical
symmetric bi-matrix game also gives evolutionary stability to a mixed
symmetric NE.

Following the approach developed for quantum RSP game, we now consider a
general form of a two-qubit initial quantum state. Our results show that for
this form of the initial quantum state, the corresponding quantum version of a
bi-matrix game can give evolutionary stability to a mixed symmetric NE when
classically it is not stable. It is interesting to observe that by ensuring
evolutionary stability to a mixed NE in a quantum form of the game, the BCT
forces out the pure ESSs present in classical form of the game.

The payoff to a player in the quantized version of the RSP game can also be
written in similar form to (\ref{PayoffsMixed}), provided the matrix
$\mathbf{M}$ is replaced with a matrix corresponding to quantum version of
this game.

For example, Alice's payoff, who plays the strategy $\mathbf{p}$ (where
$\mathbf{p}^{T}\mathbf{=}[(1-p-p_{1})$ \ \ \ $p_{1}$ \ \ \ $p]$) against Bob
who plays the strategy $\mathbf{q}$ (where $\mathbf{q}^{T}\mathbf{=}%
[(1-q-q_{1})$ \ $\ \ q_{1}$ \ \ $\ q]$), can be written as%

\begin{equation}
P_{A}(\mathbf{p,q)=p}^{T}\mathbf{\omega q}%
\end{equation}
where the matrix $\mathbf{\omega}$ is given by%

\begin{equation}
\mathbf{\omega=}\left(
\begin{array}
[c]{ccc}%
\omega_{11} & \omega_{12} & \omega_{13}\\
\omega_{21} & \omega_{22} & \omega_{23}\\
\omega_{31} & \omega_{32} & \omega_{33}%
\end{array}
\right)  \label{QMatrixMixed}%
\end{equation}
and the elements of $\mathbf{\omega}$ are given by the following matrix equation%

\begin{align}
&  \left(
\begin{array}
[c]{ccccccccc}%
\omega_{11} & \omega_{12} & \omega_{13} & \omega_{21} & \omega_{22} &
\omega_{23} & \omega_{31} & \omega_{32} & \omega_{33}%
\end{array}
\right) \nonumber\\
&  =\left(
\begin{array}
[c]{ccccccccc}%
\alpha_{11} & \alpha_{12} & \alpha_{13} & \alpha_{21} & \alpha_{22} &
\alpha_{23} & \alpha_{31} & \alpha_{32} & \alpha_{33}%
\end{array}
\right)  \times\nonumber\\
&  \left(
\begin{array}
[c]{ccccccccc}%
\left|  c_{11}\right|  ^{2} & \left|  c_{12}\right|  ^{2} & \left|
c_{13}\right|  ^{2} & \left|  c_{21}\right|  ^{2} & \left|  c_{22}\right|
^{2} & \left|  c_{23}\right|  ^{2} & \left|  c_{31}\right|  ^{2} & \left|
c_{32}\right|  ^{2} & \left|  c_{33}\right|  ^{2}\\
\left|  c_{12}\right|  ^{2} & \left|  c_{11}\right|  ^{2} & \left|
c_{12}\right|  ^{2} & \left|  c_{22}\right|  ^{2} & \left|  c_{21}\right|
^{2} & \left|  c_{22}\right|  ^{2} & \left|  c_{32}\right|  ^{2} & \left|
c_{31}\right|  ^{2} & \left|  c_{32}\right|  ^{2}\\
\left|  c_{13}\right|  ^{2} & \left|  c_{13}\right|  ^{2} & \left|
c_{11}\right|  ^{2} & \left|  c_{23}\right|  ^{2} & \left|  c_{23}\right|
^{2} & \left|  c_{21}\right|  ^{2} & \left|  c_{33}\right|  ^{2} & \left|
c_{33}\right|  ^{2} & \left|  c_{31}\right|  ^{2}\\
\left|  c_{21}\right|  ^{2} & \left|  c_{22}\right|  ^{2} & \left|
c_{23}\right|  ^{2} & \left|  c_{11}\right|  ^{2} & \left|  c_{12}\right|
^{2} & \left|  c_{13}\right|  ^{2} & \left|  c_{21}\right|  ^{2} & \left|
c_{22}\right|  ^{2} & \left|  c_{23}\right|  ^{2}\\
\left|  c_{22}\right|  ^{2} & \left|  c_{21}\right|  ^{2} & \left|
c_{22}\right|  ^{2} & \left|  c_{12}\right|  ^{2} & \left|  c_{11}\right|
^{2} & \left|  c_{12}\right|  ^{2} & \left|  c_{22}\right|  ^{2} & \left|
c_{21}\right|  ^{2} & \left|  c_{22}\right|  ^{2}\\
\left|  c_{23}\right|  ^{2} & \left|  c_{23}\right|  ^{2} & \left|
c_{21}\right|  ^{2} & \left|  c_{13}\right|  ^{2} & \left|  c_{13}\right|
^{2} & \left|  c_{11}\right|  ^{2} & \left|  c_{23}\right|  ^{2} & \left|
c_{23}\right|  ^{2} & \left|  c_{21}\right|  ^{2}\\
\left|  c_{31}\right|  ^{2} & \left|  c_{32}\right|  ^{2} & \left|
c_{33}\right|  ^{2} & \left|  c_{31}\right|  ^{2} & \left|  c_{32}\right|
^{2} & \left|  c_{33}\right|  ^{2} & \left|  c_{11}\right|  ^{2} & \left|
c_{12}\right|  ^{2} & \left|  c_{13}\right|  ^{2}\\
\left|  c_{32}\right|  ^{2} & \left|  c_{31}\right|  ^{2} & \left|
c_{32}\right|  ^{2} & \left|  c_{32}\right|  ^{2} & \left|  c_{31}\right|
^{2} & \left|  c_{32}\right|  ^{2} & \left|  c_{12}\right|  ^{2} & \left|
c_{11}\right|  ^{2} & \left|  c_{12}\right|  ^{2}\\
\left|  c_{33}\right|  ^{2} & \left|  c_{33}\right|  ^{2} & \left|
c_{31}\right|  ^{2} & \left|  c_{33}\right|  ^{2} & \left|  c_{33}\right|
^{2} & \left|  c_{31}\right|  ^{2} & \left|  c_{13}\right|  ^{2} & \left|
c_{13}\right|  ^{2} & \left|  c_{11}\right|  ^{2}%
\end{array}
\right) \nonumber\\
&  \label{QMatrix1Mixed}%
\end{align}
The matrix (\ref{QMatrixMixed}) reduces to its classical form
(\ref{Gen3StrategyMatrix}) by fixing $\left|  c_{11}\right|  ^{2}=1$.

In a symmetric game the exchange of strategies by Alice and Bob also exchanges
their respective payoffs. The quantum game corresponding to the matrix
(\ref{Gen3StrategyMatrix}), when played using the initial quantum state of Eq.
(\ref{RSPIniStat}), becomes symmetric when%

\begin{equation}
\left|  c_{ij}\right|  ^{2}=\left|  c_{ji}\right|  ^{2}\text{ for }i\neq j
\label{SymCond}%
\end{equation}
The two-player quantum game, with three pure strategies, gets a form similar
to a classical matrix game. The payoff matrix of the classical game, however,
is replaced with its quantum version (\ref{QMatrixMixed}). Also the matrix
(\ref{QMatrixMixed}) now involves the coefficients $c_{ij}$ of the initial
quantum state (\ref{RSPIniStat}).

To reduce the above mathematical formalism to two-player, two-strategy quantum
game lets fix $p_{1}=q_{1}=0$, that is, both players do not use the operator
$\hat{D}$ at all, and apply only the operators $\hat{C}$ and $\hat{I}$ on the
initial quantum state. Payoff to the player who plays the strategy vector
$\mathbf{p}$ (where $\mathbf{p}^{T}\mathbf{=}[1-p$ \ \ $p]$) against the
player who plays the strategy vector $\mathbf{q}$ (where $\mathbf{q}%
^{T}\mathbf{=}[1-q$ \ \ $q]$) can again be written as $P(\mathbf{p,q)=p}%
^{T}\mathbf{\omega q}$. Nevertheless, $\mathbf{\omega}$ is reduced to its
simpler form:
\begin{equation}
\mathbf{\omega}=\left(
\begin{array}
[c]{cc}%
\omega_{11} & \omega_{13}\\
\omega_{31} & \omega_{33}%
\end{array}
\right)
\end{equation}
where elements of the matrix are%

\begin{equation}
\left(
\begin{array}
[c]{c}%
\omega_{11}\\
\omega_{13}\\
\omega_{31}\\
\omega_{33}%
\end{array}
\right)  =\left(
\begin{array}
[c]{cccc}%
\left|  c_{11}\right|  ^{2} & \left|  c_{13}\right|  ^{2} & \left|
c_{31}\right|  ^{2} & \left|  c_{33}\right|  ^{2}\\
\left|  c_{13}\right|  ^{2} & \left|  c_{11}\right|  ^{2} & \left|
c_{33}\right|  ^{2} & \left|  c_{31}\right|  ^{2}\\
\left|  c_{31}\right|  ^{2} & \left|  c_{33}\right|  ^{2} & \left|
c_{11}\right|  ^{2} & \left|  c_{13}\right|  ^{2}\\
\left|  c_{33}\right|  ^{2} & \left|  c_{31}\right|  ^{2} & \left|
c_{13}\right|  ^{2} & \left|  c_{11}\right|  ^{2}%
\end{array}
\right)  \left(
\begin{array}
[c]{c}%
\alpha_{11}\\
\alpha_{13}\\
\alpha_{31}\\
\alpha_{33}%
\end{array}
\right)  \label{TermsMixed}%
\end{equation}
It becomes a bi-matrix game played with the initial quantum state
(\ref{RSPIniStat}). The available pure strategies are $1$ and $3$ \emph{only}
and the terms with subscripts containing $2$ disappear. Take $x=1-p$ and
$y=1-q,$ so that $x$ and $y$ are probabilities with which players apply
identity operator on the initial state $\left|  \psi_{ini}\right\rangle $. The
strategy vectors $\mathbf{p}$ and $\mathbf{q}$ can now be represented only by
the numbers $x$ and $y$, respectively. Payoff to a $x$-player against a $y
$-player is then obtained as%

\begin{equation}
P(x\mathbf{,}y\mathbf{)=p}^{T}\mathbf{\omega q=}x\left\{  \omega_{11}%
y+\omega_{13}(1-y)\right\}  +(1-x)\left\{  \omega_{31}y+\omega_{33}%
(1-y)\right\}  .
\end{equation}
Suppose $(x^{\star},x^{\star})$ is a NE, i.e.%

\begin{align}
&  P(x^{\star},x^{\star})-P(x,x^{\star})\nonumber\\
&  =(x^{\star}-x)\left\{  x^{\star}(\omega_{11}-\omega_{13}-\omega_{31}%
+\omega_{33})+(\omega_{13}-\omega_{33})\right\}  \geq0
\end{align}
for all $x\in\lbrack0,1]$. The mixed strategy $x^{\star}=x_{q}^{\star}%
=(\omega_{33}-\omega_{13})/(\omega_{11}-\omega_{13}-\omega_{31}+\omega_{33})$
makes the payoff difference $P(x^{\star},x^{\star})-P(x,x^{\star})$
identically zero. The subscript $q$ is for `quantum'. Let $\bigtriangleup
x=x^{\star}-x$ then%

\begin{equation}
P(x_{q}^{\star},x)-P(x,x)=-(\bigtriangleup x)^{2}\left\{  \omega_{11}%
-\omega_{13}-\omega_{31}+\omega_{33}\right\}  \label{ESS2}%
\end{equation}
Now $x_{q}^{\star}$ is an ESS if $\left\{  P(x_{q}^{\star},x)-P(x,x)\right\}
>0$ for all $x\neq x_{q}^{\star}$, which leads to the requirement
$(\omega_{11}-\omega_{31}-\omega_{13}+\omega_{33})<0.$

The classical game corresponds when $\left|  c_{11}\right|  ^{2}=1$ and it
gives $\omega_{11}=\alpha_{11}$, $\omega_{13}=\alpha_{13},$ $\omega
_{31}=\alpha_{31}$, and $\omega_{33}=\alpha_{33}$, in accordance with the Eq.
(\ref{QMatrix1Mixed})$.$ In case $(\alpha_{11}-\alpha_{13}-\alpha_{31}%
+\alpha_{33})>0$, the mixed NE of a classical game, i.e. $x^{\star}%
=x_{c}^{\star}=(\alpha_{33}-\alpha_{13})/(\alpha_{11}-\alpha_{13}-\alpha
_{31}+\alpha_{33})$, is not an ESS. Here the subscript $c$ is for `classical'.
Since we look for a situation where evolutionary stability of a symmetric NE
changes --- while the corresponding NE remains intact --- in a switch-over of
the game from its classical to quantum form, lets take%

\begin{equation}
x_{c}^{\star}=x_{q}^{\star}=\frac{\alpha_{33}-\alpha_{13}}{\alpha_{11}%
-\alpha_{31}-\alpha_{13}+\alpha_{33}}=\frac{\omega_{33}-\omega_{13}}%
{\omega_{11}-\omega_{31}-\omega_{13}+\omega_{33}}. \label{MixedNE1}%
\end{equation}
saying that the classical NE $x_{c}^{\star}$ is also a NE in quantum form of
the game. Now from the matrix in the Eq. (\ref{TermsMixed})%

\begin{align}
&  (\omega_{11}-\omega_{31}-\omega_{13}+\omega_{33})\nonumber\\
&  =(\alpha_{11}-\alpha_{13}-\alpha_{31}+\alpha_{33})(\left|  c_{11}\right|
^{2}-\left|  c_{13}\right|  ^{2}-\left|  c_{31}\right|  ^{2}+\left|
c_{33}\right|  ^{2}) \label{MixedNE2}%
\end{align}
and
\begin{align}
\omega_{33}-\omega_{13}  &  =\left|  c_{11}\right|  ^{2}(\alpha_{33}%
-\alpha_{13})+\left|  c_{13}\right|  ^{2}(\alpha_{31}-\alpha_{11})+\nonumber\\
&  \left|  c_{31}\right|  ^{2}(\alpha_{13}-\alpha_{33})+\left|  c_{33}\right|
^{2}(\alpha_{11}-\alpha_{31}) \label{MixedNE3}%
\end{align}
A substitution from Eqs. (\ref{MixedNE2},\ref{MixedNE3}) into the Eq.
(\ref{MixedNE1}) gives $\alpha_{33}-\alpha_{13}=\alpha_{11}-\alpha_{31}$, and
this leads to $x_{c}^{\star}=x_{q}^{\star}=1/2$. Therefore the mixed strategy
$x^{\star}=1/2$ remains a NE in both the classical and quantum form of the
game. Consider this mixed NE for a classical game with $(\alpha_{11}%
-\alpha_{13}-\alpha_{31}+\alpha_{33})>0$, a condition that assures that it is
not an ESS. The Eq. (\ref{MixedNE2}) shows an interesting possibility that one
can have $(\omega_{11}-\omega_{31}-\omega_{13}+\omega_{33})<0$ if%

\begin{equation}
(\left|  c_{11}\right|  ^{2}+\left|  c_{33}\right|  ^{2})<(\left|
c_{13}\right|  ^{2}+\left|  c_{31}\right|  ^{2}) \label{IniCondMixed}%
\end{equation}
In other words, the evolutionary stability of a mixed strategy, which is a NE
in both classical and quantum versions of the game, changes when the game
switches-over between its two forms. To have a symmetric game also in its
quantum form we need $\left|  c_{13}\right|  ^{2}=\left|  c_{31}\right|  ^{2}%
$, which reduces the inequality (\ref{IniCondMixed}) to $\frac{1}{2}(\left|
c_{11}\right|  ^{2}+\left|  c_{33}\right|  ^{2})<\left|  c_{13}\right|  ^{2}$.

Hence, a quantum version of a symmetric bi-matrix classical game with the matrix:%

\begin{equation}
\left(
\begin{array}
[c]{cc}%
(\alpha_{11},\alpha_{11}) & (\alpha_{13},\alpha_{31})\\
(\alpha_{31},\alpha_{13}) & (\alpha_{33},\alpha_{33})
\end{array}
\right)
\end{equation}
can be played if both the players have access to two unitary and Hermitian
operators and the game starts with a two-qubit quantum state of the form:%

\begin{equation}
\left|  \psi_{ini}\right\rangle =\underset{i,j=1,3}{\sum}c_{ij}\left|
ij\right\rangle \label{SymmInit}%
\end{equation}
where $\underset{i,j=1,3}{\sum}\left|  c_{ij}\right|  ^{2}=1$. In case
$\alpha_{33}-\alpha_{13}=\alpha_{11}-\alpha_{31}$ the mixed strategy
$x^{\star}=1/2$ is not an ESS in the classical game if $(\alpha_{33}%
-\alpha_{13})>0$. Nevertheless, the strategy $x^{\star}=1/2$ becomes an ESS
when $\left|  c_{11}\right|  ^{2}+\left|  c_{33}\right|  ^{2}<\left|
c_{13}\right|  ^{2}+\left|  c_{31}\right|  ^{2}$. In case $(\alpha_{33}%
-\alpha_{13})<0$ the strategy $x^{\star}=1/2$ is an ESS classically but does
not remain so if again $\left|  c_{11}\right|  ^{2}+\left|  c_{33}\right|
^{2}<\left|  c_{13}\right|  ^{2}+\left|  c_{31}\right|  ^{2}$. Now suppose
$\left|  c_{13}\right|  ^{2}=\left|  c_{31}\right|  ^{2}=0$, the Eq.
(\ref{MixedNE2}) reduces to%

\begin{equation}
(\omega_{11}-\omega_{13}-\omega_{31}+\omega_{33})=(\alpha_{11}-\alpha
_{13}-\alpha_{31}+\alpha_{33}) \label{StabilityMixedStrategyCondition}%
\end{equation}
It is observed from the equation (\ref{StabilityMixedStrategyCondition}) that
if a quantum game is played by the following simple form of the initial
quantum state:%

\begin{equation}
\left|  \psi_{ini}\right\rangle =c_{11}\left|  11\right\rangle +c_{33}\left|
33\right\rangle \label{SimpInit}%
\end{equation}
then it is not possible to influence the evolutionary stability of a mixed NE.

\subsection{Discussion}

Mixed ESSs appear in many games of interest that are played in the natural
world. The examples of the RSP and Hawks and Doves games are well known from
evolutionary game theory. The Bishop-Cannings theorem of evolutionary game
theory does not permit pure ESSs when a mixed ESS exists in a population
engaged in a bi-matrix game. The possibility of changing evolutionary
stability of a pure symmetric NE has been considered earlier with normalized
states of the form $\left|  \psi_{ini}\right\rangle =c_{11}\left|
1,1\right\rangle +c_{22}\left|  2,2\right\rangle $. Using such initial states,
however, \emph{cannot} change evolutionary stability of a mixed strategy. In
this section, following an approach developed for the quantum RSP game, we
play a bi-matrix game with a general two-qubit pure initial quantum state.
Such state allows changing evolutionary stability of a mixed NE. For a
bi-matrix game a symmetric mixed NE is found that remains intact in both the
classical and quantum versions of the game. For this mixed NE conditions are
then found allowing the change of evolutionary stability with a switch-over of
the game, between its two forms, one classical and the other quantum.

\section{Equilibria of replicator dynamics in quantum games}

\subsection{Introduction}

Maynard Smith and Price \cite{Smith Price} introduced the idea of an ESS
essentially as a static concept. Nothing in the definition of an ESS
guarantees that the dynamics of evolution in small mutational steps will
necessarily converge the process of evolution to an ESS. In fact directional
evolution may also become responsible for the establishment of strategies that
are not evolutionary stable \cite{Cressman}.

What are the advantages involved in a dynamic approach towards theory of ESSs?
A stated reason \cite{CressmanSchlag} is that dynamic approach introduces
structural stability into game theory. Historically Liapunov provided a
classic definition of stability of equilibria for general dynamical systems.
This definition can also be adapted for the stability of a NE. A pair of
strategies $(p^{\star},q^{\star})$ is \emph{Liapunov stable} when for every
trajectory starting somewhere in a small neighborhood of radius $\epsilon>0$
around a point representing the pair $(p^{\star},q^{\star})$ another small
neighborhood of radius $\delta>0$ can be defined such that the trajectory
stays in it. When every trajectory starting in a small neighborhood of radius
$\sigma>0$ around the point $(p^{\star},q^{\star})$ converges to $(p^{\star
},q^{\star})$ the strategy pair $(p^{\star},q^{\star})$ becomes an
\emph{attractor}. Trajectories are defined by the dynamics underlying the game.

Taylor and Jonker \cite{TaylorJonker} introduced a dynamics into evolutionary
games with the hypothesis that the growth rate of those playing each strategy
is proportional to the advantage of that strategy. This hypothesis is now
understood as one of many different forms of replicator dynamics\textit{\ }%
\cite{Cressman,Bomze}. In simple words assume that $p_{i}$ is the frequency
(i.e. relative proportion) of the individuals using strategy $i$ and
$\mathbf{p}$, where $\mathbf{p}^{T}=[p_{1},p_{2}...p_{i}...p_{n}]$ and $T$ is
the transpose, is a vector whose components are the frequencies with
$\overset{n}{\underset{i=1}{\Sigma}}p_{i}=1$. Let $P_{i}(\mathbf{p})$ be the
average payoff for using $i$ when the population is in the state $\mathbf{p}$.
Let $\bar{P}=\Sigma p_{j}P_{j}$ be the \emph{average success} in the
population. The replicator equation is, then, written as \cite{Sigmund}%

\begin{equation}
\dot{p}_{i}=p_{i}(P_{i}(\mathbf{p})-\bar{P}) \label{Replicator Eq}%
\end{equation}
where the dot is derivative with respect to time. Let the payoff matrix be
$A=(a_{ij})$ with $a_{ij}$ being the average payoff for strategy $i$ when the
other player uses $j$. The average payoff for the strategy $i$ in the
population (with the assumption of random encounters of the individuals) is
$(A\mathbf{p)}_{i}=a_{i1}p_{1}+...+a_{in}p_{n}$ and the Eq. (\ref{Replicator
Eq}) becomes%

\begin{equation}
\dot{p}_{i}=p_{i}((A\mathbf{p)}_{i}-\mathbf{p}^{T}A\mathbf{p})
\label{Replicator Eq1}%
\end{equation}

The population state is then given as a point in $n$ simplex $\bigtriangleup$
\cite{Zeeman}. The hypothesis of Taylor and Jonker \cite{TaylorJonker} gives a
flow on $\bigtriangleup$ whose flow lines represent the evolution of the
population. In evolutionary game theory it is agreed \cite{Weibull} that every
ESS is an attractor of the flow defined on $\bigtriangleup$ by the replicator
equation (\ref{Replicator Eq}), however, the converse does not hold: an
attractor is not necessarily an ESS.

Our motivation is to find how equilibria of replicator dynamics are affected
when a matrix game played by a population is quantized. It should, of course,
be sensitive to quantization procedure selected for the matrix game. We find
the effects when the matrix game is quantized via Marinatto and Weber's scheme.

\subsection{Equilibria and attractors of replicator dynamics}

Early studies about the attractors of replicator dynamic by Schuster, Sigmund
and Wolff \cite{Eigen,Schuster} reported the dynamics of enzymatic actions of
chemicals in a mixture when their relative proportions could be changed. For
example, in the case of a mixture of three chemicals added in a correct order,
such that corresponding initial conditions are in the basin of an interior
attractor, it becomes a stable cooperative mixture of all three chemicals. But
if they are added in a wrong order the initial conditions then lie in another
basin and only one of the chemicals survives with others two excluded. Eigen
and Schuster \cite{Eigen,Schuster,Hofbauer} also studied resulting dynamics in
the evolution of macromolecules before the advent of life.

Schuster and Sigmund \cite{Schuster1} applied the dynamic to animal behavior
in BoS and described the evolution of strategies by treating it as a dynamical
system. They wrote replicator Eqs. (\ref{Replicator Eq1}) for the payoff matrix:%

\begin{equation}%
\begin{array}
[c]{c}%
\text{Male's strategy}%
\end{array}%
\begin{array}
[c]{c}%
X_{1}\\
X_{2}%
\end{array}
\overset{%
\begin{array}
[c]{cc}%
\text{Female's} & \text{strategy}\\
Y_{1} & Y_{2}%
\end{array}
}{\left(
\begin{array}
[c]{cc}%
(a_{11},b_{11}) & (a_{12},b_{21})\\
(a_{21},b_{12}) & (a_{22},b_{22})
\end{array}
\right)  } \label{GenMatrixReplicator}%
\end{equation}
where a male can play pure strategies $X_{1}$, $X_{2}$ and a female can play
pure strategies $Y_{1}$, $Y_{2}$ respectively. Let in a population engaged in
this game the frequencies of $X_{1}$ and $X_{2}$ are $p_{1}$ and $p_{2}$
respectively. Similarly the frequencies of $Y_{1}$ and $Y_{2}$ are $q_{1}$ and
$q_{2}$ respectively. Obviously%

\begin{align}
p_{1}+p_{2}  &  =q_{1}+q_{2}=1\nonumber\\
\text{where }p_{i}  &  \geq0,\text{ }q_{i}\geq0\text{, for }i=1,2
\label{CondReplicator}%
\end{align}
the replicator equations (\ref{Replicator Eq1}) for the matrix
(\ref{GenMatrixReplicator}) with conditions (\ref{CondReplicator}) are, then,
written as%

\begin{align}
\dot{p}  &  =p(1-p)\left\{  q(a_{11}-a_{12}-a_{21}+a_{22})+(a_{12}%
-a_{22})\right\} \nonumber\\
\dot{q}  &  =q(1-q)\left\{  p(b_{11}-b_{12}-b_{21}+b_{22})+(b_{12}%
-b_{22})\right\}  \label{ReplicatorEqs}%
\end{align}
where $p_{1}=p$ and $q_{1}=q$. These are Lotka-Volterra type equations
describing the evolution of two populations identified as predator and prey
\cite{Hirsch}. Schuster and Sigmund \cite{Schuster1} simplified the problem by taking%

\begin{align}
a_{11}  &  =b_{11}=a_{22}=b_{22}=0\nonumber\\
a_{12}  &  =a\text{ \ \ }a_{21}=b\text{ \ \ and }\nonumber\\
b_{12}  &  =c\text{ \ \ }b_{21}=d \label{SimplyReplicator}%
\end{align}
which does not restrict generality of the problem and the replicator Eqs.
(\ref{ReplicatorEqs}) remain similar. Payoffs to the male $P_{M}(p,q)$ and to
the female $P_{F}(p,q)$, when the male plays $X_{1}$ with probability $p$ (he
then plays $X_{2}$ with the probability $(1-p)$) and the female plays $Y_{1}$
with the probability $q$ (she then plays $Y_{2}$ with the probability $(1-q)$)
are written as \cite{MarkBroom3}%

\begin{align}
P_{M}(p,q)  &  =\mathbf{p}^{T}\mathbf{Mq}\nonumber\\
P_{F}(p,q)  &  =\mathbf{q}^{T}\mathbf{Fp} \label{PayoffsReplicator}%
\end{align}
where%

\begin{equation}
\mathbf{M=}\left(
\begin{array}
[c]{cc}%
a_{11} & a_{12}\\
a_{21} & a_{22}%
\end{array}
\right)  ,\text{ \ \ and \ \ }\mathbf{F=}\left(
\begin{array}
[c]{cc}%
b_{11} & b_{12}\\
b_{21} & b_{22}%
\end{array}
\right)
\end{equation}
also%

\begin{equation}
\mathbf{p=}\left(
\begin{array}
[c]{c}%
p\\
1-p
\end{array}
\right)  ,\text{ \ \ and \ \ }\mathbf{q=}\left(
\begin{array}
[c]{c}%
q\\
1-q
\end{array}
\right)
\end{equation}
and $T$ is for transpose.

Now a quantum form of the matrix game (\ref{GenMatrixReplicator}) can be
played using Marinatto and Weber's scheme \cite{Marinatto1}. The players have
at their disposal an initial quantum state:%

\begin{equation}
\left|  \psi_{ini}\right\rangle =\underset{i,j=1,2}{\sum}c_{ij}\left|
ij\right\rangle
\end{equation}
with%

\begin{equation}
\underset{i,j=1,2}{\sum}\left|  c_{ij}\right|  ^{2}=1 \label{NormRp}%
\end{equation}
In quantum version the male and female players apply the identity $\hat{I}$ on
$\left|  \psi_{ini}\right\rangle $ with probabilities $p$ and $q$
respectively. Also they apply $\hat{\sigma}_{x}$ with probabilities $(1-p)$
and $(1-q)$, respectively. Payoffs to the players are written in a similar
form, as in the Eq. (\ref{PayoffsReplicator}):%

\begin{align}
P_{M}(p,q)  &  =\mathbf{p}^{T}\mathbf{\omega q}\nonumber\\
P_{F}(p,q)  &  =\mathbf{q}^{T}\mathbf{\chi p}%
\end{align}
$\mathbf{\omega}$ and $\mathbf{\chi}$ are quantum forms of the payoff matrices
$\mathbf{M}$ and $\mathbf{F}$ respectively i.e.%

\begin{equation}
\mathbf{\omega=}\left(
\begin{array}
[c]{cc}%
\omega_{11} & \omega_{12}\\
\omega_{21} & \omega_{22}%
\end{array}
\right)  \text{ \ \ and\ \ \ }\mathbf{\chi=}\left(
\begin{array}
[c]{cc}%
\chi_{11} & \chi_{12}\\
\chi_{21} & \chi_{22}%
\end{array}
\right)
\end{equation}
where%

\begin{align}
\omega_{11}  &  =a_{11}\left|  c_{11}\right|  ^{2}+a_{12}\left|
c_{12}\right|  ^{2}+a_{21}\left|  c_{21}\right|  ^{2}+a_{22}\left|
c_{22}\right|  ^{2}\nonumber\\
\omega_{12}  &  =a_{11}\left|  c_{12}\right|  ^{2}+a_{12}\left|
c_{11}\right|  ^{2}+a_{21}\left|  c_{22}\right|  ^{2}+a_{22}\left|
c_{21}\right|  ^{2}\nonumber\\
\omega_{21}  &  =a_{11}\left|  c_{21}\right|  ^{2}+a_{12}\left|
c_{22}\right|  ^{2}+a_{21}\left|  c_{11}\right|  ^{2}+a_{22}\left|
c_{12}\right|  ^{2}\nonumber\\
\omega_{22}  &  =a_{11}\left|  c_{22}\right|  ^{2}+a_{12}\left|
c_{21}\right|  ^{2}+a_{21}\left|  c_{12}\right|  ^{2}+a_{22}\left|
c_{11}\right|  ^{2}%
\end{align}
similarly%

\begin{align}
\chi_{11}  &  =b_{11}\left|  c_{11}\right|  ^{2}+b_{12}\left|  c_{12}\right|
^{2}+b_{21}\left|  c_{21}\right|  ^{2}+b_{22}\left|  c_{22}\right|
^{2}\nonumber\\
\chi_{12}  &  =b_{11}\left|  c_{12}\right|  ^{2}+b_{12}\left|  c_{11}\right|
^{2}+b_{21}\left|  c_{22}\right|  ^{2}+b_{22}\left|  c_{21}\right|
^{2}\nonumber\\
\chi_{21}  &  =b_{11}\left|  c_{21}\right|  ^{2}+b_{12}\left|  c_{22}\right|
^{2}+b_{21}\left|  c_{11}\right|  ^{2}+b_{22}\left|  c_{12}\right|
^{2}\nonumber\\
\chi_{22}  &  =b_{11}\left|  c_{22}\right|  ^{2}+b_{12}\left|  c_{21}\right|
^{2}+b_{21}\left|  c_{12}\right|  ^{2}+b_{22}\left|  c_{11}\right|  ^{2}
\label{termsF}%
\end{align}
For the initial state $\left|  \psi_{ini}\right\rangle =\left|
11\right\rangle $ the matrices $\mathbf{\omega}$ and $\mathbf{\chi}$ are same
as $\mathbf{M}$ and $\mathbf{F}$ respectively. The classical game is,
therefore, embedded in the quantum game. Simplified matrices $\mathbf{\omega}$
and $\mathbf{\chi}$ can be obtained by the assumption of Eq.
(\ref{SimplyReplicator}), that is%

\begin{align}
\omega_{11}  &  =a\left|  c_{12}\right|  ^{2}+b\left|  c_{21}\right|
^{2}\text{, \ \ }\omega_{12}=a\left|  c_{11}\right|  ^{2}+b\left|
c_{22}\right|  ^{2}\nonumber\\
\omega_{21}  &  =a\left|  c_{22}\right|  ^{2}+b\left|  c_{11}\right|
^{2}\text{, \ \ }\omega_{22}=a\left|  c_{21}\right|  ^{2}+b\left|
c_{12}\right|  ^{2}\nonumber\\
\chi_{11}  &  =c\left|  c_{12}\right|  ^{2}+d\left|  c_{21}\right|
^{2}\text{, \ \ }\chi_{12}=c\left|  c_{11}\right|  ^{2}+d\left|
c_{22}\right|  ^{2}\nonumber\\
\chi_{21}  &  =c\left|  c_{22}\right|  ^{2}+d\left|  c_{11}\right|
^{2}\text{, \ \ }\chi_{22}=c\left|  c_{21}\right|  ^{2}+d\left|
c_{12}\right|  ^{2}%
\end{align}
The replicator Eqs. (\ref{ReplicatorEqs}) can now be written in the following
`quantum' form:%

\begin{align}
\dot{x}  &  =x(1-x)[aK_{1}+bK_{2}-(a+b)(K_{1}+K_{2})y]\nonumber\\
\dot{y}  &  =y(1-y)[cK_{1}+dK_{2}-(c+d)(K_{1}+K_{2})x] \label{QRpEq}%
\end{align}
where $K_{1}=\left|  c_{11}\right|  ^{2}-\left|  c_{21}\right|  ^{2}$ and
$K_{2}=\left|  c_{22}\right|  ^{2}-\left|  c_{12}\right|  ^{2}$. These
equations reduce to Eqs. (\ref{ReplicatorEqs}) for $\left|  \psi
_{ini}\right\rangle =\left|  11\right\rangle $ i.e. $\left|  c_{11}\right|
^{2}=1$. Similar to the classical version \cite{Schuster1} the dynamics
(\ref{QRpEq}) has five rest or equilibrium points $x=0,$ $y=0$;$\qquad x=0,$
$y=1$;$\qquad x=1,$ $y=0$;$\qquad x=1,$ $y=1$; and an interior equilibrium point:%

\begin{equation}
x=\frac{cK_{1}+dK_{2}}{(c+d)(K_{1}+K_{2})}\text{, \ \ }y=\frac{aK_{1}+bK_{2}%
}{(a+b)(K_{1}+K_{2})} \label{IntEqRp}%
\end{equation}
This equilibrium point is the same as in the classical game \cite{Schuster1}
for $\left|  \psi_{ini}\right\rangle =\left|  11\right\rangle $ i.e.%

\begin{equation}
x=\frac{c}{c+d}\text{, \ \ }y=\frac{a}{a+b}%
\end{equation}
We use the method of linear approximation \cite{Hirsch} at equilibrium points
to find the general character of phase diagram of the system (\ref{QRpEq}).
Write the system (\ref{QRpEq}) as%

\begin{equation}
\dot{x}=\mathbf{X(}x,y\mathbf{)}\text{, \ \ }\dot{y}=\mathbf{Y}(x,y)
\end{equation}
The matrix for linearization \cite{Hirsch} is%

\begin{equation}
\left[
\begin{array}
[c]{cc}%
\mathbf{X}_{x} & \mathbf{X}_{y}\\
\mathbf{Y}_{x} & \mathbf{Y}_{y}%
\end{array}
\right]  \label{linztn}%
\end{equation}
where, for example, $\mathbf{X}_{x}$ denotes $\frac{\partial\mathbf{X}%
}{\partial x}$. The matrix (\ref{linztn}) is evaluated at each equilibrium
point in turn. Write now these terms as%

\begin{align}
\mathbf{X}_{x}  &  =(1-2x)\left\{  aK_{1}+bK_{2}-(a+b)(K_{1}+K_{2})y\right\}
\nonumber\\
\mathbf{X}_{y}  &  =-x(1-x)(a+b)(K_{1}+K_{2})\nonumber\\
\mathbf{Y}_{x}  &  =-y(1-y)(c+d)(K_{1}+K_{2})\nonumber\\
\mathbf{Y}_{y}  &  =(1-2y)\left\{  cK_{1}+dK_{2}-(c+d)(K_{1}+K_{2})x\right\}
\label{TermML}%
\end{align}
and the characteristic equation \cite{Hirsch} at an equilibrium point is
obtained from%

\begin{equation}
\left|
\begin{array}
[c]{cc}%
(\mathbf{X}_{x}-\lambda) & \mathbf{X}_{y}\\
\mathbf{Y}_{x} & (\mathbf{Y}_{y}-\lambda)
\end{array}
\right|  =0 \label{CharEqRp}%
\end{equation}
The patterns of phase paths around equilibrium points classify the points into
a few principal cases. Suppose $\lambda_{1},\lambda_{2}$ are roots of the
characteristic Eq. (\ref{CharEqRp}). A few cases are as follows:

\begin{enumerate}
\item $\lambda_{1},\lambda_{2}$ are real, different, non-zero, and of same
sign. If $\lambda_{1},\lambda_{2}>0$ then the equilibrium point is an
\emph{unstable node} or a repeller. If $\lambda_{1},\lambda_{2}<0$ the node is
stable or an \emph{attractor}.

\item $\lambda_{1},\lambda_{2}$ are real, different, non-zero, and of opposite
sign. The equilibrium point is a \emph{saddle point}.

\item $\lambda_{1},\lambda_{2}=\alpha\pm i\beta$, and $\beta\neq0$. The
equilibrium is a \emph{stable spiral} (attractor) if $\alpha<0$, an
\emph{unstable spiral} (repeller) if $\alpha>0$, a \emph{center} if $\alpha=0$.
\end{enumerate}

Consider an equilibrium or rest point $x=1,$ $y=0$, written simply as $(1,0)$.
At this point the characteristic Eq. (\ref{CharEqRp}) has the roots:%

\begin{equation}
\lambda_{1}=-aK_{1}-bK_{2}\text{, \ \ }\lambda_{2}=-cK_{2}-dK_{1}
\label{RootsRp}%
\end{equation}
For the classical game, i.e. $\left|  \psi_{ini}\right\rangle =\left|
11\right\rangle $, these roots are $\lambda_{1}=-a$, $\lambda_{2}=-d$.
Therefore in case $a,d>0$ the equilibrium point $(1,0)$ is an attractor in the
classical game.

Consider the interior equilibrium point $(x,y)$ of Eq. (\ref{IntEqRp}). The
terms of the matrix of linearization of Eq. (\ref{TermML}) are:%

\begin{align}
\mathbf{X}_{x}  &  =0\text{, \ \ }\mathbf{Y}_{y}=0\nonumber\\
\mathbf{X}_{y}  &  =\frac{-(cK_{1}+dK_{2})(cK_{2}+dK_{1})(a+b)}{(c+d)^{2}%
(K_{1}+K_{2})}\nonumber\\
\mathbf{Y}_{x}  &  =\frac{-(aK_{1}+bK_{2})(aK_{2}+bK_{1})(c+d)}{(a+b)^{2}%
(K_{1}+K_{2})}%
\end{align}
the roots of the characteristic Eq. (\ref{CharEqRp}) are numbers $\pm\lambda$ where%

\begin{equation}
\lambda=\sqrt{\frac{(aK_{1}+bK_{2})(aK_{2}+bK_{1})(cK_{1}+dK_{2}%
)(cK_{2}+dK_{1})}{(a+b)(c+d)(K_{1}+K_{2})^{2}}}%
\end{equation}
the term in square root can be a positive or negative real number. Therefore:

\begin{itemize}
\item  A saddle (center) in classical game can be a center (saddle) in certain
quantum form of the game.

\item  A saddle or a center in a classical (quantum) game can not be an
attractor or a repeller in quantum (classical) form of the game.
\end{itemize}

\chapter{Relevance of evolutionary stability in quantum games}

Evolutionary game theory considers attractors of a dynamics and ESSs with
reference to population models. Extending these ideas to a quantum setting
requires an assumption of population of individuals, or entities, with access
to quantum states and quantum mechanical operators. What is the possible
relevance of such an assumption in the real world? To answer it we observe
that the concept of evolutionary stability is based on the following
assumptions, that also define its population setting:

\begin{itemize}
\item  There are random and pair-wise interactions between the participating
players forming a population. These interactions can be re-expressed in
game-theoretic language by constructing symmetric bi-matrices.

\item  A step-wise selection mechanism that ensures that a successful strategy
has better chance to spread itself in the population at the expense of other strategies.
\end{itemize}

While bringing the ESS concept to quantum games, we retain the population
setting of evolutionary game theory as well as the step-wise selection
mechanism. However, the games played among the players, during pair-wise
interactions, are replaced with their quantum counterparts. Questions now
naturally arise how such a switch-over to a quantum game changes the
evolutionary outcome.

Following are the some suggestions where a relevance of quantization of games
may affect, and even decide, an evolutionary outcome.

\section{Quantum mechanics deciding an evolutionary outcome}

Evolutionary game theory was developed to provide game-theoretic models of
animal conflicts that occur in our macro-world. However, recent work in
biology \cite{Turner} suggests that nature also plays classical games at
micro-level. Bacterial infections by viruses are classical game-like
situations where nature prefers dominant strategies. The game-theoretical
explanation of stable states in a population of interacting individuals can be
considered a model of rationality which is physically grounded in natural selection.

A motivation to study evolutionary stability in quantum games exists because
the population setting of evolutionary game theory can also be introduced to
quantum games. It can be done on the \emph{same} ground as it is done in the
classical games. The notion of a Nash equilibrium, that became the topic of
pioneering work on quantum games, was itself motivated by a population setting.

Consideration of evolutionary stability in quantum games shows how
quantization of games, played in a population, can lead to new stable states
of the population. It shows that the presence of quantum interactions, in a
population undergoing evolution, can alter its stable states resulting from
the evolutionary dynamics. When quantum effects decide the evolutionary
outcomes, the role for quantum mechanics clearly increases, from just keeping
atoms together, to deciding the outcome of an evolutionary dynamics.

\section{Development of complexity and self-organization}

This new role for quantum mechanics can be to define and maintain complexity
emerging from quantum interactions among a collection of molecules. Eigen,
Schuster, Sigmund and Wolf \cite{Eigen,Schuster} consider an example of a
mixture in which an equilibrium is achieved from competing chemical reactions.
Such an equilibrium can also be an outcome of quantum interactions occurring
at molecular level. When quantum nature of molecular interactions can decide
an equilibrium state, there is a clear possibility for the quantum mechanical
role in the models of self-organization in matter. These considerations seem
quite relevant to the evolution of macromolecules before the advent of life.
The possibility that stability of solutions (or equilibria) can be affected by
quantum interactions provides a new approach towards understanding of rise of
complexity in groups of quantum-interacting entities.

Physicists have expressed opinions \cite{Frohlich} about the possibility of
quantum mechanics `fast tracking' a chemical soup to states that are
biological and complex, and the debate continues. We suggest that quantum game
theory also have contributions to make towards the attempts to understand
quantum mechanical role in life, especially evolution and development of self
organization and complexity in molecular systems, and possibly the origin of consciousness.

Considering development of quantum mechanical models of life, in a recent
paper Flitney and Abbott\ \cite{FlitneyAbbott3}\ studied a version of John
Conway's game of Life \cite{Game of Life}\ where the normal binary values of
the cells are replaced by oscillators which can represent a superposition of
states. They showed that the original game of Life is reproduced in the
classical limit, but in general additional properties not seen in the original
game are present that display some of the effects of a quantum mechanical Life.

\section{Genetic code evolution}

Genetic code is the relationship between sequence of bases in DNA and the
sequence of amino acids in proteins. Suggestions have been made earlier about
quantum mechanical role in the genetic code. For example, supersymmetry in
particle physics, giving a unified description of fermions and bosons, have
been suggested \cite{Bashford} to provide an explanation of coding assignments
in genetic code. Recent work \cite{Knight} about evolvability of the genetic
code suggests that the code, like all other features of organisms, was shaped
by natural selection. The question about the process and evolutionary
mechanism by which the genetic code was optimized is still unanswered. Two
suggested possibilities are:

\begin{itemize}
\item  A large number of codes existed out of which the adaptive one was selected.

\item  Adaptive and error-minimizing constraints gave rise to an adaptive code
via code expansion and simplification.
\end{itemize}

The second possibility of code expansion from earlier simpler forms is now
thought to be supported by much empirical and genetic evidence \cite{Knight1}
and results suggest that the present genetic code was strongly influenced by
natural selection for error minimization.

Patel \cite{Patel} suggested quantum dynamics played a role in the DNA
replication and in the optimization criteria involved in genetic information
processing. He considers the criteria as a task similar to an unsorted
assembly operation, with possible connection to the Grover's database search
algorithm \cite{Grover}, given different optimal solutions result from the
classical and quantum dynamics.

The assumption in this approach is that an adaptive code was selected out of a
large numbers that existed earlier. The suggestion of natural selection being
the process, for error minimization in the mechanism of adaptive code
evolution, puts forward an evolutionary approach for this optimization
problem. We believe that, in the evolution and expansion of the code from its
earlier simpler forms, quantum dynamics has played a role. The mechanism
leading to this optimization will be, however, different. Our result that
stable outcomes, of an evolutionary process based on natural selection, may
also depend on the quantum nature of interactions clearly implies the
possibility that such interactions may decide the optimal outcome of evolution.

We believe that the code optimization is a problem having close similarities
to the problem of evolutionary stability. And this optimization was probably
achieved by interactions that were quantum mechanical in nature.

\section{Quantum evolutionary algorithms}

A polynomial time algorithm that can solve an NP problem is not known yet. A
viable alternative approach, shown to find acceptable solutions within a
reasonable time period, is the evolutionary search \cite{Back}. Iteration of
selection based on competition, random variation or mutation, and exploration
of the fitness landscape of possible solutions, are the basic ingredients of
many distinct paradigms of evolutionary computing \cite{Back1}. On the other
hand superposition of all possible solution states, unitary operators
exploiting interference to enhance the amplitude of the desired states, and
final measurement extracting the solution are the components of quantum
computing. These two approaches in computing are believed to represent
different philosophies \cite{Greenwood}.

Finding ESSs can also be formulated as an evolutionary algorithm. The view
that quantum mechanics has a role in the theory of ESSs suggests that the two
philosophies -- considered different so far -- may have common grounds uniting
them. It also hints the possibility of evolutionary algorithms that utilize,
or even exploit, quantum effects. In such an evolutionary algorithm, we may
have, for example, fitness functions depending on the amount of entanglement
present. The natural question to ask is then how the population will evolve
towards an equilibrium state in relation to the amount of entanglement.

\section{Evolutionary quantum optimization and computation}

The perspective that matrix game theory provides, on what should be an outcome
of evolution, has been studied in this thesis. Another perspective is provided
by optimization models \cite{Meszena}. In evolutionary matrix games a
frequency-dependent selection takes place and all alternative strategies
become equally fit when an ESS establishes itself. On the other hand, in
optimization models the selection is frequency-independent and evolution is
imagined as a hill-climbing process. Optimal solution is obtained when fitness
is maximized. Evolutionary optimization is the basis of evolutionary and
genetic algorithms and is generally considered to be a different approach from
ESSs in matrix games. These are not, however, in direct contradiction and give
different outlooks on evolutionary process. We suggest that evolutionary
optimization is another area where a role for quantum mechanics exists and
quantum game theory provides hints to find it.

It seems appropriate to mention here the evolutionary quantum computation
(EQC) described in the Ref \cite{Goertzel}. In EQC an ensemble of quantum
subsystems is considered changing continually such a way as to optimize some
measure of emergent patterns between the system and its environment. It seems
reasonable that this optimization is related to an equilibrium or some of its
properties. When quantum interactions decide the equilibria and their
stability properties, it implies that the optimization itself depends on it.
Brain also has been proposed \cite{Goertzel} as an evolutionary quantum computer.

\chapter{Cooperation in quantum games}

\section{Introduction}

In contrast to non-cooperative games the players in cooperative games are not
able to form binding agreements even if they may communicate. The distinctive
feature of cooperative games is a strong incentive to work together to receive
the largest total payoff. These games allow players to form coalitions,
binding agreements, pay compensations, make side payments etc. Von Neumann and
Morgenstern \cite{Neumann} in their pioneering work on the theory of games
offered models of coalition formation where the strategy of each player
consists of choosing the coalition s/he wishes to join. In coalition games,
that are part of cooperative game theory, the possibilities for the players
are described by the available resources of different groups (coalitions) of
players. Joining a group or remaining outside is part of strategy of a player
affecting his/her payoff.

Recent work in quantum games arises a natural and interesting question: what
is the possible quantum mechanical role in cooperative games that are
considered an integral part of the classical game theory? In our view it may
be quite interesting, and fruitful as well, to investigate coalitions in
quantum versions of cooperative games. Our present motivation is to
investigate what might happen to the advantage of forming a coalition in a
quantum game compared to its classical analogue. We rely on the concepts and
ideas of von Neumann's cooperative game theory \cite{Neumann} and consider a
three-player coalition game in a quantum form. We then compare it to the
classical version of the game and see how the advantage of forming a coalition
can be affected.

In classical analysis of coalition games the notion of a strategy disappears;
the main features are those of a coalition and the value or worth of the
coalition. The underlying assumption is that each coalition can guarantee its
members a certain amount called the `\emph{value of a coalition}'\textit{.}
``The value of coalition measures the worth the coalition possesses and is
characterized as the payoff which the coalition can assure for itself by
selecting an appropriate strategy, whereas the `odd man' can prevent the
coalition from getting more than this amount''\textit{\ }\cite{Burger}. Using
this definition we study cooperative games in quantum settings to see how
advantages of making coalitions can be influenced in the new setting.

Within the framework of playing a quantum game given by Marinatto and Weber,
we find a quantum form of a symmetric cooperative game played by three
players. In classical form of this game any two players, out of three, get an
advantage when they successfully form a coalition and play the same strategy.
We investigate how the advantage for forming a coalition are affected when the
game switches its form from classical to quantum.

\section{A three-player symmetric cooperative game}

\subsection{Classical form}

A classical three-person normal form game \cite{Burger} is given by:

\begin{itemize}
\item  Three non-empty sets $\Sigma_{A}$, $\Sigma_{B}$, and $\Sigma_{C}$.
These are the strategy sets of the players $A$, $B$, and $C$.

\item  Three real valued functions $P_{A}$, $P_{B}$, and $P_{C}$ defined on
$\Sigma_{A}\times\Sigma_{B}\times\Sigma_{C}$.
\end{itemize}

The \emph{product space} $\Sigma_{A}\times\Sigma_{B}\times\Sigma_{C}$ is the
set of all tuples $(\sigma_{A},\sigma_{B},\sigma_{C})$ with $\sigma_{A}%
\in\Sigma_{A}$, $\sigma_{B}\in\Sigma_{B}$ and $\sigma_{C}\in\Sigma_{C}$. A
strategy is understood as such a tuple $(\sigma_{A},\sigma_{B},\sigma_{C})$
and $P_{A}$, $P_{B}$, $P_{C}$ are payoff functions of the three players. The
game is denoted as $\Gamma=\left\{  \Sigma_{A},\Sigma_{B},\Sigma_{C}%
;P_{A},P_{B},P_{C}\right\}  $. Let $\Re=\left\{  A,B,C\right\}  $ be the set
of players and $\wp$ be an arbitrary subset of $\Re$. The players in $\wp$ may
form a coalition so that, for all practical purposes, the coalition $\wp$
appears as a single player. It is expected that players in $(\Re-\wp)$ will
form an opposing coalition and the game has two opposing ``coalition players''
i.e. $\wp$ and $(\Re-\wp)$.

We study quantum version of an example of a classical three player cooperative
game discussed in Ref. \cite{Burger}. Each of the three players $A$, $B$, and
$C$ chooses one of the two strategies $1$, $2$. If the three players choose
the same strategy there is no payoff; otherwise, the two players who have
chosen the same strategy receive one unit of money each from the `odd man.'
Payoff functions $P_{A}$, $P_{B}$ and $P_{C}$ for players $A$, $B$ and $C$,
respectively, are given as \cite{Burger}:%

\begin{align}
P_{A}(1,1,1)  &  =P_{A}(2,2,2)=0\nonumber\\
P_{A}(1,1,2)  &  =P_{A}(2,2,1)=P_{A}(1,2,1)=P_{A}(2,1,2)=1\nonumber\\
P_{A}(1,2,2)  &  =P_{A}(2,1,1)=-2 \label{PayoffsCoop}%
\end{align}
with similar expressions for $P_{B}$ and $P_{C}$. Suppose $\wp=\left\{
B,C\right\}  $, hence $\Re-\wp=\left\{  A\right\}  $. The coalition game
represented by $\Gamma_{\wp}$ is given by the payoff matrix \cite{Burger}:%

\begin{equation}%
\begin{array}
[c]{c}%
\left[  11\right] \\
\left[  12\right] \\
\left[  21\right] \\
\left[  22\right]
\end{array}
\overset{%
\begin{array}
[c]{cc}%
\left[  1\right]  & \left[  2\right]
\end{array}
}{\left(
\begin{array}
[c]{cc}%
0 & 2\\
-1 & -1\\
-1 & -1\\
2 & 0
\end{array}
\right)  }%
\end{equation}
Here the strategies $\left[  12\right]  $ and $\left[  21\right]  $ are
dominated by $\left[  11\right]  $ and $\left[  22\right]  $. After
eliminating these dominated strategies the payoff matrix becomes%

\begin{equation}%
\begin{array}
[c]{c}%
\left[  11\right] \\
\left[  22\right]
\end{array}
\overset{%
\begin{array}
[c]{cc}%
\left[  1\right]  & \left[  2\right]
\end{array}
}{\left(
\begin{array}
[c]{cc}%
0 & 2\\
2 & 0
\end{array}
\right)  }%
\end{equation}
It is seen that the mixed strategies:%

\begin{align}
&  \frac{1}{2}\left[  11\right]  +\frac{1}{2}\left[  22\right]  \text{,}%
\label{cltCoop}\\
&  \frac{1}{2}\left[  1\right]  +\frac{1}{2}\left[  2\right]  \text{.}
\label{lftCoop}%
\end{align}
are optimal for $\wp$ and $(\Re-\wp)$ respectively. With these strategies a
payoff $1$ for players $\wp$ is assured for all strategies of the opponent;
hence, the value of the coalition $\upsilon(\Gamma_{\wp})$ is $1$ i.e.
$\upsilon(\left\{  B,C\right\}  )=1$. Since $\Gamma$ is a zero-sum game
$\upsilon(\Gamma_{\wp})$ can also be used to find $\upsilon(\Gamma_{\Re-\wp})$
as $\upsilon(\left\{  A\right\}  )=-1$. The game is symmetric and one can write%

\begin{align}
\upsilon(\Gamma_{\wp})  &  =1\text{, \ \ and\ \ \ }\upsilon(\Gamma_{\Re-\wp
})=-1\text{ or}\nonumber\\
\upsilon(\left\{  A\right\}  )  &  =\upsilon(\left\{  B\right\}
)=\upsilon(\left\{  C\right\}  )=-1\nonumber\\
\upsilon(\left\{  A,B\right\}  )  &  =\upsilon(\left\{  B,C\right\}
)=\upsilon(\left\{  C,A\right\}  )=1 \label{VcltC}%
\end{align}

\subsection{Quantum form}

In quantum form of this three-player game the players -- identified as $A$,
$B$ and $C$ -- play their strategies by applying the identity operator
$\hat{I}$ with probabilities $p$, $q$, and $r$, respectively, on a three-qubit
initial quantum state. The players apply the operator $\hat{\sigma}_{x}$ with
probabilities $(1-p)$, $(1-q)$, and $(1-r)$ respectively. If $\rho_{in}$ is
the initial state, the final state, after players have played their
strategies, becomes%

\begin{equation}
\rho_{fin}=\underset{\hat{U}=\hat{I},\hat{\sigma}_{x}}{\sum}\Pr(\hat{U}%
_{A})\Pr(\hat{U}_{B})\Pr(\hat{U}_{C})\hat{U}_{A}\otimes\hat{U}_{B}\otimes
\hat{U}_{C}\rho_{in}\hat{U}_{A}^{\dagger}\otimes\hat{U}_{B}^{\dagger}%
\otimes\hat{U}_{C}^{\dagger} \label{FinstatCoop}%
\end{equation}
where the unitary and Hermitian operator $\hat{U}$ can be either $\hat{I}$ or
$\hat{\sigma}_{x}$. $\Pr(\hat{U}_{A})$, $\Pr(\hat{U}_{B})$ and $\Pr(\hat
{U}_{C})$ are the probabilities with which players $A$, $B$, and $C$ apply the
operator $\hat{U}$ on the initial state respectively. $\rho_{fin}$ corresponds
to a convex combination of players' quantum operations. Let the arbiter
prepares a three-qubit pure initial quantum state:%

\begin{equation}
\left|  \psi_{in}\right\rangle =\underset{i,j,k=1,2}{\sum}c_{ijk}\left|
ijk\right\rangle \text{, \ \ where \ \ }\underset{i,j,k=1,2}{\sum}\left|
c_{ijk}\right|  ^{2}=1 \label{InstateCoop}%
\end{equation}
where the basis vectors of the quantum state are $\left|  ijk\right\rangle $
for $i,j,k=1,2$. The state (\ref{InstateCoop}) is in $2\otimes2\otimes2$
dimensional Hilbert space and corresponds to three qubits.

Assume the matrix of the three-player game is given by $24$ constants
$\alpha_{t},\beta_{t},\gamma_{t}$ with $1\leq t\leq8$. Write the payoff
operators for players $A$, $B$ and $C$ as%

\begin{align}
(P_{A,B,C})_{oper}  &  =\alpha_{1},\beta_{1},\gamma_{1}\left|
111\right\rangle \left\langle 111\right|  +\alpha_{2},\beta_{2},\gamma
_{2}\left|  211\right\rangle \left\langle 211\right|  +\nonumber\\
&  \alpha_{3},\beta_{3},\gamma_{3}\left|  121\right\rangle \left\langle
121\right|  +\alpha_{4},\beta_{4},\gamma_{4}\left|  112\right\rangle
\left\langle 112\right|  +\nonumber\\
&  \alpha_{5},\beta_{5},\gamma_{5}\left|  122\right\rangle \left\langle
122\right|  +\alpha_{6},\beta_{6},\gamma_{6}\left|  212\right\rangle
\left\langle 212\right|  +\nonumber\\
&  \alpha_{7},\beta_{7},\gamma_{7}\left|  221\right\rangle \left\langle
221\right|  +\alpha_{8},\beta_{8},\gamma_{8}\left|  222\right\rangle
\left\langle 222\right|  \label{PayoperCoop}%
\end{align}
Payoffs to the players $A$, $B$ and $C$\ are then obtained as mean values of
these operators:%

\begin{equation}
P_{A,B,C}(p,q,r)=\text{Tr}\left[  (P_{A,B,C})_{oper}\rho_{fin}\right]
\end{equation}%

\begin{figure}
[ptb]
\begin{center}
\includegraphics[
height=4.6216in,
width=5.3333in
]%
{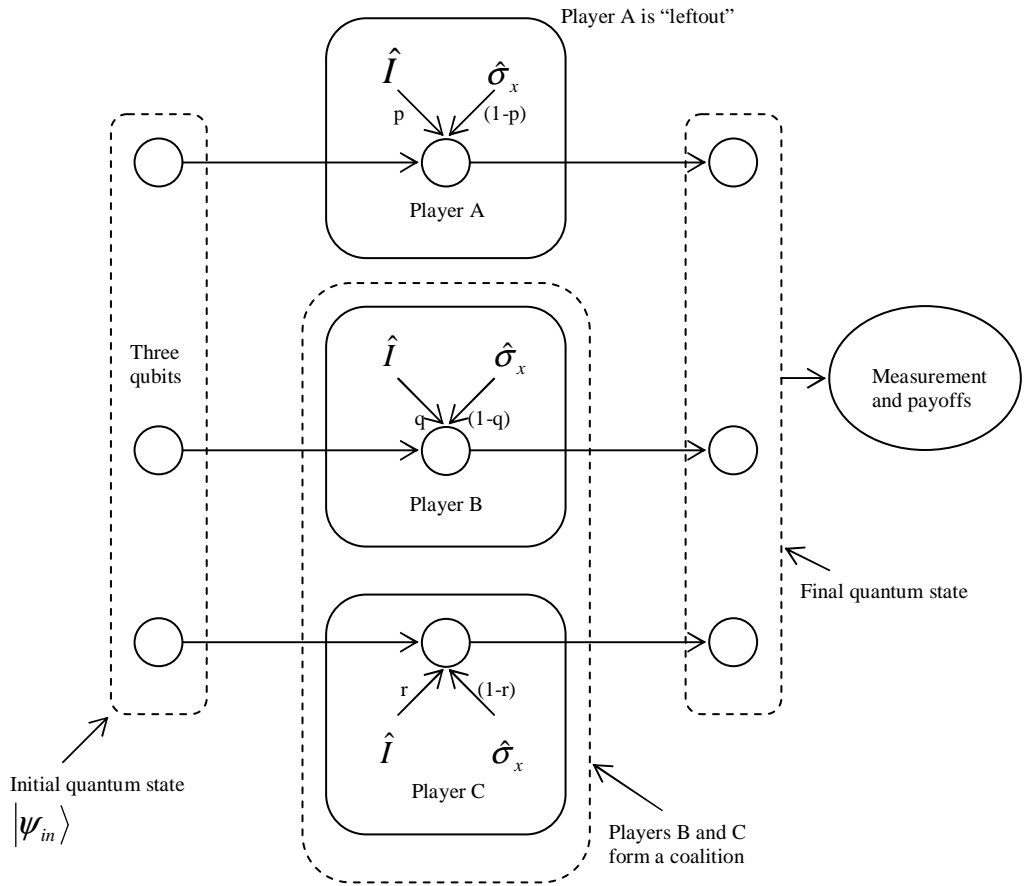}%
\caption{A three-player quantum game played with Marinatto and Weber's scheme.
Players $B$ and $C$ form a coalition. $\hat{I}$ is the identity and
$\hat{\sigma}_{x}$ is the inversion operator.}%
\label{Fig3}%
\end{center}
\end{figure}
Where the players' moves are identified by the numbers $p$, $q$ and $r$,
respectively. Fig. (\ref{Fig3}) shows the three-player quantum game. The
cooperative game of Eq. (\ref{PayoffsCoop}) with the classical payoff
functions $P_{A}$, $P_{B}$ and $P_{C}$ for players $A$, $B$ and $C$
respectively, together with the definition of payoff operators for these
players in Eq. (\ref{PayoperCoop}), imply that%

\begin{equation}
\alpha_{1}=\alpha_{8}=0\text{, \ \ \ \ }\alpha_{3}=\alpha_{4}=\alpha
_{6}=\alpha_{7}=1\text{, \ \ and\ \ \ }\alpha_{2}=\alpha_{5}=-2
\end{equation}
With these constants, in quantum version of the game, the payoff to player $A
$, for example, can be found as%

\begin{equation}
P_{A}(p,q,r)=%
\begin{array}
[c]{c}%
(-4rq-2p+2pr+2pq+r+q)(\left|  c_{111}\right|  ^{2}+\left|  c_{222}\right|
^{2})\\
+(-4rq+2p-2pr-2pq+3r+3q-2)(\left|  c_{211}\right|  ^{2}+\left|  c_{122}%
\right|  ^{2})\\
+(4rq+2pr-2pq-3r-q+1)(\left|  c_{121}\right|  ^{2}+\left|  c_{212}\right|
^{2})\\
+(4rq-2pr+2pq-r-3q+1)(\left|  c_{112}\right|  ^{2}+\left|  c_{221}\right|
^{2})
\end{array}
\label{PoffCoopQ}%
\end{equation}
Similarly, payoffs to players $B$ and $C$ can be obtained. Classical mixed
strategy payoffs can be recovered from the Eq. (\ref{PoffCoopQ}) by taking
$\left|  c_{111}\right|  ^{2}=1$. The classical game is therefore imbedded in
its quantum form.

The classical form of this game is symmetric in the sense that payoff to a
player depends on his/her strategy and not on his/her identity. These
requirements that result in a three-player symmetric game are written as%

\begin{align}
P_{A}(p,q,r)  &  =P_{A}(p,r,q)=P_{B}(q,p,r)=P_{B}(r,p,q)\nonumber\\
&  =P_{C}(r,q,p)=P_{A}(q,r,p) \label{RqmntsCoop}%
\end{align}
Now in this quantum form of the game, $P_{A}(p,q,r)$ becomes same as
$P_{A}(p,r,q)$ when \cite{DaiChen}:%

\begin{equation}
\left|  c_{121}\right|  ^{2}+\left|  c_{212}\right|  ^{2}=\left|
c_{112}\right|  ^{2}+\left|  c_{221}\right|  ^{2} \label{Rqmnts1Coop}%
\end{equation}
Similarly $P_{B}(q,p,r)=P_{B}(r,p,q)$ and $P_{C}(r,q,p)=P_{C}(q,r,p)$ when the
following conditions hold \cite{DaiChen}:%

\begin{align}
\left|  c_{211}\right|  ^{2}+\left|  c_{122}\right|  ^{2}  &  =\left|
c_{112}\right|  ^{2}+\left|  c_{221}\right|  ^{2}\nonumber\\
\left|  c_{211}\right|  ^{2}+\left|  c_{122}\right|  ^{2}  &  =\left|
c_{121}\right|  ^{2}+\left|  c_{212}\right|  ^{2}
\label{Requrmnts2Cooperativegames}%
\end{align}
Combining Eq. (\ref{Rqmnts1Coop}) and Eq. (\ref{Requrmnts2Cooperativegames}) give%

\begin{equation}
\left|  c_{211}\right|  ^{2}+\left|  c_{122}\right|  ^{2}=\left|
c_{121}\right|  ^{2}+\left|  c_{212}\right|  ^{2}=\left|  c_{112}\right|
^{2}+\left|  c_{221}\right|  ^{2}%
\end{equation}
and then payoff to a $p$-player remains same when other two players
interchange their strategies. The symmetry conditions (\ref{RqmntsCoop}) hold
if, together with Eqs. (\ref{Rqmnts1Coop}), the following relations are
\emph{also} true%

\begin{equation}%
\begin{array}
[c]{cc}%
\alpha_{1}=\beta_{1}=\gamma_{1}, & \alpha_{5}=\beta_{6}=\gamma_{7}\\
\alpha_{2}=\beta_{3}=\gamma_{4}, & \alpha_{6}=\beta_{5}=\gamma_{6}\\
\alpha_{3}=\beta_{2}=\gamma_{3}, & \alpha_{7}=\beta_{7}=\gamma_{5}\\
\alpha_{4}=\beta_{4}=\gamma_{2}, & \alpha_{8}=\beta_{8}=\gamma_{8}%
\end{array}
\end{equation}
These form the extra restrictions on the constants of payoff matrix and,
together with the conditions (\ref{Rqmnts1Coop}), give a three player
symmetric game in a quantum form. No subscript in a payoff expression is then
needed and $P(p,q,r)$ represents the payoff to a $p$-player against two other
players playing $q$ and $r$. The payoff $P(p,q,r)$ is found as%

\begin{equation}
P(p,q,r)=(\left|  c_{111}\right|  ^{2}+\left|  c_{222}\right|  ^{2}-\left|
c_{211}\right|  ^{2}-\left|  c_{122}\right|  ^{2})(-4rq-2p+2pr+2pq+r+q)
\label{QpayoffCoop}%
\end{equation}
Assume now that the pure strategies $[1]$ and $[2]$ correspond to $p=0$ and
$p=1$, respectively. The mixed strategy $n\left[  1\right]  +(1-n)\left[
2\right]  $, where $0\leq n\leq1$, means that the strategy $\left[  1\right]
$ is played with probability $n$ and $\left[  2\right]  $ with probability
$(1-n)$. Also \emph{suppose} that coalition $\wp$ plays the mixed
strategy\footnote{In a Comment on ``Quantum cooperative games'' that appeared
in Physics Letters A, Volume 328, Issues 4-5, Pages 414-415, 2 August
2004,\ Liang Dai and Qing Chen have pointed out that because the mixed
strategies $[12]$ and $[21]$ are not always dominated by $[11]$ and $[22]$ in
quantum form, there is no ground for assuming that the coalition $\wp$ always
plays the mixed strategy $l[11]+(1-l)[22]$.}:%

\begin{equation}
l[11]+(1-l)[22] \label{cltQ}%
\end{equation}
where the strategy $[11]$ means that both players in the coalition $\wp$ apply
the identity operator $\hat{I}$ with zero probability. Similarly the strategy
$[22]$ can be defined. The strategy in the Eq. (\ref{cltQ}) is such that the
coalition $\wp$ plays $[11]$ with probability $l$ and $[22]$ with probability
$(1-l)$. Similarly assume that the player in $(\Re-\wp)$ plays the mixed strategy:%

\begin{equation}
m[1]+(1-m)[2] \label{lftQ}%
\end{equation}
The payoff to the coalition $\wp$ is then obtained as%

\begin{align}
P_{\wp}  &  =(lm)P_{\wp\lbrack111]}+l(1-m)P_{\wp\lbrack112]}+\nonumber\\
&  (1-l)mP_{\wp\lbrack221]}+(1-l)(1-m)P_{\wp\lbrack222]} \label{cltPQ}%
\end{align}
where $P_{\wp\lbrack111]}$ is the payoff to $\wp$ when all three players play
$p=0$ i.e. the strategy $[1]$. Similarly $P_{\wp\lbrack221]}$ is the coalition
payoff when the coalition players play $p=1$ and the player in $(\Re-\wp)$
plays $p=0$. Now from Eq. (\ref{QpayoffCoop}) we get%

\begin{align}
P_{\wp\lbrack111]}  &  =2P(0,0,0)=0\nonumber\\
P_{\wp\lbrack112]}  &  =2P(0,0,1)=2(\left|  c_{111}\right|  ^{2}+\left|
c_{222}\right|  ^{2}-\left|  c_{211}\right|  ^{2}-\left|  c_{122}\right|
^{2})\nonumber\\
P_{\wp\lbrack221]}  &  =2P(1,1,0)=2(\left|  c_{111}\right|  ^{2}+\left|
c_{222}\right|  ^{2}-\left|  c_{211}\right|  ^{2}-\left|  c_{122}\right|
^{2})\nonumber\\
P_{\wp\lbrack222]}  &  =2P(1,1,1)=0
\end{align}
Also from Eq. (\ref{cltPQ}):%

\begin{equation}
P_{\wp}=2(\left|  c_{111}\right|  ^{2}+\left|  c_{222}\right|  ^{2}-\left|
c_{211}\right|  ^{2}-\left|  c_{122}\right|  ^{2})\left\{
l(1-m)+(1-l)m\right\}
\end{equation}
To find the value of coalition $\upsilon(\Gamma_{\wp})$ in the quantum game we
find $\frac{\partial P_{\wp}}{\partial m}$ and equate it to zero i.e. $P_{\wp
}$ is such a payoff to $\wp$ that the player in $(\Re-\wp)$ cannot change it
by changing his/her strategy given in Eq. (\ref{lftQ}). It gives,
interestingly, $l=\frac{1}{2}$ and the classical optimal strategy of the
coalition $\frac{1}{2}\left[  11\right]  +\frac{1}{2}\left[  22\right]  $
becomes optimal in the quantum game as well. In the quantum game the coalition
then secures following payoff, which is also termed as the value of the coalition:%

\begin{equation}
\upsilon(\Gamma_{\wp})=(\left|  c_{111}\right|  ^{2}+\left|  c_{222}\right|
^{2})-(\left|  c_{211}\right|  ^{2}+\left|  c_{122}\right|  ^{2})
\end{equation}
Similarly we get the value of coalition for $(\Re-\wp)$:%

\begin{equation}
\upsilon(\Gamma_{\Re-\wp})=-\left\{  \left|  c_{111}\right|  ^{2}+\left|
c_{222}\right|  ^{2}+\left|  c_{211}\right|  ^{2}+\left|  c_{122}\right|
^{2}\right\}
\end{equation}
Note that these values reduce to their classical counterparts of Eq.
(\ref{VcltC}) when the initial quantum state becomes unentangled and is given
by $\left|  \psi_{in}\right\rangle =\left|  111\right\rangle $. Classical form
of the coalition game is, therefore, a subset of its quantum version.

Suppose the arbiter now has at his disposal a quantum state:%

\begin{gather}
\left|  \psi_{in}\right\rangle =c_{111}\left|  111\right\rangle +c_{222}%
\left|  222\right\rangle +c_{211}\left|  211\right\rangle +c_{122}\left|
122\right\rangle \nonumber\\
\text{with }(\left|  c_{211}\right|  ^{2}+\left|  c_{122}\right|
^{2})>(\left|  c_{111}\right|  ^{2}+\left|  c_{222}\right|  ^{2})
\end{gather}
If we \emph{assume} that, with this initial state, the coalition $\wp$ still
plays the mixed strategy\footnote{Liang Dai and Qing Chan have also indicated
that when $(\left|  c_{211}\right|  ^{2}+\left|  c_{122}\right|
^{2})>(\left|  c_{111}\right|  ^{2}+\left|  c_{222}\right|  ^{2})$ the
strategies $\frac{1}{2}[12]+\frac{1}{2}[21]$ and $\frac{1}{2}[1]+\frac{1}%
{2}[2]$ are optimal.} $l[11]+(1-l)[22]$ of the classical case, then
$\upsilon(\Gamma_{\wp})$ becomes a negative quantity and $\upsilon
(\Gamma_{(\Re-\wp)})=-1$ because of the normalization given in Eq.
(\ref{InstateCoop}). Another possible case is when the arbiter has the state:%

\begin{equation}
\left|  \psi_{in}\right\rangle =c_{211}\left|  211\right\rangle +c_{122}%
\left|  122\right\rangle
\end{equation}
at his disposal. Because now both $\upsilon(\Gamma_{\wp})$ and $\upsilon
(\Gamma_{\Re-\wp})$ are $-1$ and the players are left with no motivation to
form the \emph{same} coalition as they do in the classical game.

Liang Dai and Qing Chen \cite{DaiChen} have observed \footnote{Liang Dai and
Qing Chen \cite{DaiChen} pointed out a flaw in the calculations in the Ref.
\cite{CooperativeGames} by Iqbal and Toor. We argue that even after the
detection of the indicated flaw by Liang Dai and Qing Chen \cite{DaiChen} the
main conclusion of the Ref. \cite{CooperativeGames} remains intact. It can be
seen as follows.
\par
In their comment Liang Dai and Qing Chen \cite{DaiChen} wrote ``In quantum
form, the authors (Iqbal and Toor \cite{CooperativeGames}) concluded that the
game was not zero-sum, and, in some cases, the players had no motivation to
make a coalition because the advantage was lost. In this comment we argue that
the conclusions in Ref. \cite{CooperativeGames} are incorrect and led to
invalid conclusions.'' Now we refer to the main conclusion of the Ref.
\cite{CooperativeGames}, written in its abstract, saying ``In its classical
form (of a three-player game) making a coalition gives advantage to players
and they are motivated to do so. However, in its quantum form the advantage is
lost and players are left with no motivation to make a coalition.'' We argue
that this conclusion remains intact because:
\par
\begin{enumerate}
\item  Consider the quote from the page $108$ of the Ref.
\cite{CooperativeGames} ``The underlying assumption in this approach is that
because the arbiter, responsible for providing three-qubit pure quantum
initial states to be later unitarily manipulated by the players, can forward a
quantum state that corresponds to the classical game, therefore, other games
corresponding to different initial pure quantum states are quantum forms of
the classical game.This assumption makes possible to translate the problem of
finding a quantum version of the classical coalition game, having the property
that the advantage of making a coalition is lost, to finding some pure initial
quantum states. We showed that such quantum states can be found and,
therefore, there are quantum versions of the three-player coalition game where
the motivation for coalition formation is lost.''
\par
\item  In view of this quote along with Liang Dai and Qing Chen's finding that
when $\left|  c_{111}\right|  ^{2}+\left|  c_{222}\right|  ^{2}=\left|
c_{211}\right|  ^{2}+\left|  c_{122}\right|  ^{2}$ their remains no motivation
to form a coalition, it can be observed that, even after the indicated flaw in
the calculation, the main conclusion of the Ref. \cite{CooperativeGames}
remain intact. It is because the assumption made in the Ref.
\cite{CooperativeGames}, which is quoted above in detail, allows to consider
the corresponding game when $\left|  c_{111}\right|  ^{2}+\left|
c_{222}\right|  ^{2}=\left|  c_{211}\right|  ^{2}+\left|  c_{122}\right|
^{2}$ as a quantum form of the classical game. So that, Liang Dai and Qing
Chen \cite{DaiChen} main conclusion is same as in the Ref.
\cite{CooperativeGames}, apart from their identification of the correct
mathematical conditions that are required to find the particular quantum form
of the game in which the advantages of forming a coalition are lost.
\end{enumerate}
} that in case $\left|  c_{111}\right|  ^{2}+\left|  c_{222}\right|  ^{2}%
\neq\left|  c_{211}\right|  ^{2}+\left|  c_{122}\right|  ^{2}$, the motivation
to form a coalition remains in the quantum form. In case $\left|
c_{111}\right|  ^{2}+\left|  c_{222}\right|  ^{2}=\left|  c_{211}\right|
^{2}+\left|  c_{122}\right|  ^{2}$, every player's payoff becomes zero
whatever strategies they adapt, and hence the motivation to form a coalition
is lost.

\section{Discussion}

There may appear several guises in which the players can cooperate in a game.
One possibility is that they are able to communicate and, hence, able to
correlate their strategies. In certain situations players can make binding
commitments before or during the play of a game. Even in the post-play
behavior the commitments can make players to redistribute their final payoffs.

Two-player games are different from multi-player games in an important aspect.
In two-player games the question before the players is whether to cooperate or
not. In multi-player case the players are faced with a more difficult task.
Each player has to decide which coalition to join. There is also certain
uncertainty that the player faces about the extent to which players outside
his coalition may coordinate their actions.

Recent developments in quantum games provide a motivation to see how forming a
coalition, and its associated advantages, can be influenced in quantum
versions of these classical cooperative games. To study this we selected an
interesting, but simple, cooperative game and a recently proposed scheme
telling how to play a quantum game. We allowed the players in the quantum
version of the game to form a coalition similar to the classical game.

The \emph{underlying assumption} in this approach is that because the arbiter,
responsible for providing three-qubit pure initial quantum states, to be later
manipulated by the players, can forward a quantum state that corresponds to
the classical game, therefore, other games that result from different initial
pure quantum states are quantum forms of the classical game. This assumption
reduces the problem of finding a quantum version of the classical coalition
game to finding some pure initial quantum states. It is shown that a quantum
version of the three-player coalition game can be found where the motivation
for coalition formation is lost.

\chapter{Backwards-induction outcome in quantum games}

\section{Introduction}

The notion of NE, the central solution-concept in non-cooperative game theory,
was developed by John Nash in early 1950s. In fact Cournot (1838)
\cite{Cournot} anticipated Nash's definition of equilibrium by over a century
but only in the context of a particular model of a market which is dominated
by only two producers. In economics an \textit{oligopoly} is a form of market
in which a number $n$ of producers, say, $n\geq2$, \emph{and no others},
provide the market with a certain commodity. In the special case where $n=2$
it is called a duopoly. Cournot's work \cite{Cournot} is one of the classics
of game theory and also a cornerstone of the theory of industrial organization
\cite{Tirole}.

In Cournot model of duopoly two-firms simultaneously put certain quantities of
a homogeneous product in the market. Cournot obtained an equilibrium value for
the quantities both firms will decide to put in the market. This equilibrium
value was based on a rule of behavior which says that if all the players
except one abide by it, the remaining player cannot do better than to abide by
it too. Nash gave a general concept of an equilibrium point in a
noncooperative game but existence of an equilibrium in duopoly game was known
much earlier. The ``Cournot equilibrium'' refers to NE in non-cooperative form
of duopoly that Cournot considered.

In an interesting later development, Stackelberg (1934) \cite{Stackelberg,
Gibbons} proposed a \textit{dynamic }model of duopoly in which -- contrary to
Cournot's assumption of simultaneous moves -- a leader (or dominant) firm
moves first and a follower (or subordinate) firm moves second. A well known
example is the General Motors playing this leadership role in the early
history of US automobile industry when more than one firms like Ford and
Chrysler acted as followers. In this sequential game a ``Stackelberg
equilibrium'' is obtained using the \textit{backwards-induction outcome} of
the game. Stackelberg equilibrium refers to sequential nature of the game and
it is a stronger solution-concept than the NE because sequential move games
sometimes have multiple NE , only one of which is associated with the
backwards-induction outcome of the game \cite{Gibbons}.

In this chapter we present a quantum perspective on the interesting game of
Stackelberg duopoly. We start with the same assumption that a game is decided
only by players' unitary manipulations, payoff operators, and the measuring
apparatus deciding payoffs. When these are same a different input quantum
initial state gives a different form of the same game. With this assumption we
studied evolutionary stability of a mixed NE in the RSP game. Hence, a game
obtained by using a general two-qubit pure state is a quantum form of the
classical game provided the rest of the procedures in playing the quantum game
remain same\textit{.}

We now present an analysis of the Stackelberg duopoly by raising a question:
Is it possible to find a two-qubit pure quantum state that generates the
classical Cournot equilibrium as a backwards-induction outcome of the quantum
form of Stackelberg duopoly? Why can this question be of interest? It is
interesting because in case the answer is yes, then, quantization can
potentially be a useful element for the follower in the leader-follower model
of the Stackelberg duopoly \cite{Gibbons}. It is due to a known result that,
in classical setting, when static duopoly changes to a dynamic form, the
follower becomes worse-off compared to the leader who becomes better-off. We
find that, under certain restrictions, it is possible to find the needed
two-qubit quantum states. Hence a quantum form of a dynamic game of complete
information has an equilibrium that corresponds to classical static form of
the same game. The leader, thus, does not become better-off in the quantum
form of the dynamic duopoly.

\section{Backwards-induction outcome}

Consider a simple three step game:

\begin{enumerate}
\item  Player $1$ chooses an action $a_{1}$ from the feasible set $A_{1}$.

\item  Player $2$ observes $a_{1}$ and then chooses an action $a_{2}$ from the
feasible set $A_{2}$.

\item  Payoffs are $u_{1}(a_{1},a_{2})$ and $u_{2}(a_{1},a_{2})$.
\end{enumerate}

This game is an example of the dynamic games of complete and perfect
information. Key features of such games are:

\begin{enumerate}
\item  The moves occur in sequence.

\item  All previous moves are known before next move is chosen, and

\item  The players' payoffs are common knowledge.
\end{enumerate}

Given the action $a_{1}$ is previously chosen, at the second stage of the
game, when player $2$ gets the move s/he faces the problem:%

\begin{equation}
\underset{a_{2}\in A_{2}}{Max}\text{ }u_{2}(a_{1},a_{2}) \label{max2}%
\end{equation}
Assume that for each $a_{1}$ in $A_{1}$, player $2$'s optimization problem has
a unique solution $R_{2}(a_{1})$, which is also known as the \textit{best
response} of player $2$. Now player $1$ can also solve player $2$'s
optimization problem by anticipating player $2$'s response to each action
$a_{1}$ that player $1$ might take. So that player $1$ faces the problem:%

\begin{equation}
\underset{a_{1}\in A_{1}}{Max}\text{ }u_{1}(a_{1},R_{2}(a_{1})) \label{max1}%
\end{equation}
Suppose this optimization problem also has a unique solution for player $1$
and is denoted by $a_{1}^{\star}$. The solution $(a_{1}^{\star},R_{2}%
(a_{1}^{\star}))$ is the backwards-induction outcome of this game.

In a simple version of the Cournot's model two firms simultaneously decide the
quantities $q_{1}$ and $q_{2}$ respectively of a homogeneous product they want
to put into market. Suppose $Q$ is the aggregate quantity i.e. $Q=q_{1}+q_{2}$
and $P(Q)=a-Q$ be the \emph{market-clearing price}, which is the price at
which all products or services available in a market will find buyers. Assume
the total cost to a firm producing quantity $q_{i}$ is $cq_{i}$ i.e. there are
no \emph{fixed costs} and the \emph{marginal cost} is a constant $c$ with
$c<a$. In a two-player game theoretical model of this situation a firm's
payoff or profit can be written as \cite{Gibbons}%

\begin{equation}
P_{i}(q_{i},q_{j})=q_{i}\left[  P(Q)-c\right]  =q_{i}\left[  a-c-(q_{i}%
+q_{j})\right]  =q_{i}\left[  k-(q_{i}+q_{j})\right]  \label{PayoffEq}%
\end{equation}
Solving for the NE easily gives the Cournot equilibrium:%

\begin{equation}
q_{1}^{\star}=q_{2}^{\star}=\frac{k}{3} \label{Ceqbrm}%
\end{equation}
At this equilibrium the payoffs to both the firms from Eq. (\ref{PayoffEq}) are%

\begin{equation}
P_{1}(q_{1}^{\star},q_{2}^{\star})_{Cournot}=P_{2}(q_{1}^{\star},q_{2}^{\star
})_{Cournot}=\frac{k^{2}}{9} \label{CourPayoffs}%
\end{equation}
Consider now the classical form of duopoly game when it becomes dynamic. In
dynamic form of the game the payoffs to players are given by Eq.
(\ref{PayoffEq}) as they are for the Cournot's game. We find
backwards-induction outcome in classical and a quantum form of the
Stackelberg's duopoly. Taking advantage from a bigger picture given to this
dynamic game, by the Hilbert structure of its strategy space, we then find
two-qubit pure quantum states that give classical Cournot's equilibrium as the
backwards-induction outcome of the quantum game of Stackelberg's duopoly.

\section{Stackelberg duopoly}

\subsection{Classical form}

A leader (or dominant) firm moves first and a follower (or subordinate) firm
moves second in Stackelberg model of duopoly \cite{Gibbons}. The sequence of
events is

\begin{enumerate}
\item  Firm $A$ chooses a quantity $q_{1}\geq0$.

\item  Firm $B$ observes $q_{1}$ and then chooses a quantity $q_{2}\geq0$.

\item  The payoffs to firms $A$ and $B$ are given by their respective profit
functions as
\end{enumerate}%

\begin{align}
P_{A}(q_{1},q_{2})  &  =q_{1}\left[  k-(q_{1}+q_{2})\right] \nonumber\\
P_{B}(q_{1},q_{2})  &  =q_{2}\left[  k-(q_{1}+q_{2})\right]  \label{ProfFunc}%
\end{align}
The backwards-induction outcome is found by first finding firm $B$'s reaction
to an arbitrary quantity by firm $A.$ Denoting this quantity as $R_{2}(q_{1})$
we find%

\begin{equation}
R_{2}(q_{1})=\underset{q_{2}\geq0}{Max}\text{ }P_{B}(q_{1},q_{2}%
)=\frac{k-q_{1}}{2} \label{bestRes}%
\end{equation}
with $q_{1}<k$. Firm $A$ can now solve also the firm $B$'s problem. Firm $A$
can anticipate that a choice of the quantity $q_{1}$ will meet a reaction
$R_{2}(q_{1})$. In the first stage of the game firm $A$ can then compute a
solution to his/her optimization problem as%

\begin{equation}
\underset{q_{1}\geq0}{Max}\text{ }P_{A}\left[  q_{1},R_{2}(q_{1})\right]
=\underset{q_{1}\geq0}{Max}\text{ }\frac{q_{1}(k-q_{1})}{2}%
\end{equation}
It gives%

\begin{equation}
q_{1}^{\star}=\frac{k}{2}\text{ \ \ and \ \ }R_{2}(q_{1}^{\star})=\frac{k}{4}
\label{StkEq}%
\end{equation}
It is the classical backwards-induction outcome of dynamic form of the duopoly
game. At this equilibrium payoffs to the players $A$ and $B$ are given by Eqs.
(\ref{ProfFunc}) and (\ref{StkEq})%

\begin{equation}
P_{A}\left[  q_{1}^{\star},R_{2}(q_{1}^{\star})\right]  _{Stackelberg}%
=\frac{k^{2}}{8},\text{ \ \ \ \ }P_{B}\left[  q_{1}^{\star},R_{2}(q_{1}%
^{\star})\right]  _{Stackelberg}=\frac{k^{2}}{16} \label{StkPayffs}%
\end{equation}
From Eq. (\ref{StkPayffs}) find the ratio:%

\begin{equation}
\frac{P_{A}\left[  q_{1}^{\star},R_{2}(q_{1}^{\star})\right]  _{Stackelberg}%
}{P_{B}\left[  q_{1}^{\star},R_{2}(q_{1}^{\star})\right]  _{Stackelberg}}=2
\label{ratio}%
\end{equation}
showing that in comparison with the Cournot game, the leader firm becomes
better-off and the follower firm becomes worse-off in the Stackelberg game.
This aspect also hints an important difference between single and multi-person
decision problems. In single-person decision theory having more information
can never make the decision maker worse-off. In game theory, however, having
more information (or, more precisely, having it made public that one has more
information) can make a player worse-off \cite{Gibbons}.

Now we look at the backwards-induction outcome in a quantum perspective. Our
motivation is an interesting aspect that quantum form can bring into the
backwards-induction outcome. It is the possibility of firm $B$ not becoming
worse-off because of having extra information.

\subsection{Quantum form}

Stackelberg duopoly is a two-player sequential game. Meyer \cite{MeyerDavid}
considered a quantum form of the sequential game of PQ penny flip by unitary
operations on single qubit. Important difference between Meyer's game and
Stackelberg duopoly is that at the second stage player in PQ penny flip
doesn't know the previous move but in Stackelberg duopoly he knows that.

We prefer Marinatto and Weber's scheme to play the sequential game of
Stackelberg duopoly for two reasons:

\begin{enumerate}
\item  The classical game is obtained for a product initial state.

\item  When players' actions and payoff-generating measurement are
\emph{exactly} the same, we assume other games, corresponding to every pure
two-qubit initial state, are \emph{quantum forms} of the classical game.
\end{enumerate}

As discussed earlier, the second assumption originates from the fact that the
classical game corresponds to a pure two-qubit initial product state. The
assumption reduces the problem of finding a quantum form of Stackelberg
duopoly, with the property that its equilibrium is the same as in Cournot's
duopoly, to the problem of finding conditions on the parameters of two-qubit
pure initial quantum state. If the conditions are realistic then the
corresponding quantum game gives Cournot's equilibrium as the
backwards-induction outcome.

Stackelberg duopoly is a dynamic game of complete information. Its quantum
form in Marinatto and Weber's scheme starts by preparing a pure two-qubit
initial quantum state. Suppose Alice plays first and she announces her move
immediately, so that Bob knows Alice's move before playing his move. Bob plays
his move and both players forward their qubits for measurement.

Information about the previous moves is crucial for the game considered here.
A comparison of the sequential game of Stackelberg duopoly with the
simultaneous-move game of BoS makes evident different information structure in
these games. For example let BoS be played sequentially. But Alice does not
announce her first move to Bob before he makes his. It makes the game
sequential but the information structure is still the same as in its static
form. Hence, a sequential BoS in the above form has the same NE as in its
static form. An unobserved-action form of a game has the same NE as its
simultaneous-move form. This observation led us to play a quantum form of
Stackelberg duopoly while keeping intact the original structure of a scheme
designed for simultaneous moves. A consideration of playing a sequential game
in a quantum way brings to mind the Meyer's PQ penny-flip \cite{MeyerDavid}
where only one qubit is used. Contrary to this, in present section we use the
two-qubit system of a simultaneous-move game, to play a sequential game.

Why to use two qubits when a quantum form of this sequential game can also be
played by only one qubit, in similar way as Meyer's PQ penny-flip. We prefer
two qubits because in this case a comparison between classical and a quantum
form of the game translates itself to comparing two games resulting from using
a product and an entangled initial quantum state. We do not rule out the
possibility that a consideration of the dynamic game using only single qubit
gives equally, or even more, interesting results. We let classical payoffs in
Stackelberg duopoly, given by Eq. (\ref{ProfFunc}), reproduced when the
initial state $\left|  \psi_{in}\right\rangle =\left|  11\right\rangle $ is
used to play the game. The upper state of a qubit is then represented by $2$.
The state $\left|  \psi_{in}\right\rangle $ in density matrix notation is%

\begin{equation}
\rho_{in}=\left|  11\right\rangle \left\langle 11\right|  \label{IniDenMat}%
\end{equation}
Assume the player Alice and Bob apply $\hat{I}$ with probabilities $x$ and $y
$ respectively. The state (\ref{IniDenMat}) changes to%

\begin{align}
\rho_{fin}  &  =xy\hat{I}_{A}\otimes\hat{I}_{B}\rho_{in}\hat{I}_{A}^{\dagger
}\otimes\hat{I}_{B}^{\dagger}+x(1-y)\hat{I}_{A}\otimes\hat{\sigma}_{xB}%
\rho_{in}\hat{I}_{A}^{\dagger}\otimes\hat{\sigma}_{xB}^{\dagger}+\nonumber\\
&  y(1-x)\hat{\sigma}_{xA}\otimes\hat{I}_{B}\rho_{in}\hat{\sigma}%
_{xA}^{\dagger}\otimes\hat{I}_{B}^{\dagger}+(1-x)(1-y)\hat{\sigma}_{xA}%
\otimes\hat{\sigma}_{xB}\rho_{in}\hat{\sigma}_{xA}^{\dagger}\otimes\hat
{\sigma}_{xB}^{\dagger}\nonumber\\
&  \label{FinDenMat}%
\end{align}
where $x,y\in\lbrack0,1]$ are identified as the players' moves. The moves by
Alice and Bob in classical duopoly game are given by quantities $q_{1}$ and
$q_{2}$ where $q_{1},q_{2}\in\lbrack0,\infty)$. We assume that Alice and Bob
agree on a function that can \emph{uniquely }define a real positive number in
the range $(0,1]$ for every quantity $q_{1},q_{2}$ in $[0,\infty)$. A simple
such function is $1/(1+q_{i})$. So that, Alice and Bob find $x$ and $y$,
respectively, as%

\begin{equation}
x=\frac{1}{1+q_{1}},\text{ \ \ \ \ \ }y=\frac{1}{1+q_{2}} \label{DefFuns}%
\end{equation}
and use these real positive numbers as the probabilities with which they apply
the identity operator $\hat{I}$ on the quantum state at their disposal. With a
substitution from Eqs. (\ref{IniDenMat}, \ref{DefFuns}) the final state
(\ref{FinDenMat}) becomes%

\begin{equation}
\rho_{fin}=\frac{1}{(1+q_{1})(1+q_{2})}\left[  \left|  11\right\rangle
\left\langle 11\right|  +q_{1}q_{2}\left|  22\right\rangle \left\langle
22\right|  +q_{1}\left|  21\right\rangle \left\langle 21\right|  +q_{2}\left|
12\right\rangle \left\langle 12\right|  \right]
\end{equation}
We now assume that in the measurement and payoffs-finding phase the quantities
$q_{1}$ and $q_{2}$ are also known to the referee. The referee applies the
following payoff operators on the final quantum state:%

\begin{align}
(P_{A})_{oper}  &  =(1+q_{1})(1+q_{2})q_{1}\left[  k\left|  11\right\rangle
\left\langle 11\right|  -\left|  21\right\rangle \left\langle 21\right|
-\left|  12\right\rangle \left\langle 12\right|  \right] \nonumber\\
(P_{B})_{oper}  &  =(1+q_{1})(1+q_{2})q_{2}\left[  k\left|  11\right\rangle
\left\langle 11\right|  -\left|  21\right\rangle \left\langle 21\right|
-\left|  12\right\rangle \left\langle 12\right|  \right]  \label{PayOpers}%
\end{align}
Note that the classical payoffs of Eq. (\ref{ProfFunc}) are reproduced with
the initial state $\left|  \psi_{in}\right\rangle =\left|  11\right\rangle $ as%

\begin{equation}
P_{A,B}(q_{1},q_{2})=\text{Tr}\left[  (P_{A,B})_{oper}\rho_{fin}\right]
\label{PlayersPayoffs}%
\end{equation}
A more general form of quantum duopoly can now be played by keeping the payoff
operators of Eq. (\ref{PayOpers}) in the referee's possession and preparing an
initial two-qubit pure state:%

\begin{equation}
\left|  \psi_{in}\right\rangle =\underset{i,j=1,2}{\sum}c_{ij}\left|
ij\right\rangle \text{, \ \ \ with }\underset{i,j=1,2}{\sum}\left|
c_{ij}\right|  ^{2}=1 \label{BIOIniStat}%
\end{equation}
Payoffs to Alice and Bob can now be obtained, in this quantum game, from Eqs.
(\ref{PlayersPayoffs}) that use the payoff operators of Eqs. (\ref{PayOpers}).
The payoffs to Alice and Bob are written as%

\begin{align}
\left[  P_{A}(q_{1},q_{2})\right]  _{qtm}  &  =\frac{(\omega_{11}+\omega
_{12}q_{2})+q_{1}(\omega_{21}+\omega_{22}q_{2})}{(1+q_{1})(1+q_{2}%
)}\nonumber\\
\left[  P_{B}(q_{1},q_{2})\right]  _{qtm}  &  =\frac{(\chi_{11}+\chi_{12}%
q_{2})+q_{1}(\chi_{21}+\chi_{22}q_{2})}{(1+q_{1})(1+q_{2})} \label{QPayoffs}%
\end{align}
where the subscript $qtm$ is for `quantum' and%

\begin{align}
\left[
\begin{array}
[c]{c}%
\omega_{11}\\
\omega_{12}\\
\omega_{21}\\
\omega_{22}%
\end{array}
\right]   &  =\left[
\begin{array}
[c]{cccc}%
\left|  c_{11}\right|  ^{2} & \left|  c_{12}\right|  ^{2} & \left|
c_{21}\right|  ^{2} & \left|  c_{22}\right|  ^{2}\\
\left|  c_{12}\right|  ^{2} & \left|  c_{11}\right|  ^{2} & \left|
c_{22}\right|  ^{2} & \left|  c_{21}\right|  ^{2}\\
\left|  c_{21}\right|  ^{2} & \left|  c_{22}\right|  ^{2} & \left|
c_{11}\right|  ^{2} & \left|  c_{12}\right|  ^{2}\\
\left|  c_{22}\right|  ^{2} & \left|  c_{21}\right|  ^{2} & \left|
c_{12}\right|  ^{2} & \left|  c_{11}\right|  ^{2}%
\end{array}
\right]  \left[
\begin{array}
[c]{c}%
kq_{1}(1+q_{1})(1+q_{2})\\
-q_{1}(1+q_{1})(1+q_{2})\\
-q_{1}(1+q_{1})(1+q_{2})\\
0
\end{array}
\right] \nonumber\\
\left[
\begin{array}
[c]{c}%
\chi_{11}\\
\chi_{12}\\
\chi_{21}\\
\chi_{22}%
\end{array}
\right]   &  =\left[
\begin{array}
[c]{cccc}%
\left|  c_{11}\right|  ^{2} & \left|  c_{12}\right|  ^{2} & \left|
c_{21}\right|  ^{2} & \left|  c_{22}\right|  ^{2}\\
\left|  c_{12}\right|  ^{2} & \left|  c_{11}\right|  ^{2} & \left|
c_{22}\right|  ^{2} & \left|  c_{21}\right|  ^{2}\\
\left|  c_{21}\right|  ^{2} & \left|  c_{22}\right|  ^{2} & \left|
c_{11}\right|  ^{2} & \left|  c_{12}\right|  ^{2}\\
\left|  c_{22}\right|  ^{2} & \left|  c_{21}\right|  ^{2} & \left|
c_{12}\right|  ^{2} & \left|  c_{11}\right|  ^{2}%
\end{array}
\right]  \left[
\begin{array}
[c]{c}%
kq_{2}(1+q_{1})(1+q_{2})\\
-q_{2}(1+q_{1})(1+q_{2})\\
-q_{2}(1+q_{1})(1+q_{2})\\
0
\end{array}
\right] \nonumber\\
&  \label{consts}%
\end{align}
The classical payoffs of duopoly game given in Eqs. (\ref{ProfFunc}) are
recovered from the Eqs. (\ref{QPayoffs}) when the initial quantum state is
$\left|  \psi_{ini}\right\rangle =\left|  11\right\rangle $. Classical duopoly
is, therefore, a subset of its quantum version.%

\begin{figure}
[ptb]
\begin{center}
\includegraphics[
height=3.2984in,
width=5.5988in
]%
{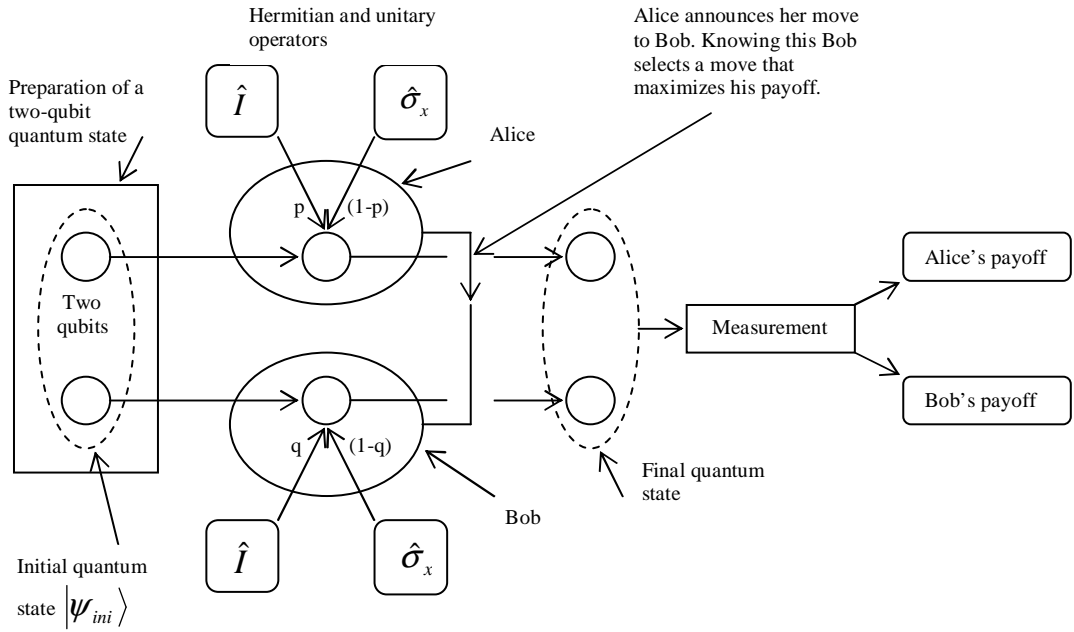}%
\caption{Playing a quantum form of Stackelberg duopoly.}%
\label{Fig4}%
\end{center}
\end{figure}

We now find the backwards-induction outcome in this quantum form of
Stackelberg duopoly. Fig. (\ref{Fig4}) shows the overall idea to play the
game. We proceed in exactly the same way as it is done in the classical game
except that players' payoffs are now given by Eqs. (\ref{QPayoffs}) and
\emph{not} by Eqs. (\ref{ProfFunc}). The first step in the backwards-induction
in quantum game is to find Bob's reaction to an arbitrary quantity $q_{1}$
chosen by Alice. Denoting this quantity as $\left[  R_{2}(q_{1})\right]
_{qtm}$ we find%

\begin{equation}
\left[  R_{2}(q_{1})\right]  _{qtm}=\underset{q_{2}\geq0}{Max}\left[
P_{B}(q_{1},q_{2})\right]  _{qtm}=\frac{q_{1}\triangle_{1}+\triangle_{2}%
}{-2\left\{  q_{1}\triangle_{3}+\triangle_{4}\right\}  } \label{QbestRes}%
\end{equation}
where%

\begin{gather}
\left|  c_{11}\right|  ^{2}+\left|  c_{22}\right|  ^{2}-k\left|
c_{21}\right|  ^{2}=\triangle_{1}\text{, \ \ \ \ \ \ \ \ }\left|
c_{12}\right|  ^{2}+\left|  c_{21}\right|  ^{2}-k\left|  c_{11}\right|
^{2}=\triangle_{2}\nonumber\\
\left|  c_{12}\right|  ^{2}+\left|  c_{21}\right|  ^{2}-k\left|
c_{22}\right|  ^{2}=\triangle_{3}\text{, \ \ \ \ \ \ \ \ }\left|
c_{11}\right|  ^{2}+\left|  c_{22}\right|  ^{2}-k\left|  c_{12}\right|
^{2}=\triangle_{4}%
\end{gather}
This reaction reduces to its classical value of Eq. (\ref{bestRes}) when
$\left|  c_{11}\right|  ^{2}=1$. Similar to the classical game Alice can now
solve Bob's problem as well. Alice can anticipate that a choice of quantity
$q_{1}$ will meet a reaction $\left[  R_{2}(q_{1})\right]  _{qtm}$. In the
first stage of the game, like its classical version, Alice can compute a
solution to her optimization problem as%

\begin{equation}
\underset{q_{1}\geq0}{Max}\left[  P_{A}\left\{  q_{1},\left\{  R_{2}%
(q_{1})\right\}  _{qtm}\right\}  \right]  _{qtm}%
\end{equation}
To find it Alice calculates the quantity:%

\begin{align}
\frac{d\left[  P_{A}(q_{1},q_{2})\right]  _{qtm}}{dq_{1}}  &  =\frac{(\left|
c_{11}\right|  ^{2}+\left|  c_{22}\right|  ^{2}-\left|  c_{12}\right|
^{2}-\left|  c_{21}\right|  ^{2})}{(1+q_{1})}\left\{  -2q_{1}^{2}%
+q_{1}(k-2)+k\right\} \nonumber\\
&  +(1+2q_{1})\left\{  (k-1)\left|  c_{21}\right|  ^{2}-\left|  c_{12}\right|
^{2}\right\}  +k(\left|  c_{12}\right|  ^{2}-\left|  c_{22}\right|
^{2})\nonumber\\
&  -q_{1}\frac{dq_{2}}{dq_{1}}\left\{  \triangle_{4}+q_{1}\triangle
_{3}\right\}  -q_{2}\left\{  2q_{1}\triangle_{3}+\triangle_{4}\right\}
\nonumber\\
&  \label{derivative}%
\end{align}
and replaces $q_{2}$ in Eq. (\ref{derivative}) with $\left[  R_{2}%
(q_{1})\right]  _{qtm}$ given by Eq. (\ref{QbestRes}) and then equates Eq.
(\ref{derivative}) to zero to find a $q_{1}^{\star}$ that maximizes her payoff
$\left[  P_{A}(q_{1},q_{2})\right]  _{qtm}$. For a maxima she would ensure
that the second derivative of $P_{A}\left\{  q_{1},\left\{  R_{2}%
(q_{1})\right\}  _{qtm}\right\}  $ with respect to $q_{1}$ at $q_{1}%
=q_{1}^{\star}$ is a negative quantity. The quantity $q_{1}^{\star}$ together
with $\left[  R_{2}(q_{1}^{\star})\right]  _{qtm}$ will form the
backwards-induction outcome of the quantum game.

An interesting situation is when the backwards-induction outcome in quantum
version of Stackelberg duopoly becomes same as the classical Cournot
equilibrium of duopoly. The classical situation of leader becoming better-off,
while the follower becomes worse-off, is then avoided in the quantum form of
Stackelberg duopoly. To look for this possibility we need such an initial
state $\left|  \psi_{ini}\right\rangle $ that at $q_{1}^{\star}=q_{2}^{\star
}=k/3$ we have the following relations to be true, along with the
normalization condition given in Eq. (\ref{BIOIniStat}):%

\begin{gather}
\frac{d\left[  P_{A}\left\{  q_{1},\left\{  R_{2}(q_{1})\right\}
_{qtm}\right\}  \right]  _{qtm}}{dq_{1}}\mid_{q_{1}=q_{1}^{\star}%
}=0\label{Condn1}\\
\left[  \frac{d^{2}\left[  P_{A}\left\{  q_{1},\left\{  R_{2}(q_{1})\right\}
_{qtm}\right\}  \right]  _{qtm}}{dq_{1}^{2}}\mid_{q_{1}=q_{1}^{\star}}\right]
<0\label{Condn2}\\
q_{2}^{\star}=\left\{  R_{2}(q_{1}^{\star})\right\}  _{qtm} \label{Condn3}%
\end{gather}
The conditions (\ref{Condn1}, \ref{Condn2}) simply say that the
backwards-induction outcome of the quantum game is the same as Cournot
equilibrium in classical game. The condition (\ref{Condn3}) says that Bob's
reaction to Alice's choice of $q_{1}^{\star}=k/3$ is $q_{2}^{\star}=k/3$. To
show that such quantum states can exist for which the conditions
(\ref{Condn1}, \ref{Condn2}, \ref{Condn3}), with the normalization
(\ref{BIOIniStat}), to be true, we give an example where $\left|
c_{11}\right|  ^{2},\left|  c_{12}\right|  ^{2}$ and $\left|  c_{21}\right|
^{2}$ are written as functions of $k$, with our assumption that $\left|
c_{22}\right|  ^{2}=0$. Though this assumption puts its own restriction, on
the possible range of $k$ for which the above conditions hold for these
functions, but still it shows clearly the possibility of finding the required
initial quantum states. The functions are found as%

\begin{align}
\left|  c_{12}(k)\right|  ^{2}  &  =\frac{-f(k)+\sqrt{f(k)^{2}-4g(k)h(k)}%
}{2g(k)}\text{ \ \ where}\nonumber\\
f(k)  &  =j(k)\left\{  \frac{-7}{18}k^{2}+\frac{1}{3}k+\frac{1}{2}\right\}
+\left\{  \frac{k^{2}}{9}+\frac{k}{3}+\frac{1}{2}\right\} \nonumber\\
g(k)  &  =j(k)^{2}\left\{  \frac{-1}{9}k^{3}+\frac{7}{18}k^{2}-\frac{1}%
{2}\right\}  +j(k)\left\{  \frac{2}{9}k^{3}+\frac{5}{18}k^{2}-\frac{1}%
{2}k-1\right\}  +\nonumber\\
&  \left\{  \frac{-1}{9}k^{2}-\frac{1}{2}k-\frac{1}{2}\right\} \nonumber\\
h(k)  &  =\frac{-1}{6}k\text{, \ \ \ \ \ \ }j(k)=\frac{9-4k^{2}}{k^{2}-9}%
\end{align}
also%

\begin{align}
\left|  c_{21}(k)\right|  ^{2}  &  =j(k)\left|  c_{12}(k)\right|  ^{2}\\
\left|  c_{11}(k)\right|  ^{2}  &  =1-\left|  c_{12}(k)\right|  ^{2}-\left|
c_{21}(k)\right|  ^{2}%
\end{align}

Now, interestingly, given that allowed range of $k$ is $1.5\leq k\leq1.73205$,
all of the conditions (\ref{BIOIniStat}, \ref{Condn1}, \ref{Condn2},
\ref{Condn3}) hold at $q_{1}^{\star}=q_{2}^{\star}=k/3$. So that in this range
of $k$ a quantum form of Stackelberg duopoly exists that gives the classical
Cournot equilibrium as the backwards-induction outcome. The restriction on
allowed range of $k$ is the result of our assumption that $\left|
c_{22}(k)\right|  ^{2}=0$, which is introduced to simplify the calculations.
Nevertheless it does not forbid obtaining a quantum form of Stackelberg
duopoly, without the mentioned restriction of the range of $k$, when the
quantum game still possesses the same properties.

\section{Discussion}

What can be a possible relevance of considering a quantum form of a game that
models a competition between two firms in macroscopic world of economics?
Quantum mechanics was developed to understand phenomena in the regime of
atomic and subatomic interactions and it is still mostly used in that domain.
What is of interest in extending a game-theoretical model of interaction
between firms towards quantum domain? These questions naturally arise not only
with reference to Stackelberg duopoly considered in this paper but also other
related works in quantum games.

We believe that like other notions of game theory, finding some relevance in
quantum information, a consideration of backwards-induction can be of interest
for exactly the same reasons. It does not seem hard to imagine situations in
quantum information where moves occur in sequence, all previous moves are
observed before the next move is chosen, and players' payoffs from each
feasible combination of moves are common knowledge. Interesting questions then
arise about how a quantum version of dynamic game of complete information can
influence the outcome.

The duopoly game models economic competition between firms and applied
economics is the area where it is studied in detail. The fact that quantum
game theory can give entirely new views on games, which are important in
economics, is apparent in recent interesting papers by Piotrowski and
Sladkowski \cite{Piotrowski1, Piotrowski2} proposing a quantum-like
description of markets and economies where players' strategies belong to
Hilbert space. It shows that quantum games certainly have features of interest
to applied economists. Reciprocating with it we showed that games played by
firms in economic competition can give counter-intuitive solutions when played
in the quantum world.

\chapter{Quantum repeated games}

\section{Introduction}

The PD attracted early attention \cite{Eisert} in recent studies in quantum
game theory. In classical game theory \cite{Gibbons} a two-stage repeated
version of this game consists of two players playing the game twice, observing
the outcome of the first play before the second play begins. The payoffs for
the entire game are simply taken as the sum of the payoffs from the two
stages. Generally a two-stage repeated game has more complex strategic
structure than its one-stage counterpart and players' strategic choices in the
second stage are affected by the outcome of their moves in the first stage.
For the classical one-stage PD the strategy of `defection' by both the players
is a unique NE. In its two-stage version the same NE appears again at the
second stage because the first stage payoffs are added as constants to the
second stage. In fact in all of finitely repeated versions of the PD
`defection' by both the players appears as unique NE at every stage
\cite{Gibbons}.

Eisert et al.'s study \cite{Eisert} of the one-stage quantum PD raises a
question: what can possibly be a role for quantum mechanics when the game is
played twice? It seems that this role should be relevant to the new feature
showing itself in the game i.e. the two-stages. A role for quantum mechanics
exists if it inter-links the two stages of the game in some way of interest.
Classically both the players `defect' at each stage and strategic choices
remain the same because of uniqueness of the NE at each stage. In our search
for a quantum role we find useful the idea of \textit{subgame-perfect outcome}
(SGPO) \cite{Gibbons} in a two-stage repeated bi-matrix game in its quantum form.

For a two-stage repeated game the idea of a SGPO is natural analog of the
backwards-induction outcome (BIO) \cite{Gibbons} from the games of complete
and perfect information. In the last chapter we considered the BIO idea in a
quantum form of duopoly game and showed how a quantum version of this game can
give an outcome corresponding to the static form of the duopoly, even when the
game is played dynamically. In present chapter we study the natural analogue
of BIO for a two-stage repeated PD quantum game, i.e., the idea of SGPO in a
situation that can be said to lie in quantum domain. We solve the two-stage PD
quantum game in the spirit of backwards-induction from the last section; but
now the first step in working backwards from the end of the game involves
solving a real game rather than solving a single-person optimization problem.

In game theory the idea of SGPO comes out as a stronger solution concept
especially when multiple NE appear in a stage. Our motivation is the
observation that a quantization scheme for the PD is known in which the NE in
a stage is not unique -- thus making relevant a consideration of the concept
of SGPO in the two-stage game played in a quantum setting. For the purpose of
completeness, we will first describe how SGPO works for the classical
two-stage PD. Afterwards, we quantize the game using a known scheme, and then,
show how a SGPO can exist that is counter-intuitive compared to the classical
SGPO for the two-stage repeated PD.

\section{Two-stage games of complete but imperfect information}

Like dynamic game of complete and perfect information -- for example the
Stackelberg duopoly -- the play in a two-stage game of complete but imperfect
information proceeds in a sequence, with the moves in the first stage observed
before the next stage begins. The new feature is that within each stage now
there are simultaneous moves. The simultaneity of moves within each stage
means that information is imperfect in the game. A two-stage game of complete
but imperfect information consists of the steps \cite{Gibbons}:

\begin{enumerate}
\item  Players $A$ and $B$ simultaneously choose actions $p$ and $q$ from
feasible sets $\mathbf{P}$ and $\mathbf{Q}$, respectively.

\item  Players $A$ and $B$ observe outcome of the first stage, $(p,q)$, and
then simultaneously choose actions $p_{1}$ and $q_{1}$ from feasible sets
$\mathbf{P}$ and $\mathbf{Q}$, respectively.

\item  Payoffs are $P_{i}(p,q,p_{1},q_{1})$ for $i=A,$ $B$.
\end{enumerate}

A usual approach to solve a game from this class uses the method of
backwards-induction. In the last section the first step in working backwards
involves solving a single-person optimization problem. Now the first step
involves solving a simultaneous-move game between players $A$ and $B$ in the
second stage, given the outcome from stage one. If the players $A$ and $B$
anticipate that their second-stage behavior will be given by $(p_{1}^{\star
}(p,q),q_{1}^{\star}(p,q))$, then the first-stage interaction between them
amounts to the simultaneous-move game:

\begin{enumerate}
\item  Players $A$ and $B$ simultaneously choose actions $p$ and $q$ from
feasible sets $\mathbf{P}$ and $\mathbf{Q}$, respectively.

\item  Payoffs are $P_{i}(p,q,p_{1}^{\star}(p,q),q_{1}^{\star}(p,q))$ for
$i=A,B$.
\end{enumerate}

When $(p^{\star},q^{\star})$ is the unique NE of this simultaneous-move game,
the set of four numbers $(p^{\star},q^{\star},p_{1}^{\star}(p,q),q_{1}^{\star
}(p,q))$ is known as the SGPO \cite{Gibbons} of this two-stage game. This
outcome is the natural analog of BIO in games of complete and perfect information.

\section{Two-stage Prisoners' Dilemma}

\subsection{Classical form}

We use a normal form of the PD given by the matrix:%

\begin{equation}%
\begin{array}
[c]{c}%
\text{A}%
\end{array}%
\begin{array}
[c]{c}%
C\\
D
\end{array}
\overset{\overset{%
\begin{array}
[c]{c}%
\text{B}%
\end{array}
}{%
\begin{array}
[c]{cc}%
C & D
\end{array}
}}{\left(
\begin{array}
[c]{cc}%
(3,3) & (5,0)\\
(0,5) & (1,1)
\end{array}
\right)  } \label{Matrix1Repeated}%
\end{equation}
The players play this simultaneous-move game twice. The outcome of the first
play is observed before the second stage begins. Payoff for the entire game is
simply the sum of the payoffs from the two stages. It is a two-stage game of
complete but imperfect information \cite{Gibbons}.

Assume players $A$ and $B$ play the pure strategy $C$ with probabilities $p$
and $q$, respectively, in the stage $1$. Also assume the players $A$ and $B$
play the strategy $C$ with probabilities $p_{1}$ and $q_{1}$, respectively, in
the stage $2$. Call $\left[  P_{A1}\right]  _{cl}$ and $\left[  P_{B1}\right]
_{cl}$ the payoffs to players $A$ and $B$, respectively, in the stage $1$,
where the symbol $cl$ is for `classical'. These payoffs can be found from the
matrix (\ref{Matrix1Repeated}) as%

\begin{equation}
\left[  P_{A1}\right]  _{cl}=-pq+4q-p+1\text{, \ \ \ \ \ }\left[
P_{B1}\right]  _{cl}=-pq+4p-q+1 \label{PayoffsC1}%
\end{equation}
The NE conditions for this stage are%

\begin{equation}
\left[  P_{A1}(p^{\star},q^{\star})-P_{A1}(p,q^{\star})\right]  _{cl}%
\geq0\text{, \ \ \ \ \ }\left[  P_{B1}(p^{\star},q^{\star})-P_{B1}(p^{\star
},q)\right]  _{cl}\geq0 \label{NEconds1}%
\end{equation}
giving $p^{\star}=q^{\star}=0$ (i.e. defection for both the players) as the
unique NE in this stage. Likewise, in the second stage the payoffs to players
$A$ and $B$ are written as $\left[  P_{A2}\right]  _{cl}$ and $\left[
P_{B2}\right]  _{cl}$ respectively, where%

\begin{equation}
\left[  P_{A2}\right]  _{cl}=-p_{1}q_{1}+4q_{1}-p_{1}+1\text{, \ \ \ \ \ }%
\left[  P_{B2}\right]  _{cl}=-p_{1}q_{1}+4p_{1}-q_{1}+1 \label{PayoffsC2}%
\end{equation}
and once again the strategy of defection, i.e. $p_{1}^{\star}=q_{1}^{\star}%
=0$, comes out as the unique NE in the second stage. To compute SGPO of this
two-stage game, we analyze its first stage given that the second-stage outcome
is also the NE of that stage ---namely $p_{1}^{\star}=q_{1}^{\star}=0$. For
this NE the players' payoffs in the second stage are%

\begin{equation}
\left[  P_{A2}(0,0)\right]  _{cl}=1,\text{ \ \ \ \ \ }\left[  P_{B2}%
(0,0)\right]  _{cl}=1 \label{PDefC}%
\end{equation}
The players' first-stage interaction, therefore, in the two-stage PD amounts
to a one-shot game, in which the payoff pair $(1,1)$ from the second stage is
added to their first-stage payoff pair. Write the players' payoffs in the
one-shot game as%

\begin{align}
\left[  P_{A(1+2)}\right]  _{cl}  &  =\left[  P_{A1}+P_{A2}(0,0)\right]
_{cl}=-pq+4q-p+2\nonumber\\
\left[  P_{B(1+2)}\right]  _{cl}  &  =\left[  P_{B1}+P_{B2}(0,0)\right]
_{cl}=-pq+4p-q+2 \label{PayoffsTotal}%
\end{align}
It has again $(0,0)$ as the unique NE. Therefore, the unique SGPO of the
two-stage PD is $(0,0)$ in the first stage, followed by $(0,0)$ in the second
stage. The strategy of defection in both the stages comes out as SGPO for the
two stage classical PD.

It is now shown how it becomes possible, in a quantum form of this two-stage
PD, to achieve a SGPO in which the players decide to cooperate in the first
stage while knowing that they both will defect in the second. The quantum form
of the two-stage PD is played using a system of four qubits. Players
manipulate these qubits in Marinatto and Weber's scheme to play a quantum form
of a matrix game.

\subsection{Quantum form}

\begin{figure}
[ptb]
\begin{center}
\includegraphics[
height=6.0027in,
width=4.5792in
]%
{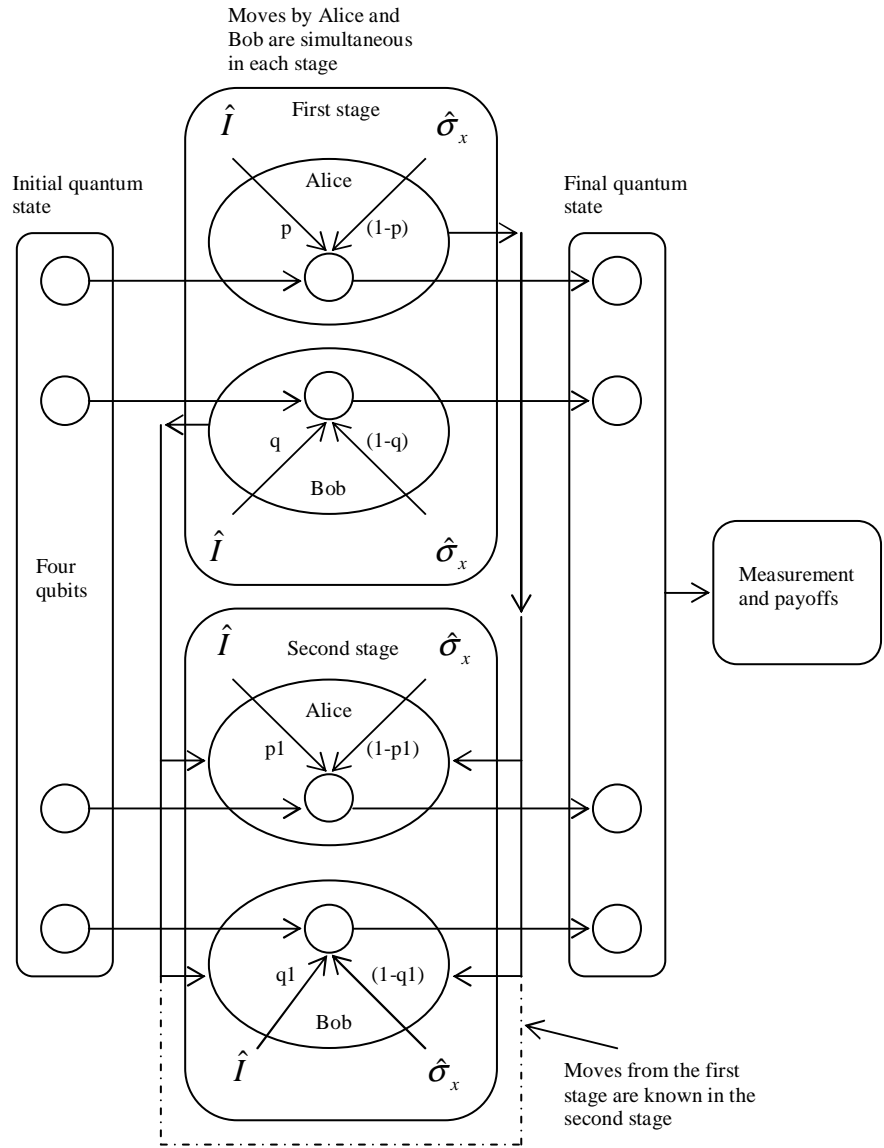}%
\caption{Playing a two-stage quantum game of Prisoners' Dilemma.}%
\label{Fig6}%
\end{center}
\end{figure}

Marinatto and Weber's scheme can be extended to play a two-stage version of a
bi-matrix game. For example, a quantum version of the two-stage PD starts by
making available a four-qubit pure quantum state to the players. This state
can be written as%

\begin{equation}
\left|  \psi_{in}\right\rangle =\underset{i,j,k,l=1,2}{\sum}c_{ijkl}\left|
ijkl\right\rangle \text{ \ \ where \ \ }\underset{i,j,k,l=1,2}{\sum}\left|
c_{ijkl}\right|  ^{2}=1 \label{IniStatRepeated}%
\end{equation}
where $i,j,k$ and $l$ are identifying symbols for four qubits. The upper and
lower states of a qubit are $1$ and $2$ respectively; and $c_{ijkl}$ are
complex numbers. It is a quantum state in $2\otimes2\otimes2\otimes
2$-dimensional Hilbert space. We suppose the qubits $i$ and $j$ are
manipulated by the players in the first stage of the game and, similarly, the
qubits $k$ and $l$ are manipulated in the second stage. Let $\rho_{in}$ denote
the initial state (\ref{IniStatRepeated}) in the density matrix formalism.
Assume during their moves in the first stage of the game, the players $A$ and
$B$ apply the identity operator $\hat{I}$ on the initial state with
probabilities $p$ and $q$, respectively. Also they apply the operator
$\hat{\sigma}_{x}$ with probabilities $(1-p)$ and $(1-q)$, respectively. The
players' actions in the first stage changes $\rho_{in}$ to%

\begin{align}
\rho_{fin}  &  =pq\hat{I}_{A}\otimes\hat{I}_{B}\rho_{in}\hat{I}_{A}^{\dagger
}\otimes\hat{I}_{B}^{\dagger}+p(1-q)\hat{I}_{A}\otimes\hat{\sigma}_{xB}%
\rho_{in}\hat{I}_{A}^{\dagger}\otimes\hat{\sigma}_{xB}^{\dagger}+\nonumber\\
&  q(1-p)\hat{\sigma}_{xA}\otimes\hat{I}_{B}\rho_{in}\hat{\sigma}%
_{xA}^{\dagger}\otimes\hat{I}_{B}^{\dagger}+(1-p)(1-q)\hat{\sigma}_{xA}%
\otimes\hat{\sigma}_{xB}\rho_{in}\hat{\sigma}_{xA}^{\dagger}\otimes\hat
{\sigma}_{xB}^{\dagger}\nonumber\\
&
\end{align}
The players' actions in this stage are simultaneous and they remember their
moves (i.e. the numbers $p$ and $q$) also in the next stage. In the second
stage, players $A$ and $B$ apply the identity operator with probabilities
$p_{1}$ and $q_{1}$, respectively, on $\rho_{fin}$. The operator $\hat{\sigma
}_{x}$ is, then, applied with probabilities $(1-p_{1})$ and $(1-q_{1})$ on
$\rho_{fin}$, respectively. Fig. (\ref{Fig6}) shows the overall idea of
playing the two-stage game. Players' moves in the two stages of the game are
done on two different pairs of qubits.%

After the moves performed in the second stage the quantum state changes to%

\begin{align}
\rho_{ffin}  &  =p_{1}q_{1}\hat{I}_{A}\otimes\hat{I}_{B}\rho_{fin}\hat{I}%
_{A}^{\dagger}\otimes\hat{I}_{B}^{\dagger}+p_{1}(1-q_{1})\hat{I}_{A}%
\otimes\hat{\sigma}_{xB}\rho_{fin}\hat{I}_{A}^{\dagger}\otimes\hat{\sigma
}_{xB}^{\dagger}+\nonumber\\
&  q_{1}(1-p_{1})\hat{\sigma}_{xA}\otimes\hat{I}_{B}\rho_{fin}\hat{\sigma
}_{xA}^{\dagger}\otimes\hat{I}_{B}^{\dagger}+\nonumber\\
&  (1-p_{1})(1-q_{1})\hat{\sigma}_{xA}\otimes\hat{\sigma}_{xB}\rho_{fin}%
\hat{\sigma}_{xA}^{\dagger}\otimes\hat{\sigma}_{xB}^{\dagger}%
\end{align}
which is ready for measurement, giving payoffs for the two stages of the game.
If classically the bi-matrix game (\ref{Matrix1Repeated}) is played at each
stage, the possession of the following four payoff operators by the referee
corresponds to a `quantum version' of the two-stage game:%

\begin{align}
\left[  \left(  P_{A}\right)  _{oper}\right]  _{1}  &  =\underset
{k,l=1,2}{\sum}\left\{  3\left|  11kl\right\rangle \left\langle 11kl\right|
+5\left|  21kl\right\rangle \left\langle 21kl\right|  +\left|
22kl\right\rangle \left\langle 22kl\right|  \right\} \nonumber\\
\left[  \left(  P_{A}\right)  _{oper}\right]  _{2}  &  =\underset
{i,j=1,2}{\sum}\left\{  3\left|  ij11\right\rangle \left\langle ij11\right|
+5\left|  ij21\right\rangle \left\langle ij21\right|  +\left|
ij22\right\rangle \left\langle ij22\right|  \right\} \nonumber\\
\left[  \left(  P_{B}\right)  _{oper}\right]  _{1}  &  =\underset
{k,l=1,2}{\sum}\left\{  3\left|  11kl\right\rangle \left\langle 11kl\right|
+5\left|  12kl\right\rangle \left\langle 12kl\right|  +\left|
22kl\right\rangle \left\langle 22kl\right|  \right\} \nonumber\\
\left[  \left(  P_{B}\right)  _{oper}\right]  _{2}  &  =\underset
{i,j=1,2}{\sum}\left\{  3\left|  ij11\right\rangle \left\langle ij11\right|
+5\left|  ij12\right\rangle \left\langle ij12\right|  +\left|
ij22\right\rangle \left\langle ij22\right|  \right\} \nonumber\\
&
\end{align}

The corresponding payoffs are, then, obtained as mean values of these
operators. For example, Alice's payoff in stage $1$ is%

\begin{equation}
\left[  P_{A1}\right]  _{qu}=\text{Tr}\left\{  \left[  \left(  P_{A}\right)
_{oper}\right]  _{1}\rho_{ffin}\right\}
\end{equation}
We consider a two-stage quantum PD played with pure four-qubit initial state
in the form:%

\begin{equation}
\left|  \psi_{ini}\right\rangle =c_{1}\left|  1111\right\rangle +c_{2}\left|
1122\right\rangle +c_{3}\left|  2211\right\rangle +c_{4}\left|
2222\right\rangle
\end{equation}
with $\underset{t=1}{\overset{4}{\sum}}\left|  c_{t}\right|  ^{2}=1$. For this
state the payoffs to the players $A$ and $B$ in the two stages are found as%

\begin{align}
\left[  P_{A1}\right]  _{qu}  &  =(\left|  c_{1}\right|  ^{2}+\left|
c_{2}\right|  ^{2})(-pq-p+4q+1)+\nonumber\\
&  (\left|  c_{3}\right|  ^{2}+\left|  c_{4}\right|  ^{2}%
)(-pq+2p-3q+3)\nonumber\\
\left[  P_{A2}\right]  _{qu}  &  =(\left|  c_{1}\right|  ^{2}+\left|
c_{3}\right|  ^{2})(-p_{1}q_{1}-p_{1}+4q_{1}+1)+\nonumber\\
&  (\left|  c_{2}\right|  ^{2}+\left|  c_{4}\right|  ^{2})(-p_{1}q_{1}%
+2p_{1}-3q_{1}+3)\nonumber\\
\left[  P_{B1}\right]  _{qu}  &  =(\left|  c_{1}\right|  ^{2}+\left|
c_{2}\right|  ^{2})(-pq-q+4p+1)+\nonumber\\
&  (\left|  c_{3}\right|  ^{2}+\left|  c_{4}\right|  ^{2}%
)(-pq+2q-3p+3)\nonumber\\
\left[  P_{B2}\right]  _{qu}  &  =(\left|  c_{1}\right|  ^{2}+\left|
c_{3}\right|  ^{2})(-p_{1}q_{1}-q_{1}+4p_{1}+1)+\nonumber\\
&  (\left|  c_{2}\right|  ^{2}+\left|  c_{4}\right|  ^{2})(-p_{1}q_{1}%
+2q_{1}-3p_{1}+3) \label{Payoffs12q}%
\end{align}
The players' payoffs in the classical two-stage PD given by Eqs.
(\ref{PayoffsC1}, \ref{PayoffsC2}) can now be recovered from the Eq.
(\ref{Payoffs12q}) by taking $\left|  c_{1}\right|  ^{2}=1$. The classical
game is, therefore, a subset of its quantum version.

One now proceeds, in the spirit of backwards-induction, to find a NE in the
second stage of the quantum game. Suppose $(p_{1}^{\star},q_{1}^{\star})$ is a
NE in the second stage, then%

\begin{equation}
\left[  P_{A2}(p_{1}^{\star},q_{1}^{\star})-P_{A2}(p_{1},q_{1}^{\star
})\right]  _{qu}\geq0,\text{ \ \ \ \ \ \ }\left[  P_{B2}(p_{1}^{\star}%
,q_{1}^{\star})-P_{B2}(p_{1}^{\star},q_{1})\right]  _{qu}\geq0
\label{NashIneq}%
\end{equation}
With the players' payoffs of the two stages given by Eq. (\ref{Payoffs12q}),
the Nash inequalities (\ref{NashIneq}) can be written as%

\begin{align}
(p_{1}^{\star}-p_{1})\left\{  -q_{1}^{\star}+2(\left|  c_{2}\right|
^{2}+\left|  c_{4}\right|  ^{2})-(\left|  c_{1}\right|  ^{2}+\left|
c_{3}\right|  ^{2})\right\}   &  \geq0\nonumber\\
(q_{1}^{\star}-q_{1})\left\{  -p_{1}^{\star}+2(\left|  c_{2}\right|
^{2}+\left|  c_{4}\right|  ^{2})-(\left|  c_{1}\right|  ^{2}+\left|
c_{3}\right|  ^{2})\right\}   &  \geq0
\end{align}
and the strategy of defection by both the players, i.e. $p_{1}^{\star}%
=q_{1}^{\star}=0,$ becomes a NE in the second stage of the quantum game, if%

\begin{equation}
\left\{  2(\left|  c_{2}\right|  ^{2}+\left|  c_{4}\right|  ^{2})-(\left|
c_{1}\right|  ^{2}+\left|  c_{3}\right|  ^{2})\right\}  \leq0 \label{Cond1}%
\end{equation}
Similar to the classical analysis, players' payoffs can be found when both
decide to defect in the second stage:%

\begin{equation}
\left[  P_{A2}(0,0)\right]  _{qu}=\left[  P_{B2}(0,0)\right]  _{qu}=3(\left|
c_{2}\right|  ^{2}+\left|  c_{4}\right|  ^{2})+(\left|  c_{1}\right|
^{2}+\left|  c_{3}\right|  ^{2}) \label{PDefQ}%
\end{equation}
The classical payoffs, when both players defect, of the Eq. (\ref{PDefC}) can
be recovered from Eq. (\ref{PDefQ}) when $\left|  c_{1}\right|  ^{2}=1$, i.e.
for an unentangled initial state.

Similar to the classical case the players' first-stage interaction, in the
two-stage quantum PD, amounts to a one-shot game. In this one-shot game the
payoff $3(\left|  c_{2}\right|  ^{2}+\left|  c_{4}\right|  ^{2})+(\left|
c_{1}\right|  ^{2}+\left|  c_{3}\right|  ^{2})$, from the second stage, is
added to their first-stage payoffs:%

\begin{align}
\left[  P_{A(1+2)}\right]  _{qu}  &  =\left[  P_{A1}+P_{A2}(0,0)\right]
_{qu}=\left|  c_{1}\right|  ^{2}(-pq+4q-p+2)+\nonumber\\
&  \left|  c_{2}\right|  ^{2}(-pq+4q-p+4)+\left|  c_{3}\right|  ^{2}%
(-pq-3q+2p+4)+\nonumber\\
&  \left|  c_{4}\right|  ^{2}(-pq-3q+2p+6)\nonumber\\
\left[  P_{B(1+2)}\right]  _{qu}  &  =\left[  P_{B1}+P_{B2}(0,0)\right]
_{qu}=\left|  c_{1}\right|  ^{2}(-pq+4p-q+2)+\nonumber\\
&  \left|  c_{2}\right|  ^{2}(-pq+4p-q+4)+\left|  c_{3}\right|  ^{2}%
(-pq-3p+2q+4)+\nonumber\\
&  \left|  c_{4}\right|  ^{2}(-pq-3p+2q+6)
\end{align}
The strategy of cooperation (that is $p_{1}^{\star}=q_{1}^{\star}=1$) can now
be a NE for the first-stage interaction in this two-stage quantum game, if%

\begin{equation}
\left\{  2(\left|  c_{1}\right|  ^{2}+\left|  c_{2}\right|  ^{2})-(\left|
c_{3}\right|  ^{2}+\left|  c_{4}\right|  ^{2})\right\}  \leq0 \label{Cond2}%
\end{equation}
The inequalities (\ref{Cond1}) and (\ref{Cond2}) are the conditions on the
initial state when the players cooperate in their first-stage interaction
while both defect in the next stage. These conditions can be rewritten as%

\begin{equation}
\left|  c_{1}\right|  ^{2}+\left|  c_{2}\right|  ^{2}\leq\frac{1}{3}\text{,
\ \ \ \ \ \ \ }\left|  c_{2}\right|  ^{2}+\left|  c_{4}\right|  ^{2}\leq
\frac{1}{3} \label{Conds}%
\end{equation}
For example, at $\left|  c_{1}\right|  ^{2}=\left|  c_{2}\right|  ^{2}=\left|
c_{4}\right|  ^{2}=\frac{1}{6}$ and $\left|  c_{3}\right|  ^{2}=\frac{1}{2}$
these conditions hold. Because for the classical game the inequalities
(\ref{Conds}) together cannot hold, showing why classically it is not possible
that players cooperate in the first stage knowing that they both will defect
in the second.

\section{Discussion}

Essentially, the repeated games differ from one-shot games in that players'
current actions can depend on their past behavior. In a repeated bi-matrix
game the same matrix game is played repeatedly, over a number of stages that
represent the passing of time. The payoffs are accumulated over time.
Accumulation of information about the ``history'' of the game changes the
structure of the game with time. With each new stage the information at the
disposal of the players changes and, since strategies transform this
information into actions, the players' strategic choices are affected. If a
game is repeated twice, the players' moves at the second stage depend on the
outcome of the first stage. This situation becomes more and more complex as
the number of stages increases, since the players can base their decisions on
histories represented by sequences of actions and outcomes observed over
increasing number of stages.

Recent findings in quantum game theory motivate a study of repeated games in
the new quantum setting. It is because useful and extensive analysis of
repeated games is already found in literature of classical game theory. In
present chapter -- to look for a quantum role in repeated games -- we consider
a quantum form of a well-known bi-matrix game of PD.

Classical analysis of the PD has been developed in many different formats,
including its finitely and infinitely repeated versions. In the history of
quantum games the PD became a focus of early and important study \cite{Eisert}
telling how to play a quantum form of a bi-matrix game. To play a quantum form
of repeated PD we select Marinatto and Weber's scheme. In this scheme a
repeated game is played when the players select positive numbers in the range
$\left[  0,1\right]  $, giving the probabilities with which they apply the
identity operator $\hat{I}$ on a four-qubit pure initial quantum state. The
players' actions in each stage are performed on two different pairs of qubits.
The classical two-stage PD corresponds to an unentangled initial state, and
the classical SGPO consists of players defecting in both the stages. It is
shown that a SGPO where the players go for cooperation in a stage is a
non-classical feature that can be made to appear in quantum setting.

The argument presented here is based on the assumption that other games,
resulting from a play starting with a four-qubit quantum state of the form of
the Eq. (\ref{IniStatRepeated}), are `quantum forms' of the classical
two-stage game. This assumption originates from the fact that the classical
game corresponds to a particular four-qubit pure quantum state which is
unentangled. The assumption makes possible to translate the desired appearance
of cooperation in a stage to certain conditions on the parameters of the
initial state, thus giving a SGPO\ where players decide to cooperate in their
first-stage interaction while they both defect in the next stage.

One may ask about the compelling reason to choose a $2\otimes2\otimes
2\otimes2$ dimensional Hilbert space instead of a $2\otimes2$ dimensional one.
A $2\otimes2$ dimensional treatment of this problem, in the same quantization
scheme, involves denominator terms in the expressions for payoff operators,
when these are obtained under the condition that classical game corresponds to
an unentangled initial state. It then leads to many `if-then' conditions
before one gets finally the payoffs. On the contrary, a treatment in
$2\otimes2\otimes2\otimes2$ dimensions is much smoother. Also a study of the
concept of SGPO in a two-stage repeated quantum game, then, becomes a logical
extension of the backwards-induction procedure proposed in the last chapter.

\chapter{New proposals to play quantum games}

\section{\label{IntNewProp}Introduction}

Meyer \cite{MeyerDavid} and Du et al. \cite{Du1} have shown that entanglement
may not be essential for a quantum game. Eisert et al.'s quantum PD was the
first proposal where entanglement was used as a resource. There has been
noticeably greater attention paid in exploiting entanglement for quantum
games. It is not unusual and can be traced back to exciting and
counter-intuitive properties of this phenomenon, as well as to its recent
enthusiastic investigation in quantum information theory \cite{Nielsen}.

Local unitary manipulations of entangled qubits to play matrix games is indeed
an interesting concept that gives new dimensions to classical game theory. But
it does not forbid the use of other quantum mechanical effects to play other
`quantum forms' of matrix games -- games for which extensive analysis in the
classical domain already exists in literature \cite{Burger,Gibbons}. A look at
the Eisert et al.'s set-up \cite{Eisert,Eisert1} makes apparent some of its
similarities to well-known Young's double-slit apparatus \cite{Hecht}.
Simultaneous and local unitary manipulation of a maximally entangled two-qubit
quantum state, and its later measurement, is the essential feature of Eisert
et al.'s set-up. In Young's double-slit set-up, however, coherent light passes
through two slits to form a diffraction pattern on a screen facing the slits.
Similarity between these setups becomes noticeable if a comparison is made between:

\begin{itemize}
\item  The properties of entanglement and coherence.

\item  Players' moves in manipulations of qubits and the process of opening or
closing the slits.

\item  Wavefunction-collapsing measurement and the appearance of the
diffraction pattern.
\end{itemize}

Such a comparison, in its turn, asks for a quantum feature that can be
exploited to give new dimension to a matrix game, when it is played using a
Young's double-slit like apparatus. In Eisert et al.'s set-up this quantum
feature is obviously the quantum phenomenon of entanglement. In Young's
apparatus this feature is the association of wave-like properties to material
objects like electrons, producing a diffraction pattern on a screen. Section
(\ref{diffraction}) exploits such association of waves as a resource to play
quantum versions of classical games.

Playing of a game requires resources for its physical implementation. For
example, to play a bi-matrix game the resources may consist of pairs of
two-valued `objects', like coins, distributed between the players. The players
perform their moves on the objects and later a referee decides payoffs after
observing the objects. Game theory usually links players' actions
\emph{directly }to their payoffs, without a reference to the nature of the
objects on which the players have made their moves. However, playing quantum
games \cite{Eisert} show that radically different `solutions' can emerge when
the \emph{same} game is physically implemented on distributed objects which
are quantum mechanically correlated.

In Section (\ref{EnkPikeComment}) Enk and Pike's argument \cite{EnkPike} is
briefly discussed. Essentially it says that the emergence of new equilibrium
in quantum Prisoners' Dilemma can also be understood as an equilibrium in a
modified form of the game. They constructed \emph{another} matrix game, in
which players have access to three pure classical strategies instead of the
usual two, commenting that it `captures' everything quantum Prisoners' Dilemma
has to offer. Constructing an extended matrix with an extra pure classical
move, in their view, is justified because \emph{also} in quantum Prisoners'
Dilemma players can play moves which are superpositions of the two classical moves.

Truly quantum pairs of objects possess non-local correlations. Though it is
impossible to have a local model of a quantum game set-up, producing
\emph{exactly} the same data, but how such unusual correlations may explicitly
affect solutions of a game when implemented with quantum objects. To how far
extent solutions of a quantum game themselves can be called `truly quantum' in
nature. Section ($10.3$) tries to address these questions.

\section{Quantum games with a diffraction set-up\label{diffraction}}

Historically speaking, the De Broglie's original idea \cite{Hecht,Debroglie}
-- that travelling material particles have waves associated with them -- was
undoubtedly the key concept behind the development of quantum physics in early
part of the twentieth century. Soon afterwards, Davisson and Germer
\cite{Hecht} verified the idea in their experimental demonstration of the
diffraction of electrons by crystals. De Broglie's proposed that a travelling
electron with momentum $p$ has an associated wave of wavelength $\lambda=h/p$,
where $h$ is the Plank's constant. To make $\lambda$ a measurable quantity,
under normal laboratory conditions, the momentum $p$ should have similar order
of magnitude as $h$. $h$ being a very small quantity shows why it is very hard
to detect waves associated with macroscopic objects. Our motivation is to take
this quantum feature -- associating wave-like properties to micro objects --
as a resource that can be used to play a quantum game. Such a quantum game can
be realized using an apparatus consisting of travelling electrons, multiple
slits intercepting them, and a resulting diffraction pattern. In this set-up a
player's choice of a `pure strategy' consists of opening or closing slits at
his/her disposal. Suppose the apparatus is adjusted such that when $\lambda$
approaches zero the classical game is reproduced. It can then be argued that
because an observation of a value of $\lambda$ quite away from zero is
entirely a quantum feature, therefore, the resulting different payoffs for the
players correspond to a quantum form of the classical game. In this setup the
players' payoffs are to be found from the diffraction pattern formed on the
screen. We show the possibility of finding a value for $\lambda$ that makes
appear a non-classical equilibrium in the PD when the players play only the
pure strategies. The classical game remains a subset of its quantum version
because with $\lambda$ approaching zero the classical game is reproduced.

The motivation to play a quantum form of PD, without using the phenomenon of
entanglement, also derives from Feynman's excellent exposition \cite{Feynman}
of quantum behavior of atomic objects. He describes and compares the
diffraction patterns in two similar set-ups that are imaginary but
experimentally realizable. The two set-ups consist of bullets and electrons
passing through a wall with two slits. Feynman then describes the well-known
quantum property -- associating waves to all material particles -- to
distinguish the diffraction patterns of bullets and electrons. The
disappearance of a pattern for the bullets, he explains, is due to tiny
wavelengths of the associated waves. For such waves the pattern becomes very
fine and, with a detector of finite size, one cannot distinguish the separate
maxima and minima. We ask why not to play a game, in the Feynman's imaginary
experimental set-up, such that the classical game corresponds when, in
Feynman's words, bullets are fired; and a quantum game corresponds when
electrons replace the bullets.

\subsection{Playing Prisoners' Dilemma}

\begin{figure}
[ptb]
\begin{center}
\includegraphics[
height=4.3007in,
width=4.5844in
]%
{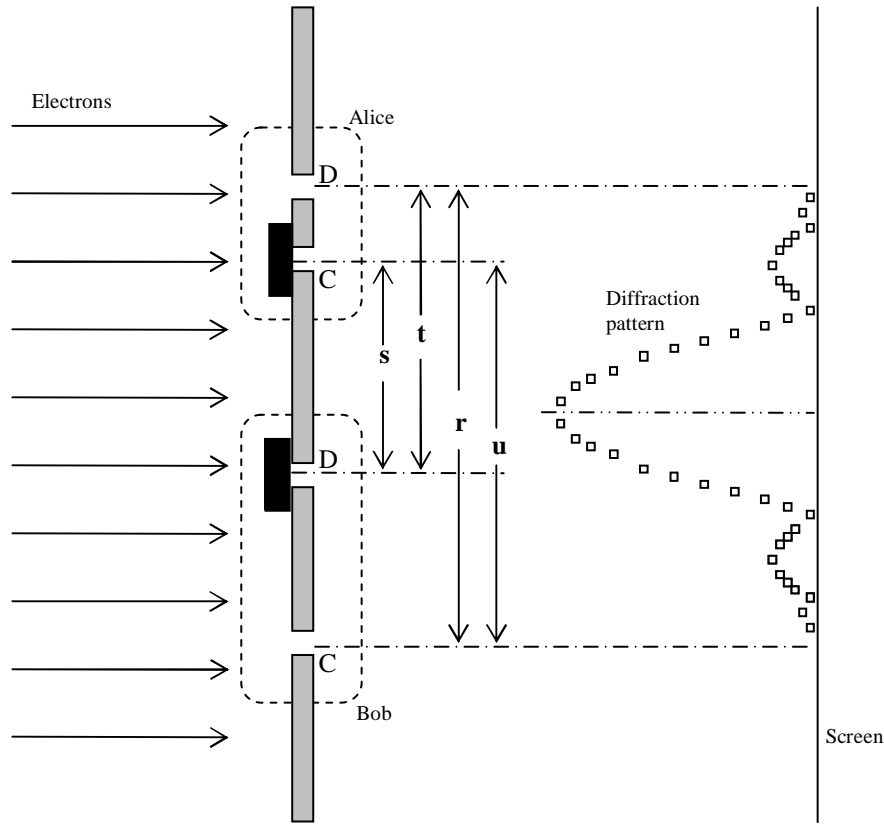}%
\caption{A multi-slit diffraction set-up to play a quantum form of Prisoners'
Dilemma. A window with four slits faces an electron source. Each player has
access to two slits. A player plays a pure strategy by opening a slit and
closing the other. Referee finds the players' payoffs by measuring the
peak-to-peak distance on the diffraction pattern formed on the screen.}%
\label{Fig7}%
\end{center}
\end{figure}

We select the PD to be played in a diffraction set-up. The classical PD, in
its general form, is represented by the following matrix:%

\begin{equation}%
\begin{array}
[c]{c}%
\text{Alice}%
\end{array}%
\begin{array}
[c]{c}%
C\\
D
\end{array}
\overset{%
\begin{array}
[c]{c}%
\text{Bob}%
\end{array}
}{\overset{%
\begin{array}
[c]{cc}%
C & D
\end{array}
}{\left(
\begin{array}
[c]{cc}%
(r,r) & (s,t)\\
(t,s) & (u,u)
\end{array}
\right)  }} \label{DiffractionPD}%
\end{equation}
where $t>r>u>s$. To make the classical game imbedded in its quantum version
the positive coefficients $u,r,s$ and $t$, appearing in the matrix
(\ref{DiffractionPD}), are translated into distances between the slits. Each
player is in control of two slits such that his/her strategy consists of
opening one of the slits and closing the other. For example, if Alice decides
to cooperate then she opens the slit $C$ and closes the slit $D$. Because Bob
has a similar choice, therefore, all possible moves by the players leads to
the opening of two slits and closure of the other two, with the separation
between the two open slits depending on the moves of the players. It happens
when only the so-called `pure-strategies' can be played by the players, which
in the present setup means to open a slit and close the other. Now, at the
final stage of the game, the action of the arbiter -- responsible for finding
the payoffs when the players have made their moves -- consists of measuring
the distance between two peaks of the diffraction pattern. This peak-to-peak
distance is known \cite{Hecht} to be $\lambda/d$, where $d$ is the separation
between the open sits and $\lambda$ is the wavelength associated with the
bombarded material objects, like electrons. Payoffs to the players are
functions of $\lambda/d$ and it, then, explains why it is useful to translate
the coefficients of the matrix of the classical game into the separations $d$
between the slits. When bullets are fired, which means the particles become
heavier and corresponding $\lambda$ is very nearly zero, the payoffs become
classical and depend only on $d$ i.e. the separation between the slits. A
payoff representation in terms of $\lambda/d$ contains both the classical and
quantum aspects of the matrix game played in this set-up. The experimental
set-up shown in the Fig. (\ref{Fig7}) sketches the diffraction set-up to play
a quantum game.%

For PD the payoffs are symmetric for the players and a single equation can
describe the payoffs to both the players when their strategies are known. A
usual way to express it is to write $P(s_{1},s_{2})$ for the payoff to the
$s_{1}$-player against the $s_{2}$-player. Such a single equation
representation is usually used in evolutionary games \cite{Weibull} consisting
of symmetric bi-matrix conflicts. The $s_{1}$-player is referred to as the
`focal' player and the $s_{2}$-player as just the `other' player. The PD is
one such example for which a single payoff equation can capture the essence of
the idea of a symmetric NE. A strategy $s^{\star}$ is a symmetric NE if%

\begin{equation}
P(s^{\star},s^{\star})-P(s,s^{\star})\geq0,\text{ \ \ \ for all }s
\label{NEDiffraction}%
\end{equation}
saying that the focal player cannot be better off by diverging away from
$s^{\star}$. Because the set-up of the Fig. (\ref{Fig7}) involves coefficients
in the classical payoff matrix corresponding to the first player, therefore,
finding a symmetric NE with Eq. (\ref{NEDiffraction}) becomes immediately
possible when the first player is taken as focal. It also shows why writing
payoff as $P(s_{1},s_{2})$ is relevant to the set-up of the Fig. (\ref{Fig7}).
For example, classically the strategy of defection $D$ is a symmetric NE
because $P(D,D)-P(C,D)=(u-s)>0$, where the players' moves consist of only the
pure strategies.

In the set-up of Fig. (\ref{Fig7}) for every pure strategy move the players
have option to play, a unique separation $d$ between the slits is obtained
that can have four possible values i.e. $u,r,s$ or $t$. Classically
$P(C,C)=r$, $P(C,D)=s$, $P(D,C)=t$ and $P(D,D)=u$. It is observed in the Fig.
(\ref{Fig7}) that the classical payoff to the focal player, against the other,
can be equated to the separation between the two open slits $d$.

Now assume that instead of simply $P(s_{1},s_{2})=d$ the arbiter uses the
payoff equation:%

\begin{equation}
P(s_{1},s_{2})=d+k(\lambda/d)
\end{equation}
where $k$ is a positive constant that can be called a \textit{scaling factor}.
$P(s_{1},s_{2})$, obviously, reduces to its classical counterpart when
$\lambda$ is very nearly zero. Suppose the strategy of cooperation $C$ is a
symmetric NE:%

\begin{equation}
P(C,C)-P(D,C)=\left\{  k\lambda(1/r-1/t)-(t-r)\right\}  \geq0
\end{equation}
It requires $\lambda\geq rt/k$. For electrons of mass $m$ travelling with
velocity $v$ it gives $v\leq(kh/mrt).$ Supposing $r$ and $t$ are both
non-zero, the arbiter's problem consists of finding an appropriate value for
the scaling factor $k$ that brings $v$ into a reasonable range from
experimental point of view. When the electrons have associated wavelength
$\lambda\geq rt/k$ the strategy of cooperation becomes a symmetric NE, and
each player gets a payoff $r+k\lambda/r$. Similarly when the pure strategy of
defection is a symmetric NE in the quantum game, we have%

\begin{equation}
P(D,D)-P(C,D)=\left\{  -k\lambda(1/s-1/u)+(u-s)\right\}  \geq0
\end{equation}
It requires $\lambda\leq su/k$. After the scaling factor $k$ is decided the
wavelength $\lambda$ determines which pure strategy should be a symmetric NE.
Two ranges for $\lambda$ can be indicated as $\lambda\leq su/k$ and
$\lambda\geq rt/k$. Defection and cooperation are symmetric NE for these
ranges, respectively. Because the constants $t,r,u,s$ and $k$ are all positive
the classical game is in the first range of $\lambda$. Non-classical
equilibrium of cooperation shows itself in the second range of $\lambda$.

Du et al.'s recent analysis \cite{Du} of the quantum PD, with players' access
to Eisert's two-parameter set of unitary operators, has shown an intriguing
structure in the game as a function of the amount of entanglement. The game
becomes classical when the entanglement vanishes.

In the set-up of Fig. (\ref{Fig7}) the quantity $\lambda$ behaves in a way
similar to the amount of entanglement in Du et al.'s analysis \cite{Du,Du2}.
But this set-up is devoid of the notion of entanglement and relies instead on
a different quantum aspect. An aspect which is as much `quantum' in nature as
the phenomenon of entanglement for qubit systems.

There is however a difference, to be noticed, between the set-ups of Eisert et
al. and that of the Fig. (\ref{Fig7}). Players' actions in Eisert et al.'s
set-up are quantum mechanical in nature in that they make moves with quantum
operators. In the present set-up, on the contrary, the players' actions are
entirely classical consisting of opening or closing slits. In a sense it is
similar to the players' actions in Marinatto and Weber's scheme. In their
scheme players possess quantum operators but they apply them on an initial
quantum state with classical probabilities; so that the players' moves can be
considered classical as well. It can be said that, apart from the work of
Eisert et al., the set up of Fig. (\ref{Fig7}) is also motivated, to an almost
equal extent, by the Marinatto and Weber's idea of playing a quantum version
of a matrix game.

\section{\label{HullSect}Performing EPR type experiments to play a bi-matrix
game\label{EPR type expts}}

To address the question raised in the Section (\ref{IntNewProp}), i.e. to how
much extent a quantum game can be called `truly quantum', following two
\emph{constraints} are suggested \cite{iqbalweigert} which a quantization
scheme should follow:

(C1). In both classical and quantum version of the game the \emph{same} set of
moves should be made available to the players.

(C2). The players agree together on explicit expressions for their payoffs
which \emph{must} not be modified when introducing the quantized version of
the game.

With these constraints one can hope that only the nature of correlations,
existing between the objects the players receive, will decide whether the
resulting game is classical or quantum.

Consider a symmetric bi-matrix game between two players Alice and Bob with the
matrix representation:%

\begin{equation}%
\begin{array}
[c]{c}%
\text{Alice}%
\end{array}
\overset{%
\begin{array}
[c]{c}%
\text{Bob}%
\end{array}
}{%
\begin{array}
[c]{c}%
S_{1}\\
S_{2}%
\end{array}
\overset{%
\begin{array}
[c]{cc}%
S_{1} & S_{2}%
\end{array}
}{%
\begin{array}
[c]{cc}%
(r,r) & (s,t)\\
(t,s) & (u,u)
\end{array}
}} \label{CorrelationMatrix}%
\end{equation}
For which the mixed strategy payoffs for the players can be written as%

\begin{align}
P_{A}(p,q)  &  =Kpq+Lp+Mq+N\nonumber\\
P_{B}(p,q)  &  =Kpq+Mp+Lq+N \label{CorrelClassPayoffs}%
\end{align}
where the constants $K,L,M$ and $N$ can be found in terms of $r,s,t$ and $u$,
the coefficients of the bi-matrix (\ref{CorrelationMatrix}). The NE defining
conditions are%

\begin{align}
P_{A}(p^{\ast},q^{\ast})-P_{A}(p,q^{\ast})  &  \geq0\nonumber\\
P_{B}(p^{\ast},q^{\ast})-P_{B}(p^{\ast},q)  &  \geq0 \label{CorrelClassNE}%
\end{align}
For example, for PD we may have $r=3,s=0,t=5$ and $u=1$ that reduce the
inequalities (\ref{CorrelClassNE}) to%

\begin{align}
(p^{\ast}-p)(1+q^{\ast})  &  \leq0\nonumber\\
(q^{\ast}-q)(1+p^{\ast})  &  \leq0 \label{CorrelPDNE}%
\end{align}
It produces $p^{\ast}=q^{\ast}=0$ or $(D,D)$ as the unique equilibrium.

\subsection{Quantum correlation games}

The idea of a `correlation game' in the Ref. \cite{iqbalweigert} was
introduced to put forward a scheme to play a quantum version of a bi-matrix
game that respects the constraints C1 and C2 of the Section (\ref{EPR type
expts}). Its motivation comes from EPR type experiments performed on singlet
states involving \emph{correlations} of the measurement outcomes. In such
experiments the Bell's inequalities \cite{AsherPeres} are well-known to be the
constraints -- derived under the \emph{principle of local causes} -- on
correlations of measurement outcomes of two-valued (dichotomic) variables.
Truly quantum correlations are non-local in character and violate the inequalities.

The two parties involved in the usual \ EPR type setting are recognized as the
players. Repeated measurements are performed on correlated pairs of objects by
the two players, each receiving one half.

Players Alice and Bob share a Cartesian coordinate system between them and
each player's move consists of deciding a direction in a given plane. For
Alice and Bob these are the $x$-$z$ and $y$-$z$ planes respectively.\ Call
$\alpha$ and $\beta$ the unit vectors representing the players' moves. Both
players have a choice between two different orientations i.e. $\alpha$ and $z$
for Alice and $\beta$ and $z$ for Bob. Each player measures the angular
momentum or spin of his/her respective half in one of two directions. Let the
vectors $\alpha$ and $\beta$ make angles $\theta_{A}$ and $\theta_{B}$,
respectively, with the $z$-axis.

To link the players' moves, represented now by angles $\theta_{A}$ and
$\theta_{B}$, to the usual probabilities $p$ and $q$ appearing in a bi-matrix
game, an invertible function $g$ is made public at the start of a game. The
$g$-function maps $[0,\pi]$ to $[0,1]$ and allows to translate the players'
moves to the probabilities $p$ and $q$.

The results of measurements performed on dichotomic variables may take only
the values $\pm1$. These are represented by $a,b$ and $c$ for the directions
$\alpha,\beta$ and the $z$-axis respectively. Correlations $\langle
ac\rangle,\langle cb\rangle$ and $\langle ab\rangle$ can then be found from
the measurement outcomes, where the two entries in a bracket represent the
players' chosen directions.

In a correlation experiment in which the $z$-axis is the common direction for
the players, the Bell's inequality\footnote{For perfectly anticorrelated pairs
the right hand side of the inequality is $1+\left\langle bc\right\rangle $.}
is written \cite{AsherPeres} as%

\begin{equation}
\left|  \left\langle ab\right\rangle -\left\langle ac\right\rangle \right|
\leq1-\left\langle bc\right\rangle \label{BellsInequality}%
\end{equation}

The classical correlations corresponding to the above situation, when written
in terms of $\theta_{A}$ and $\theta_{B}$, are known \cite{AsherPeres} to be
invertible. This invertibility allows to express $\theta_{A}$ and $\theta_{B}$
in terms of the correlations $\langle ac\rangle$ and $\langle cb\rangle$. The
$g$-function allows now to translate $\theta_{A}$ and $\theta_{B}$ to $p$ and
$q$, respectively. In effect the classical bi-matrix payoffs are
\emph{re-expressed} in terms of the classical correlations $\langle ac\rangle$
and $\langle cb\rangle$.

One can now claim that the classical game is given \emph{by definition} in
terms of the correlations. The point for such a re-expression of the classical
game is that it opens the way to `quantum' version of the game. It, of course,
happens when the correlations become quantum mechanical.

In the setting of a correlation game the players' payoffs involve only the
correlations $\langle ac\rangle$ and $\langle cb\rangle$, instead of the three
correlations $\langle ac\rangle,\langle cb\rangle$ and $\langle ab\rangle$
present in the inequality (\ref{BellsInequality}), when $z$-axis is the common
direction between the players. This aspect results in \cite{iqbalweigert}
obtaining `quantum' payoffs even when the correlations are local and satisfy
the inequality (\ref{BellsInequality}).

The motivation for introducing EPR type setting to bi-matrix games is to
exploit quantum correlations to generate quantum payoffs. So that, when the
correlations are local, the classical game \emph{must} be produced. We show
below the possibility of such a connection by a different setting in which the
classical payoffs are \emph{always} obtained whenever the correlations
$\langle ac\rangle,\langle cb\rangle$ and $\langle ab\rangle$ satisfy the
Bell's inequality (\ref{BellsInequality}).

\subsection{A new approach towards defining a correlation game}

Consider an EPR type set-up to play a game between two players. Following
rules apply:

\begin{enumerate}
\item  A player's move consists of defining a direction in space by
orientating a unit vector. However, this direction is not confined to only the
$x$-$z$ or $y$-$z$ planes. A player's choice of a direction can be
\emph{anywhere} in three-dimensional space. Therefore, Alice's move is to
define a unit vector $\alpha$ and, similarly, Bob's move is to define a unit
vector $\beta$.

\item  The $z$-axis is shared between the players as the common direction.

\item  On receiving a half of a correlated pair, a player measures its spin in
one of the two directions. For Alice these directions are $\alpha$ and $z$ and
for Bob these directions are $\beta$ and $z$.

\item  Each player measures spin with \emph{equal} probability in his/her two directions.

\item  Players agree together on \emph{explicit expressions} giving their
payoffs $P_{A}$ and $P_{B}$ in terms of all three correlations i.e.
\end{enumerate}%

\begin{align}
P_{A}  &  =P_{A}(\langle ac\rangle,\langle cb\rangle,\langle ab\rangle
)\nonumber\\
P_{B}  &  =P_{B}(\langle ac\rangle,\langle cb\rangle,\langle ab\rangle)
\label{CorrelPayoffs}%
\end{align}

A game defined by these rules eliminates the need for introducing the
$g$-functions. The rules are also consistent with the constraints C1 and C2
and the idea of a correlation game essentially retains its spirit.

\subsection{Defining correlation payoffs}

A possible way is shown now to define the correlation payoffs
(\ref{CorrelPayoffs}) which reduce to the classical payoffs
(\ref{CorrelClassPayoffs}) whenever the correlations $\langle ab\rangle
,\langle ac\rangle$ and $\langle bc\rangle$ satisfy the inequality
(\ref{BellsInequality}).

Consider two quantities $\varepsilon$ and $\sigma$ defined as:%

\begin{equation}
\varepsilon=\sqrt{3+\langle bc\rangle^{2}+2\langle ab\rangle\langle ac\rangle
},\text{ \ \ \ }\sigma=\sqrt{2(1+\langle bc\rangle)+\langle ab\rangle
^{2}+\langle ac\rangle^{2}} \label{A&B}%
\end{equation}
The quantities $\varepsilon$ and $\sigma$ can adapt only real values because
the correlations $\langle ac\rangle,\langle cb\rangle$ and $\langle ab\rangle$
are always in the interval $\left[  -1,1\right]  $. Consider now the
quantities $(\varepsilon-\sigma)$ and $(\varepsilon+\sigma)$. By definition
$\varepsilon$ and $\sigma$ are non-negative, therefore, the quantity
$(\varepsilon+\sigma)$ always remains non-negative. It is observed that if
$0\leq$ $(\varepsilon-\sigma)$ then the correlations $\langle ac\rangle
,\langle cb\rangle$ and $\langle ab\rangle$ \emph{always} satisfy the
inequality (\ref{BellsInequality}). It is because if $0\leq(\varepsilon
-\sigma)$ then $0\leq(\varepsilon+\sigma)(\varepsilon-\sigma)=\varepsilon
^{2}-\sigma^{2}$. But $\varepsilon^{2}-\sigma^{2}=(1-\langle bc\rangle
)^{2}-\left|  \langle ab\rangle-\langle ac\rangle\right|  ^{2}$ so that
$\left|  \langle ab\rangle-\langle ac\rangle\right|  ^{2}\leq(1-\langle
bc\rangle)^{2}$ which results in the inequality (\ref{BellsInequality}). All
the steps in the proof can be reversed and it follows that whenever the
correlations $\langle ac\rangle,\langle cb\rangle$ and $\langle ab\rangle$
satisfy the Bell's inequality, the quantity $(\varepsilon-\sigma)$ remains non-negative

For a singlet state satisfying the inequality (\ref{BellsInequality}) both the
quantities $(\varepsilon+\sigma)$ and $(\varepsilon-\sigma)$ are non-negative
and must have maxima. Hence, it is possible to find two non-negative numbers
$\frac{(\varepsilon-\sigma)}{\max(\varepsilon-\sigma)}$ and $\frac
{(\varepsilon+\sigma)}{\max(\varepsilon+\sigma)}$ in the range $\left[
0,1\right]  $, whenever the inequality (\ref{BellsInequality}) holds. Because
$0\leq\varepsilon,\sigma\leq\sqrt{6}$ we have $\max(\varepsilon-\sigma
)=\sqrt{6}$ and $\max(\varepsilon+\sigma)=2\sqrt{6}$. The numbers
$(\varepsilon-\sigma)/\sqrt{6}$ and $(\varepsilon+\sigma)/2\sqrt{6}$ are in
the range $[0,1]$ when the inequality holds. These numbers are also
\emph{independent} from each other.

The above argument paves the way to associate a pair $(p,q)$ of independent
numbers to the players' moves $(\alpha,\beta)$, that is%

\begin{equation}
p=p(\alpha,\beta)\text{, \ \ \ \ \ \ }q=q(\alpha,\beta) \label{ProbDirLink}%
\end{equation}
where the numbers $p,q$ are in the interval $[0,1]$ when the input states do
not violate the inequality (\ref{BellsInequality}) for all direction pairs
($\alpha,\beta$). The pair $(p,q)$ is related to the two directions as%

\begin{equation}
\alpha=\alpha(p,q)\text{, \ \ \ \ \ \ }\beta=\beta(p,q) \label{ProbDirLink1}%
\end{equation}

It can be noticed that more than one pair $(\alpha,\beta)$ of directions may,
however, correspond to a given pair of numbers. The converse, although, is not
true for known input states. That is, for known input states, only one pair
$(p,q)$ can be obtained from a given pair $(\alpha,\beta)$ of directions.

Players' payoffs can now be re-expressed by making the replacements:%

\begin{equation}
p(\alpha,\beta)\sim(\varepsilon-\sigma)/\sqrt{6},\text{ \ \ }q(\alpha
,\beta)\sim(\varepsilon+\sigma)/2\sqrt{6} \label{Replacements}%
\end{equation}
which lead to re-writing the classical payoffs (\ref{CorrelClassPayoffs}) as%

\begin{align}
P_{A}(\alpha,\beta)  &  =Kp(\alpha,\beta)q(\alpha,\beta)+Lp(\alpha
,\beta)+Mq(\alpha,\beta)+N\nonumber\\
P_{B}(\alpha,\beta)  &  =Kp(\alpha,\beta)q(\alpha,\beta)+Mp(\alpha
,\beta)+Lq(\alpha,\beta)+N
\end{align}
or more explicitly:%

\begin{align}
P_{A}(\alpha,\beta)  &  =\frac{K}{12}(\varepsilon^{2}-\sigma^{2})+\frac
{L}{\sqrt{6}}(\varepsilon-\sigma)+\frac{M}{2\sqrt{6}}(\varepsilon
+\sigma)+N\nonumber\\
P_{B}(\alpha,\beta)  &  =\frac{K}{12}(\varepsilon^{2}-\sigma^{2})+\frac
{M}{\sqrt{6}}(\varepsilon-\sigma)+\frac{L}{2\sqrt{6}}(\varepsilon+\sigma)+N
\label{CorrelQpayoffs}%
\end{align}
This expression shows that a player's payoff now depends on the direction s/he
has chosen. The payoffs (\ref{CorrelQpayoffs}) are obtained under the
constraints C1 and C2 and are functions of all the three correlations.

The relations (\ref{ProbDirLink}) can also imagined as follows. When Alice
decides a direction $\alpha$ in space, it corresponds to a curve in the
$p$-$q$ plane. Similarly, Bob's decision of the direction $\beta$ defines
another curve in the $p$-$q$ plane. The relations (\ref{Replacements}) assure
that only one pair $(p,q)$ can then be obtained as the intersection between
the two curves.

The set-up assures that for input product states all of the players' moves
$(\alpha,\beta)$ result in the correlation payoffs (\ref{CorrelQpayoffs})
generating identical to the classical payoffs (\ref{CorrelClassPayoffs}). For
such input states the relations (\ref{Replacements}) give the numbers $p,q$ in
the interval $[0,1]$, which can then be interpreted as probabilities. However,
for input states for which the inequality (\ref{BellsInequality}) is violated
in some directions, a pair $(p,q)$ $\in\lbrack0,1]$ \emph{cannot} be
associated to those directions. It is because for entangled states there exist
pairs of directions for which the corresponding quantity $(\varepsilon
-\sigma)$ becomes negative. For those directions the correlation payoffs
(\ref{CorrelQpayoffs}) would generate results that can \emph{only} be
understood, within the structure of classical payoffs
(\ref{CorrelClassPayoffs}), by invoking negative probabilities.

\subsection{Nash equilibria of quantum correlation games}

Because the players' moves consist of defining directions in space, the Nash
inequalities are%

\begin{align}
P_{A}(\alpha_{0},\beta_{0})-P_{A}(\alpha,\beta_{0})  &  \geq0\nonumber\\
P_{B}(\alpha_{0},\beta_{0})-P_{B}(\alpha_{0},\beta)  &  \geq0
\label{NEdirections}%
\end{align}
where the pair $(\alpha_{0},\beta_{0})$ corresponds to the pair $(p^{\ast
},q^{\ast})$ via the relations (\ref{Replacements}). The inequalities
(\ref{NEdirections}) are same as the inequalities (\ref{CorrelClassNE}),
except their re-expression in terms of the directions.

When the correlations in the input state correspond to an entangled state, the
payoff relations (\ref{CorrelQpayoffs}) would lead to disappearance of the
classical equilibria. It can be seen, for example, by considering the Nash
inequalities for the Prisoners' Dilemma (\ref{CorrelPDNE}). Let the
directional pair $(\alpha_{D},\beta_{D})$ correspond to the equilibrium
$(D,D)$, that is, the inequalities (\ref{NEdirections}) are%

\begin{align}
P_{A}(\alpha_{D},\beta_{D})-P_{A}(\alpha,\beta_{D})  &  \geq0\nonumber\\
P_{B}(\alpha_{D},\beta_{D})-P_{B}(\alpha_{D},\beta)  &  \geq0 \label{NEDD}%
\end{align}
Assume the players receive input states that are entangled. There will now
exist pairs of players' moves $\alpha$ and $\beta$ that would make the
quantity $(\varepsilon-\sigma)<0$. The pair $(\alpha,\beta)$ will not
correspond to a point in the $p$-$q$ plane where $p,q\in\lbrack0,1]$.

It can also be noticed that for entangled input states the directional pair
$(\alpha_{D},\beta_{D})$ does not remain a NE. It is because the
pair\textit{\ }$(\alpha_{D},\beta_{D})$\textit{\ }is a NE \emph{only} if
players' choices of \emph{any} directional pair\textit{\ }$(\alpha,\beta
)$\textit{\ }corresponds to a point in the\textit{\ }$p$\textit{-}%
$q$\textit{\ }plane where\textit{\ }$p,q\in\lbrack0,1]$.

Because for entangled input states there exist pairs of players' moves
$(\alpha,\beta)$ that do not correspond to points in the $p$-$q$ plane with
$p,q\in\lbrack0,1]$. Hence, the directional pair $(\alpha_{D},\beta_{D})$ does
not remain a NE in the quantum game. Interestingly, the disappearance of the
classical equilibrium now becomes directly \textit{linked} with the violation
of the inequality (\ref{BellsInequality}) by the correlations in the input states.

\subsection{Quantum game as another classical game?}

Coming back to the questions raised in the Section (\ref{IntNewProp}), we now
try to construct a classical bi-matrix game corresponding to a quantum game
resulting from the payoff relations (\ref{CorrelQpayoffs}). The classical game
is assumed to have the \emph{same} general structure of players' payoffs as
given in Eqs. (\ref{CorrelClassPayoffs}). This assumption derives from the
hope that the quantum game, corresponding to correlations in the input states
that violate the inequality (\ref{BellsInequality}), is also equivalent to
another symmetric bi-matrix game. It is shown below that such a construction
cannot be permitted.

Suppose the correlations in the input states violate the inequality
(\ref{BellsInequality}). A pair of directions ($\alpha,\beta$) can now be
found for which the Bell's inequality is violated. For Alice's move to select
the direction $\alpha$ her payoff, given by the Eqs. (\ref{CorrelQpayoffs}), is%

\begin{equation}
P_{A}(\alpha,\beta)=K^{\prime}pq+L^{\prime}p+Mq+N \label{Alice'sNewPayoff}%
\end{equation}
where $K^{\prime}=-K$ and $L^{\prime}=-L$ and $p,q\in\lbrack0,1]$. Assuming
that the constants $K^{\prime},L^{\prime},M,$ and $N$ define a `new' symmetric
bi-matrix game the Bob's payoff is%

\begin{equation}
P_{B}(p,q)=K^{\prime}pq+Mp+L^{\prime}q+N \label{Bob'sNewPayoff1}%
\end{equation}
But in fact (\ref{Bob'sNewPayoff1}) is not obtained as the Bob's payoff in our
quantum game when he goes for the direction $\beta$. Bob's payoff is, in fact,
given as%

\begin{equation}
P_{B}(p,q)=K^{\prime}pq+M^{\prime}p+Lq+N \label{Bob'sNewPayoff2}%
\end{equation}
which may not necessarily coincide with the payoff given in the Eq.
(\ref{Bob'sNewPayoff1}). Hence, the game resulting from the presence of
quantum correlations in the input states \emph{cannot} be simply explained as
another classical symmetric bi-matrix game: a game obtained by defining new
coefficients of the matrix involved. Players' payoffs in the quantum game are
found to reside outside the structure of payoffs in a classical symmetric
bi-matrix game. The payoffs can be explained within this structure \emph{only}
by invoking negative probabilities.

An asymmetric bi-matrix game can, of course, be constructed having identical
solutions to the quantum game. In fact for \emph{any} quantum game a classical
model can \emph{always} be constructed that summarizes the complete situation
and has identical to the quantum solutions, as far as the players' payoffs are
concerned. It would be a model that relates players' moves to their payoffs in
accordance with the usual approach in game theory. But constructing such a
model is not an answer to our original question: How solutions of a game are
affected by the presence of quantum correlations between the physical objects
used to implement the game? It is because the question can then simply be
rephrased as: What if the modified classical game is played with physical
objects having quantum correlations?

\subsection{Discussion}

The idea of a correlation game is about re-expression of payoffs of a
classical bi-matrix game in terms of correlations of measurement outcomes made
on pairs of disintegrating particles. The measurement outcomes are dichotomic
variables and their correlations are obtained by averaging over a large number
of pairs. Bell's inequalities represent constraints on these correlations
obtained under the principle of local causes. A re-expression of the classical
payoffs of a bi-matrix game in terms of correlations opens the way to
explicitly see the effects of quantum correlations on the solutions of the game.

We have proposed a new setting where two players play a bi-matrix game by
repeatedly performing measurements on correlated pairs of objects. The setting
is motivated by EPR type experiments performed on singlet states. On receiving
a half of a pair, a player makes a measurement of its spin in one of the two
directions available to him/her. The measurements are performed with
\emph{equal probability} in the two directions. Both players share a common
direction and defining the \emph{other} direction is a player's \emph{move}.

We show how within this set-up a correlation version of a symmetric bi-matrix
game can be defined. The correlation game shows some interesting properties.
For example the correlation game reduces to the corresponding classical game
when the correlations in the input states are local and do not violate the
Bell's inequality (\ref{BellsInequality}). However, when the inequality is
violated, the stronger correlations generate results that can be understood,
within the structure of classical payoffs in a symmetric bi-matrix game,
\emph{only} by invoking negative probabilities. It is shown that a classical
NE is affected when the game is played with input states having quantum
correlations. The proposed set-up also provides a new perspective on the
possibility of reformulating the Bell's inequalities in terms of a bi-matrix
game played between two spatially-separated players.

\chapter{Conclusions}

To conclude a summary of the results developed in this thesis is presented
below. The results 1 -- 8 refer to Eisert et al.'s and Marinatto and Weber's
schemes of quantization of matrix games.

\begin{enumerate}
\item  In a population engaged in symmetric bi-matrix classical game of
Prisoners' Dilemma an invasion of classical ESS is possible by the mutants
exploiting Eisert's two-parameter set of quantum strategies. We presented an
example of an \emph{asymmetric} quantum game between two players in which a
strategy pair can be an ESS for either classical or quantum version of the
game, even when it remains a NE in both the versions. It shows quantization
can change evolutionary stability of Nash equilibria in certain asymmetric
bi-matrix games.

\item  ESS concept was originally defined for symmetric bi-matrix contests. We
showed that quantization can also change evolutionary stability of a NE in
certain types of \emph{symmetric} bi-matrix games. It immediately makes study
of quantum games also relevant to evolutionary game theory and conversely.
Hence, quantization not only leads to new equilibria but it also presents
itself also as \emph{another} refinement notion of the NE concept.

\item  Like pure strategies the evolutionary stability of \emph{mixed}
strategies can also change as a symmetric bi-matrix game is switched from its
classical to quantum form. However, for mixed strategies we require more
general initial quantum states.

\item  Rock-Scissors-Paper (RSP) is a two-player three-strategy game. We
consider a slightly modified form of RSP played in its classical version. A
mixed NE exists that is not an ESS. We find a quantum form of the \emph{same}
game in which the classical NE becomes an ESS. The quantum form is obtained
when the game is played with an initial \emph{entangled} state.

\item  Quantization can change properties of equilibria of replicator
dynamics. We consider a game played in a population setting with the
underlying process of replicator dynamics. We found a `quantum form' of the
replicator equations, which retain their form as that of Lotka-Volterra type.
The effects of quantization of the game on a saddle or a center of the
dynamics are then studied. It is found that a saddle (center) in the classical
game can be a center (saddle) in certain quantum form of the game. A saddle or
center in a classical (quantum) game can not be, however, an attractor or a
repeller in quantum (classical) form of the game.

\item  A symmetric cooperative game played by three players is analyzed in its
classical and quantum forms. In classical form of this game forming a
coalition gives advantage to players and players are motivated to do so. In
quantum form of the game, however, an initial quantum state can be prepared by
the arbiter such that forming the \emph{same} coalition is of no advantage.

\item  A comparison between the NE in Cournot game with the
backwards-induction outcome in classical Stackelberg duopoly shows that having
Alice (who acts first) know that Bob (who acts second) knows Alice's move
hurts Bob. In fact in classical version of the Stackelberg game Bob should not
believe that Alice has chosen its Stackelberg quantity. We have shown that
there can be a quantum version of Stackelberg duopoly where Bob is \emph{not}
hurt even if he knows the quantity chosen by Alice. The backwards-induction
outcome of this quantum game is \emph{same} as the NE in classical Cournot
game, where decisions are made simultaneously and there is no such information
that hurts a player.

\item  In infinitely repeated versions of the classical game of Prisoners'
Dilemma it is established \cite{Gibbons} that cooperation can occur in every
stage of a subgame-perfect outcome (SGPO), even though the only NE in the
stage game is defection. We find how cooperation in two-stage Prisoners'
Dilemma can be achieved by quantum means. In two-stage Prisoners' Dilemma
getting a SGPO where players cooperate in the first stage is a result with no
classical analogue. We have also introduced a possible way to study the
concept of SGPO in repeated quantum games.

\item  In the standard set-ups to play a quantum game a measure of
entanglement for a qubit system is introduced. Quantum version of the game
reduces to classical when the measure becomes zero. We suggested a set-up that
exploits another resource from quantum physics, i.e. the association of waves
with travelling material objects like electrons. We show how in this set-up
such association of waves can lead to a non-classical equilibrium in the
Prisoners' Dilemma. With associating wavelength approaching zero the quantum
aspect disappears and the classical game is reproduced.

\item  Playing a symmetric bi-matrix game is usually physical implemented by
sharing pairs of `objects' between two players. We proposed a new setting that
explicitly shows the effects of quantum correlations between the pairs on the
structure of payoff relations and `solutions' of the game. The setting is
based on a re-expression of the game such that the players play the classical
game \textit{only }if\textit{\ }their moves are performed on pairs of objects
having correlations that satisfy the Bell's inequalities. On players receiving
pairs with quantum correlations the resulting game \textit{cannot} be
considered \textit{another} classical symmetric bi-matrix game. Also the Nash
equilibria of the game are found to be decided by the nature of the correlations.
\end{enumerate}

\begin{acknowledgement}
I am grateful to Dr. A. H. Toor for supervising this thesis. I remain
particularly thankful to him for the freedom he granted me in the choice of
topics for the thesis. I am obliged to Prof. Razi Naqvi (NTNU, Trondheim,
Norway) for encouragement, as well as help, that he provided during the early
stages of research. I am thankful to Dr. Mark Broom (University of Sussex, UK)
for giving many remarks while he replied to my emails. He made available
excellent reports on his website that, in fact, aroused my interest in
evolutionarily stable strategies (ESSs). I am thankful to Dr. Jens Eisert
(University of Potsdam, Germany) for replying to all of my emails and queries.
I happened to see his pioneering paper on quantum games -- which indeed
provided much of the later motivation for interest and developments in quantum
games -- during the same time when I read about ESSs. Thanks to Dr. Luca
Marinatto (Abdus Salam ICTP, Trieste, Italy) for replying to my emails. I am
thankful to Prof. Dr. S. Azhar Abbas Rizvi (Quaid-i-Azam University) for
offering a course on probability, which helped me later to follow game theory;
to Dr. Qaiser Naqvi (Quaid-i-Azam University) for motivating discussions on a
broad range of topics; to Mr. Ahmad Nawaz (Quaid-i-Azam University) for many
exciting arguments on quantum games. During most of the research, presented in
this thesis, I was supported by the Pakistan Institute of Lasers and Optics
(PILO), Rawalpindi, Pakistan. I remain thankful to my ex-colleagues Dr. M. M.
Gualini and Dr. Mian Ashraf for encouragement as well as guidance that they
provided on many occasions.
\end{acknowledgement}


\begin{thebibliography}{99}
\bibitem{Rasmusen89}E. Rasmusen, \emph{Games and Information.} Blackwell,
Cambridge MA (1989).

\bibitem {Neumann}v. Neumann J. and O. Morgenstern. \textit{Theory of Games
and Economic Behaviour}. (Princeton 1953).

\bibitem {NeumannQM}Neumann, J. von, Mathematical Foundations of Quantum
Mechanics, Princeton, N. J., Princeton University Press 1955. First published
in German in 1932: Mathematische Grundlagen der Quantenmechank, Berlin: Springer.

\bibitem {Feynman1}R. P. Feynman. Simulating physics with computers. Int. J.
Theor. Phys., \textbf{21}, 467 (1982).

\bibitem {Deutsch}D. Deutsch. Quantum theory, the Church-Turing Principle and
the universal quantum computer. Proc. R. Soc. London. A \textbf{400}, 97 (1985).

\bibitem {Shor}P. W. Shor. Algorithm for quantum computation: discrete
logarithms and factoring. In Proceedings, 35th Annual Symposium on Foundations
of Computer Science, IEEE Press, Los Alamitos, CA, 1994.

\bibitem {Shor1}P. W. Shor. Polynomial-time algorithms for prime factorization
and discrete logarithms on a quantum computer. SIAM J. Comp., \textbf{26}(5),
1484-1509, 1997.

\bibitem {Grover}Lov K. Grover. Proceedings of 28th Annual ACM Symposium on
the Theory of Computing (STOC), Philadelphia, pages 212-219 (May 1996).

\bibitem {Grover1}L. K. Grover, Quantum mechanics helps in searching for a
needle in a haystack. Phys. Rev. Lett. \textbf{79}(2), 325 (1997).
\texttt{quant-ph/9706033}.

\bibitem {Turing}A. M. Turing. On computable numbers, with an application to
the Entscheidungsproblem. Proc. Lond. Math. Soc. 2, \textbf{42}, 230 (1936).

\bibitem {Shannon}C. E. Shannon, A mathematical theory of communication. Bell
System Tech. J., \textbf{27}, 379-423, 623-656, 1948.

\bibitem {Schumacher}B. Schumacher. Quantum coding. Phys. Rev. A \textbf{51},
2738-2747, 1995.

\bibitem {Wiesner}Wiesner S. Conjugate coding, SIGACT News. \textbf{15}/1, 78
(1983). Available at \texttt{http://kh.bu.edu/qcl/pdf/wiesners198316024137.pdf}

\bibitem {Goldenberg}L. Goldenberg, L. Vaidman, and S. Wiesner, Phys. Rev.
Lett. \textbf{82}, 3356 (1999).

\bibitem {Vaidman}Vaidman L. Foundations of Physics \textbf{29}, 615 (1999).

\bibitem {Ekert}A. K. Ekert, Phys. Rev. Lett. \textbf{67}, 661 (1991).

\bibitem {Gisin}N. Gisin and B. Huttner, Phys. Lett. A \textbf{228}, 13 (1997).

\bibitem {WernerRF}R. F. Werner, Phys. Rev. A \textbf{58}, 1827 (1998).

\bibitem {MeyerDavid}D. A. Meyer, Quantum strategies. Phys. Rev. Lett.
\textbf{82} 1052-1055 (1999).

\bibitem {Eisert}Jens Eisert, Martin Wilkins and Maciej Lewenstein. Quantum
Games and Quantum Strategies. Phy. Rev. Lett. \textbf{83}, 3077 (1999).

\bibitem {Nielsen}Michael A. Nielsen and Issac L. Chuang. \textit{Quantum
Computation and Quantum Information}. Cambridge University Press (2000). Also
Collin P. Williams, Scott H. Clearwater. \textit{Explorations in Quantum
Computing}. Springer-Verlag Inc. New York (1998).

\bibitem {Smith}Maynard Smith J. \textit{Evolution and the theory of games}.
CUP (1982).

\bibitem {JohnNash}J. Nash, Proc. of the National Academy of Sciences,
\textbf{36}, 48 (1950).

\bibitem {JohnNash1}J. Nash. Non-cooperative games. Ann. Math. \textbf{54},
287-295. MR 13: 261g (1951).

\bibitem {Cournot}A. Cournot. \textit{Researches into the Mathematical
Principles of the Theory of Wealth}. Edited by N. Bacon. New York, Macmillan (1897).

\bibitem {Broom et al}See for example in Ref. \cite{MarkBroom1}. Also M. Broom
and G. D. Ruxton, Behavioral Ecology \textbf{13}, 321-327 (2002).

\bibitem {Hofbauer}Josef Hofbauer and Karl Sigmund. \textit{Evolutionary Games
and Population Dynamics. }Cambridge University Press (1998).

\bibitem {Leonard}R. J. Leonard: Reading Cournot, reading Nash: The creation
and stabilization of the Nash equilibrium. The Economic Journal \textbf{104},
492-511 (1994).

\bibitem {Smith Price}Maynard Smith J. and Price G. R., The logic of animal
conflict. Nature, \textbf{246}, 15-18 (1973).

\bibitem {MarkBroom3}M. Broom, Patterns of evolutionarily stable strategies:
the maximal pattern conjecture revisited. J. Math. Biol. \textbf{40}, 40,
406-412 (2000).

\bibitem {CanningsOrive}See for example Cannings C. and L. Cruz Orive. Natural
selection and the sex-ratio, Optimal sex-ratios. J. Theoret. Biol.
\textbf{55}, 115-136 (1975).

\bibitem {Weibull}J. W. Weibull, \textit{Evolutionary game theory}. The MIT
Press, Cambridge (1995).

\bibitem {Prestwich}K. Prestwich. Game Theory. Department of Biology, College
of the Holy Cross, Worcester, MA, USA 01610 (1999). Available at the URL {\small http://www.holycross.edu/departments/biology/kprestwi/behavior/ESS/pdf/games.pdf}

\bibitem {TaylorJonker}Taylor P. D., and Jonker L, Evolutionarily stable
strategies and game dynamics. Math Biosc. \textbf{40}, 145-156 (1978).

\bibitem {MyersonRB}See for example Myerson R. B., Refinements of the Nash
Equilibrium Concept, International Journal of Game Theory \textbf{7}, 73-80 (1978).

\bibitem {Cressman}Cressman R, \textit{The stability concept of evolutionary
game theory}. Springer Verlag, Berlin (1992).

\bibitem {Gerard van}Gerard van der Laan and Xander Tieman, Evolutionary game
theory and the Modelling of Economic Behaviour. A Report. Research Program
``Competition and Cooperation'' of the Faculty of Economics and Econometrics.
Free University Amsterdam. November 6, 1996. Available at: \texttt{http://www.tinbergen.nl/discussionpapers/96172.pdf}

\bibitem {AsherPeres}A. Peres, \textit{Quantum Theory: Concepts and Methods.}
Kluwer Academic, Dordrecht (1993).

\bibitem {Preskill}J. Preskill. Physics 229: Advanced Mathematical Methods of
Physics -- Quantum Computation and Information. California Institute of
Technology, 1998. URL: \texttt{http://www.theory.caltech.edu/people/preskill/ph229/}

\bibitem {EPR}A. Einstein, B. Podolsky and N. Rosen, Phys. Rev. \textbf{47},
777 (1935).

\bibitem {Hoi-Kwong}\textit{Introduction to Quantum Computation and
Information}. Edited by: Hoi-Kwong Lo, Sandu Popescu and Tim Spiller. World
Scientific 1998.

\bibitem {Mermin}N. D. Mermin, Am. J. Phys. \textbf{58}, 731 (1990).

\bibitem {GHZ}D. M. Greeberger, M. Horne, A. Shimony, and A. Zeilinger, Am. J.
Phys. \textbf{58}, 1131 (1990).

\bibitem {Bell}J. S. Bell, Physics \textbf{1}, 195 (1964).

\bibitem {Elitzur-Vaidman}A. Elitzur, L. Vaidman, Found. Phys. \textbf{23},
987 (1993).

\bibitem {Eisert1}J. Eisert and M. Wilkens, Quantum Games, J. Mod. Opt.
\textbf{47}, 2543 (2000).

\bibitem {Dawkins}R. Dawkins, \textit{The Selfish Gene}, Oxford University
Press, Oxford (1976).

\bibitem {FlitneyAbbott4}Adrian P. Flitney, Derek Abbott. Advantage of a
quantum player over a classical one in 2x2 quantum games. Proc. R. Soc.
(London) A \textbf{459} (2003) 2463-74.

\bibitem {Flitney5}Adrian P. Flitney, Derek Abbott. Quantum two and three
person duels. J. Opt. B \textbf{6} (2004) S860-S866.

\bibitem {Flitney6}Adrian P. Flitney, Derek Abbott, Neil F. Johnson. Quantum
random walks with history dependence. J. Phys. A \textbf{37} (2004) 7581-7591.

\bibitem {Flitney7}Adrian P. Flitney, Derek Abbott. Quantum games with
decoherence. \texttt{quant-ph/0408070}.

\bibitem {Marinatto1}L. Marinatto and T. Weber, Phys. Lett. A \textbf{272},
291 (2000).

\bibitem {Li et al}Chuan-Feng Li, Yong-Sheng Zhang, Yun-Feng Huang and
Guang-Can Guo. Physics Letters A, Volume \textbf{280}, Issues 5-6, 5 March
2001, Pages 257-260.

\bibitem {FlitneyAbbott}Adrian P. Flitney, Derek Abbott. Phys. Rev. A
\textbf{65}, 062318 (2002).

\bibitem {DAriano}G. M. D'Ariano, R. D. Gill, M. Keyl, B. Kuemmerer, H.
Maassen, R. F. Werner. Quant. Inf. Comput. \textbf{2}, no. 5, 355-366 (2002).

\bibitem {Piotrowski1}E. W. Piotrowski, J. Sladkowski. Quantum Market Games.
Physica A \textbf{312}, 208 (2002). quant-ph/0104006.

\bibitem {Piotrowski2}E. W. Piotrowski, J. Sladkowski. Quantum Bargaining
Games. Physica A \textbf{308}, 391 (2002). quant-ph/0106140.

\bibitem {Piotrowski3}E. W. Piotrowski, J. Sladkowski. First Quantum Market
Games Site. URL: \texttt{http://alpha.uwb.edu.pl/ep/sj/index.shtml.}

\bibitem {Piotrowski}Piotrowski E. W. and Sladkowski J. The next stage:
quantum game theory. \texttt{quant-ph/0308027}. Also by the same authors: An
invitation to quantum game theory. Int. J. Theor. Phys. \textbf{42}, 1089
(2003). \texttt{quant-ph/0211191.}

\bibitem {Econophysics}See for example, URLs:
\texttt{http://www.unifr.ch/econophysics/},
\texttt{http://www.econophysics.org/}, \texttt{http://www.rsphysse.anu.edu.au/econophysics/.}

\bibitem {Blankmeyer}Eric Blankmeyer. The Heisenberg Principle in Economics.
URL: \texttt{http://econwpa.wustl.edu:8089/eps/get/papers/9904/9904004.html}

\bibitem {Lambertini}L. Lambertini. Quantum Mechanics and Mathematical
Economics are Isomorphic. John von Neumann between Physics and Economics. \texttt{http://www.dse.unibo.it/wp/370.pdf}

\bibitem {FlitneyAbbott1}Adrian P. Flitney, Derek Abbott. Physica A,
\textbf{314} (2002) 35-42.

\bibitem {FlitneyAbbott2}A. P. Flitney, Abbott D. Physica A, Volume
\textbf{324}, Number 1, 1 June 2003, pp. 152-156 (5).

\bibitem {Enk}S. J. van Enk. Phys. Rev. Lett. \textbf{84}, 789 (2000).

\bibitem {Meyer's Reply}D. A. Meyer. Phys. Rev. Lett. \textbf{84}, 790 (2000).

\bibitem {ApporvaPatel}Apoorva Patel. A Wave Implementation of the Optimal
Database Search Algorithm. \texttt{quant-ph/0401154}.

\bibitem {Khrennikov}Andrei Khrennikov. Nonexistence of a realistic model for
the two-dimensional Hilbert space. \texttt{quant-ph/0308078}.

\bibitem {Benjamin1}S. C. Benjamin, Patrick M. Hayden. Phys. Rev. Lett.
\textbf{87} (6): 069801 (2001).

\bibitem {Benjamin2}S. C. Benjamin. Physics Letters, A \textbf{277}, 180-182 (2000).

\bibitem {NawazToor}Ahmad Nawaz and A. H. Toor, J. Phys. A: Math. Gen.
\textbf{37} No 15, 4437-4443 (2004).

\bibitem {Marinatto's reply}L. Marinatto and T. Weber. Phys. Lett. A
\textbf{277}, 183-184 (2000).

\bibitem {EnkPike}S. J. van Enk, R. Pike. Phys. Rev. A \textbf{66}, 024306 (2002).

\bibitem {Oechssler}J. Oechssler and F. Riedel. On the Dynamic Foundations of
Evolutionary Stability in Continuous Models. Discussion paper 7/2000. Bonn
Graduate School of Economics, University of Bonn. Adenauerallee 24-42, D-53113
Bonn. \texttt{http://www.bgse.uni-bonn.de/papers/liste.html\#2000}

\bibitem {MarkBroom1}M. Broom, C. Canning, and G. T. Vickers. Bull. Math. Bio.
\textbf{62}, 451 (2000).

\bibitem {BroomMutiplayer}M. Broom, C. Cannings, and G. T. Vickers.
Multi-player matrix games. Bulletin of Mathematical Biology \textbf{59},
931-952 (1997).

\bibitem {LizardGame}Sinervo, B and Lively, C. M. The rock-paper-scissors game
and the evolution of alternative male strategies. Nature. Vol. \textbf{380},
no. 6571, pp. 240-243. 1996.

\bibitem {MarkBroom2}M. Broom. J. Math. Biol. \textbf{40}, 406-412 (2000).

\bibitem {BishopCanning}D. T. Bishop and C. Cannings. Adv. Appl. Prob.
\textbf{8}, 616-621 (1976).

\bibitem {Vickers}G. T. Vickers and C. Cannings. Int. J. Theor. Biol.
\textbf{132}, 387-408 (1988).

\bibitem {Cressman1}R. Cressman. The Dynamic (In)Stability of Backwards
Induction. Journal of Economic Theory. \textbf{83}, 260-285 (1998).

\bibitem {CressmanSchlag}Cressman R. and Schlag K. H. Dynamic stability in
perturbed games. Discussion paper No. B-321. Rheinische
Friedrich-Wilhelms-Universit\"{a}t D-53113 Bonn (July 1995). Available at \texttt{http://www.iue.it/Personal/Schlag/papers/dynamic.html}

\bibitem {Bomze}Bomze I. M. and P\"{o}tscher B. M. \textit{Game theoretical
foundations of evolutionary stability.} Lecture notes in Economics and
Mathematical systems. \textbf{324}. Springer Verlag. Berlin (1989).

\bibitem {Sigmund}Sigmund K. \textit{The population dynamics of conflict and
cooperation}. Interim report IR-98-102/ December. International Institute for
Applied Systems Analysis. A-2361 Laxenburg. Austria, (1998). Available at \texttt{http://www.iiasa.ac.at/Publications/Documents/IR-98-102.pdf}

\bibitem {Zeeman}Zeeman E.C. Population dynamics from game theory. Proc. Int.
Conf. Global Theory of Dynamical Systems. Northwestern, Evanston, 471-497 (1979).

\bibitem {Eigen}Eigen, M. \& Schuster, P. The Hypercycle: a principal of
natural self organization. (A) Emergence of the hypercycle.
Naturwissenschaften \textbf{64}, 541-565 (1977).

\bibitem {Schuster}Schuster P. Sigmund, K. and Wolff R. Dynamical systems
under constant organization. Bull. Math. Biophys. \textbf{40}, 743-769 (1978).

\bibitem {Schuster1}Schuster P. and Sigmund K. Coyness, Philandering and
stable strategies. Anim. Behav, \textbf{29}, 186-192 (1981).

\bibitem {Hirsch}Hirsch M. and Smale S. \textit{Differential Equations,
Dynamical Systems, and Linear Algebra}. Academic Press New York (1974).

\bibitem {Turner}P. E. Turner and L. Chao. Nature \textbf{398}, 111 (1999).

\bibitem {Frohlich}Fr\"{o}hlich H. \textit{Coherent Excitations in Biological
Systems}. Springer-Verlag, Berlin (1983).

\bibitem {FlitneyAbbott3}Adrian P. Flitney, Derek Abbott. A semi-quantum
version of the game of Life. \texttt{quant-ph/0208149}.

\bibitem {Game of Life}M. Gardiner, Sci. Am. \textbf{223}, Oct. 120 (1970).

\bibitem {Bashford}Bashford, J. D., Jarvis, P. D. \& Tsohantjis, I.
\textit{Supersymmetry in the genetic code, in Physical Applications and
Mathematical Aspects of Geometry}. Edited by H.-D. Doebner, P. Nattermann, W.
Scherer, and C. Schulte. World Scientific Press, Singapore (1998).

\bibitem {Knight}R. D. Knight, S. J. Freeland, and L. F. Landweber. Rewiring
the keyboard: evolvability of genetic code, Nature Reviews Genetics
\textbf{2}, 49 (2001).

\bibitem {Knight1}Knight, R. D., S. J. Freeland, and L. F. Landweber.
Selection, history and chemistry: the three faces of the genetic code. Trends
Biochem. Sci. \textbf{24}, 241-247 (1999).

\bibitem {Patel}A. Patel, Quantum algorithms and the genetic code. Pramana
\textbf{56} 367-381 (2001).

\bibitem {Back}T. B\"{a}ck, A. Eiben, and M. Vink. A superior evolutionary
algorithm for 3SAT. Proc. EP98 (1998).

\bibitem {Back1}T. B\"{a}ck, U. Hamel, and H. P. Schwefel. Evolutionary
computation: Comments on the history and current state. IEEE Trans. Evol.
Comp. \textbf{1}, 3-17 (1997).

\bibitem {Greenwood}G. W. Greenwood, Finding solutions to NP problems:
Philosophical differences between quantum and evolutionary search algorithms.
\texttt{quant-ph/0010021}.

\bibitem {Meszena}Meszena. G, Kidsi. E, Dieckmann. U, Geritz. S. A. H, \&
Metz. J. A. J. Evolutionary optimisation models and matrix games in the
unified perspective of adaptive dynamics. Interim report IR-00-039.
International Institute for Applied Systems Analysis. A-2361 Laxenburg.
Austria. Available at \texttt{http://www.iiasa.ac.at/Publications/Documents/IR-00-039.pdf}

\bibitem {Goertzel}Goertzel B. Evolutionary quantum computation: Its role in
the brain, Its realization in electronic hardware, and its implications for
the panpsychic theory of consciousness. IntelliGenesis Corp. Available at \texttt{http://www.goertzel.org/dynapsyc/1997/Qc.html}

\bibitem {CooperativeGames}A. Iqbal and A. H. Toor, \textit{Quantum
cooperative games}. Physics Letters, A \textbf{293}/3-4 pp 103-108 (2002).

\bibitem {Burger}Burger, E and Freund, J E. \textit{Introduction to the theory
of games}. Prentice-Hall Inc. Englewood Cliffs N. J. (1963).

\bibitem {DaiChen}Liang Dai and Qing Chen, Comment on ``Quantum cooperative
games'' Physics Letters A, Volume \textbf{328}, Issues 4-5, Pages 414-415 (2
August 2004).

\bibitem {Tirole}J. Tirole. \textit{The theory of industrial organization}.
Cambridge: MIT Press (1988).

\bibitem {Stackelberg}H. von Stackelberg. \textit{Marktform und
Gleichgewicht.} Vienna: Julius Springer (1934).

\bibitem {Gibbons}R. Gibbons. \textit{Game Theory for Applied Economists}.
Princeton University Press (1992).

\bibitem {Du1}Jiangfeng Du, Xiaodong Xu, Hui Li, Mingjun Shi, Xianyi Zhou,
Rongdian Han. Quantum Strategy Without Entanglement. \texttt{quant-ph/0011078}.

\bibitem {Hecht}Eugene Hecht. \textit{Optics}. 2nd edition. Addison-Wesley
Publishing Company (1987).

\bibitem {Debroglie}See for example Richard L. Liboff. \textit{Introductory
Quantum Mechanics}, 2nd edition. Addison-Wesley publishing company (1993).

\bibitem {Feynman}Richard P. Feynman. \textit{Lectures on Physics}. Vol 3.
Addison-Wesley publishing company (1970).

\bibitem {Du}Jiangfeng Du, Hui Li, Xiaodong Xu, Mingjun Shi, Jihui Wu, Xianyi
Zhou, Rongdian Han. Phys. Rev. Lett. \textbf{88}, 137902 (2002).

\bibitem {Du2}Jiangfeng Du, Hui Li, Xiaodong Xu, Mingjun Shi, Xianyi Zhou,
Rongdian Han. J. Phys. A \textbf{36}, 6551 (2003). quant-ph/0111138

\bibitem {iqbalweigert}A. Iqbal and S. Weigert, Quantum Correlation Games. J.
Phys. A: Math and Gen, volume \textbf{37}, issue 22, pp 5873 - 5885 (2004).
\end{thebibliography}
\end{document}